\documentclass{cernyrep} 
\usepackage{texnames}
\usepackage[T1]{fontenc}
\newcommand{\noi}{\noindent}
\newcommand{\nn}{\nonumber}
\newcommand{\barr}{\begin{array}}
\newcommand{\earr}{\end{array}}
\newcommand{\adpsi}{\overline{\psi}}
\newcommand{\adnu}{\overline{\nu}}
\newcommand{\adl}{\overline{l}}
\newcommand{\adu}{\overline{u}}
\newcommand{\add}{\overline{d}}
\newcommand{\lyu}{{\cal L}_Y}
\newcommand{\plus}{\phi^+}
\newcommand{\mis}{\phi^-}
\newcommand{\nul}{\phi^0}
\newcommand{\Gmu}{G_F}
\newcommand{\siw}{\sin^2\theta_W}
\newcommand{\cow}{\cos^2\theta_W}

\newcommand{\cosw}{\cos\theta_W}
\newcommand{\mz}{M_Z^2}
\newcommand{\mw}{M_W^2}
\newcommand{\ipi}{\frac{i}{16\pi^2}}
\newcommand{\dkm}{\mu^{4-D} \int \frac{{\rm d}^Dk}{(2\pi)^D} }
\newcommand{\Dkk}{[k^2-m_1^2+i \varepsilon]}
\newcommand{\Dkq}{[(k+q)^2-m_2^2+i \varepsilon]}
\newcommand{\m}{\mu}
\newcommand{\eps}{\epsilon}
\newcommand{\veps}{\varepsilon}
\newcommand{\g}{\gamma}

\def\NPB{{\em Nucl.\ Phys.}}
\def\PLB{{\em Phys.\ Lett.}}
\def\PRL{{\em Phys.\ Rev.\ Lett.}}
\def\PRD{{\em Phys. Rev.}}
\def\ZPC{{\em Z.\ Phys.}}

\begin{document}
\title{Quantum field theory and the Standard Model}

\author{W.~Hollik}

\institute{Max Planck Institut f\"ur Physik, Munich, Germany}

\maketitle 

\begin{abstract}
In this lecture we discuss the basic ingredients for 
gauge invariant quantum field theories.
We give an introduction to  the elements of quantum field theory,
to the construction of
the basic Lagrangian for a general gauge theory,
and proceed with the formulation of
QCD and the electroweak Standard Model 
with electroweak symmetry breaking via the Higgs mechanism.
The phenomenology of $W$ and $Z$ bosons is discussed and
implications for the Higgs boson are derived from comparison
with experimental precision data.  
\end{abstract}
 
\section{Introduction}
\label{sec:intro}

Relativistic quantum field theory is the adequate theoretical framework
to formulate the commonly accepted theory of the fundamental interactions,
the Standard Model of the strong and the
electroweak interactions~\cite{QCD,gsw,gim,cabibbo}.
The Standard Model summarizes our present knowledge of the basic
constituents of matter and their interactions. 
It is a gauge invariant quantum field theory based on
the symmetry group $SU(3)\times SU(2)\times U(1)$, 
with the colour group $SU(3)$ for the strong interaction and
with $SU(2)\times U(1)$ for the electroweak interaction
spontaneously broken by the Higgs mechanism.
 The renormalizability of this class of theories
allows us to make precise predictions for measurable quantities 
also  in higher orders of the perturbative expansion, in terms of a
few input parameters.
The higher-order terms
contain the self-coupling of the vector bosons as well as their
interactions with the Higgs field and the top quark, 
even for processes at lower energies 
involving only light fermions. 
Assuming the validity of the Standard Model, the presence of the
top quark and the Higgs boson in the loop contributions to electroweak
observables allows us to obtain indirect 
significant bounds on their masses from
precision measurements of these observables. 
The only unknown quantity at present is the Higgs boson.
Its mass is getting more and more 
constrained by a comparison of the
Standard Model predictions with the experimental data, 
preparing the ground for a crucial test at the LHC. 

In these lectures we give an introduction to the basic elements of a 
relativistic quantum field theory in the Lagrangian formulation,
involving scalar, vector, and
fermion fields, and indicate how to calculate amplitudes for
physical processes in perturbation theory with the help
of Feynman graphs.  
The principle of local gauge invariance is explained in terms of the
well-known example of Quantum Electrodynamics (QED) with an Abelian 
gauge symmetry and is then generalized to the case of non-Abelian
gauge invariance and applied to the formulation of
Quantum Chromodynamics (QCD). In the formulation of the electroweak theory 
the gauge principle has to be supplemented by the concept of
spontaneous symmetry breaking with the help of the Higgs field 
and by Yukawa interactions, for embedding massive particles in 
a gauge-invariant way.
Excellent textbooks~\cite{textbooks} are available for further reading.

The presentation of the structure of
the electroweak Standard Model
is followed by a discussion of the phenomenology of $W$ and $Z$ bosons
and of tests of the electroweak theory
at present and future colliders. 
The accurate predictions for  the
vector boson masses,
cross sections, and the $Z$ resonance observables like the width
of the $Z$ resonance, partial widths, effective neutral current
coupling constants and mixing angles at the $Z$ peak,
can be compared with precise experimental data,  
with relevant implications for the empirically 
still unexplored Higgs sector. The present situation of the Higgs sector
and expectations for the upcoming experiments 
are summarized in the final section, together with an outlook on
supersymmetric Higgs bosons.

\section{Elements of quantum field theory}
\label{sec:QFT}

\subsection{Notations and conventions}

Natural units (formally $\hbar=c=1$) are used everywhere.
Lorentz indices are always denoted by greek characters,  
 $ \mu, \nu, .. =0,1,2,3$.
Four-vectors for space--time coordinates and particle momenta have the following
contravariant components,
\begin{align}
x & = (x^\mu) = (x^0, \vec{x}), \quad x^0=t  \, , \nn \\
p & = (p^\mu) = (p^0, \vec{p}\,), \quad p^0=E 
         = \sqrt{\vec{p}^{\,2} +m^2} \, .  \nn 
\end{align}
Covariant 4-vector components are related to the contravariant 
components according to
\begin{align}
a_\mu & = g_{\mu\nu}\, a^\nu, \nn
\end{align}
with the metric tensor 
\begin{align} 
(g_{\mu\nu}) & = \left(
 \begin{array}{c c c c} 
               1& 0& 0& 0 \\
               0& -1& 0& 0 \\
               0& 0& -1& 0 \\
               0& 0& 0& -1 
 \end{array}  
\right)     \nn
\end{align} 
yielding the 4-dimensional squares resp.\ scalar products,
\begin{align}
  a^2 & = g_{\mu\nu}\, a^\mu a^\nu = 
    a_\mu a^\mu, \quad a\cdot b = a_\mu b^\mu 
    = a^0 b^0 - \vec{a}\cdot\vec{b} \, . \nn
\end{align}
Covariant and contravariant components of the
derivatives are used in the following notation,
\begin{align}
\partial_\mu & = \frac{\partial}{\partial x^\mu} = g_{\mu\nu}\, \partial^\nu,
\quad 
  \partial^\nu = \frac{\partial}{\partial x_\nu} \qquad
 [\; \partial^0 =\partial_0, \;\; \partial^k =-\partial_k\; ] \, , \nn \\
 \Box  & = \partial_\mu \partial^\mu = 
           \frac{\partial^2}{\partial t^2} - \Delta \, . \nn
\end{align}
The quantum mechanical states of spin-$s$ particles with momentum 
$p = (p^0, \vec{p})$ and helicity $\sigma = -s, -s+1, \cdots,+s$
are denoted in the conventional way by Dirac kets 
$|p\, \sigma\!\!\!>$. They are normalized according to the relativistically
invariant convention
\begin{align}
\label{eq:normalization}
 <\! p\, \sigma\, |\, p' \sigma' \!> & = 2 p^0 \, 
  \delta^3 (\vec{p} - \vec{p}\,') \, \delta_{\sigma \sigma'} \, .
\end{align}
A special state, the zero-particle state or the vacuum, respectively,  
is denoted by $|0\!>$. 
It is normalized to unity,
\begin{align}
\label{eq:vacnorm}
 <\! 0 \, | \, 0 \!> & = 1 \, .
\end{align}

\subsection{Lagrangian formalism}

The Lagrangian formalism of quantum field theory allows us
to accommodate the following basic features:

\vspace*{-0.3cm}
\begin{itemize}
 \item
space--time symmetry in terms of Lorentz invariance, as well as
internal symmetries like gauge symmetries, 
\item 
causality, 
\item
local interactions.
\end{itemize}

\vspace*{-0.3cm} \noindent
Particles are described by fields that are operators on the 
quantum mechanical Hilbert space of the particle states, acting as
creation and annihilation operators for particles and antiparticles.  
In the Standard Model, the following classes of particles appear, 
each of them described by a specific type of fields:
\begin{itemize}
\item 
spin-0 particles, described by scalar fields  $\phi(x)$,
\item
spin-1 particles, described by vector  fields  $A_\mu(x)$, 
\item
spin-1/2 fermions, described by  spinor fields  $\psi(x)$.
\end{itemize}
The dynamics of the physical system involving a set of fields, 
denoted here by a generic field variable $\phi$,
is determined by the Lorentz-invariant
Lagrangian ${\cal L}$, which yields the action
\begin{align}
 S[\phi]  & = \int {\rm d}^4 x \, {\cal L}\big(\phi(x)\big) \, ,
\end{align}
from which the equations of motions follow as Euler--Lagrange equations
from Hamilton's principle,
\begin{align}
\delta S & = S[\phi + \delta \phi] - S[\phi]   =  0 \, .
\end{align}
In particle mechanics with $n$ generalized coordinates $q_i$ 
and velocities $\dot{q}_i$,
the Lagrangian 
 $L(q_1,\dots  \dot{q}_1, \dots ) $
yields the equations of motion ($i=1,\dots n$)
\begin{align}
\frac{\rm d}{{\rm d} t} \,\frac{\partial L}{\partial {\dot{q}_i}} 
 - \frac{\partial L}{\partial q_i}  & = 0 \, .
\end{align}
Proceeding to field theory, one has to perform the replacement
\begin{align}
q_i & \rightarrow \phi(x) \, ,\quad 
\dot{q}_i \rightarrow \partial_\mu \phi(x)\, , \quad
L(q_1,\dots q_n,  \dot{q}_1, \dots \dot{q}_n) \rightarrow
 {\cal L} (\phi(x), \partial_\mu \phi(x)) 
\end{align}
and obtains the equations of motion as field equations,
\begin{align}
\label{eq:LagrangeEq}
  \partial_\mu \, \frac{\partial{\cal L}}{\partial (\partial_\mu \phi)}
    - \frac{\partial{\cal L}}{\partial \phi} & =  0 \, ,
\end{align}
for each field (or field component), which is indicated here by the
generic variable $\phi$.

\subsection{Free quantum fields}

\subsubsection{Scalar fields}

The Lagrangian for a free real scalar field, describing neutral spinless
particles with mass $m$,
\begin{align}
{\cal L} & = \frac{1}{2}\, (\partial_\mu \phi)^2 - \frac{m^2}{2} \, \phi^2 
\end{align}
yields the field equation according to~(\ref{eq:LagrangeEq}), known as the
{\it Klein--Gordon equation},
\begin{align}
\label{eq:KleinGordon}
( \Box \,  + m^2) \, \phi = 0 \, .
\end{align}
The solution can be expanded in terms of the complete set of plane waves
$e^{\pm ikx}$,
\begin{align}
\phi(x) = \frac{1}{(2\pi)^{3/2}}\,
  \int \frac{{\rm d}^3 k}{2 k^0} \,
 [ a(k) \, e^{-ikx} \,+\, a^\dagger(k)\, e^{ikx} \,  ]
\end{align}
with $k^0 =\sqrt{\vec{k\,}^2 + m^2}$. 
Constituting a quantum field,
the coefficients $a$ and the Hermitian adjoint $a^\dagger$
are operators that
annihilate and create one-particle states
(see Appendix~\ref{sec:AppA}),
\begin{align}
 a^\dagger(k) \,  |0\!>  & =  |k\!>   \nn \\
 a(k)\, |k'\!>  & =  2 k^0 \, \delta^3(\vec{k} - \vec{k}\,') \, |0\!> \, . 
\end {align}
The wave functions of one-particle states are given by the amplitudes
of the field operator  between the one-particle states and the vacuum,
\begin{align}
 <\! 0 | \phi(x) | k> & =  \frac{1}{(2\pi)^{3/2}}\, e^{-ikx} \, , \quad
 <\! k | \phi(x) | 0>  =  \frac{1}{(2\pi)^{3/2}}\, e^{ikx} \, ,
\end{align}
distinguishing between states of incoming (first) and outgoing (second)
particles.

A complex scalar field $\phi^\dagger \neq \phi$
has two degrees of freedom. It describes spinless particles which carry
a charge $\pm 1$ and can be interpreted as particles and antiparticles.
The Lagrangian
\begin{align}
\label{eq:Lscalar}
{\cal L} & = (\partial_\mu \phi)^\dagger (\partial^\mu \phi)
              - m^2 \, \phi^\dagger \phi 
\end{align}
yields the field equation~(\ref{eq:KleinGordon}) as above, but in the 
Fourier expansion one has to distinguish between the annihilation and creation 
operators $a,\, a^\dagger$ for particle states $|+, k\!>$ and $b,\, b^\dagger$
for antiparticle states $|-, k\!>$,
\begin{align}
\label{eq:scalarfieldFourier}
\phi(x) = \frac{1}{(2\pi)^{3/2}}\,
  \int \frac{{\rm d}^3 k}{2 k^0} \,
 [ a(k) \, e^{-ikx} \,+\, b^\dagger(k)\, e^{ikx} \,  ]
\end{align}
where
\begin{equation}
\begin{array}{l l}
 a^\dagger(k) \,  |0\!> \,  = \, |+, k\!> \, , &
 b^\dagger(k) \,  |0\!> \,   = \, |-, k\!>   \\
 a(k)\, |+, k'\!> \, = \,  2 k^0 \, \delta^3(\vec{k} - \vec{k}\,') \, |0\!> 
                      \, ,\qquad\qquad & 
 b(k)\, |-, k'\!> \, = \, 2 k^0 \, \delta^3(\vec{k} - \vec{k}\,') \, |0\!> \, . 
\end{array}
\end {equation}

Whereas wave functions describe free particles without space--time
limitations, the important concept of the {\it propagator} or 
{\it Green function} is required whenever the propagation from  a
point-like source at a given space--time point is considered. 
Such a Green function $D(x-y)$
is a solution of the inhomogeneous field equation
\begin{align}
\label{eq:GreenScalar}
( \Box \,  + m^2) \, D(x-y) & = - \delta^4(x-y) \, .
\end{align}
The solution can easily be determined by a Fourier transformation
\begin{align}
\label{eq:FourierGreen}
D(x-y) = \int \, \frac{{\rm d}^4k}{(2\pi)^4} \, D(k) \, e^{-ik(x-y)} 
\end{align}
yielding Eq.~(\ref{eq:GreenScalar}) in momentum space,
\begin{align}
(k^2  - m^2) \, D(k) & = 1 \, .
\end{align}
The solution
\begin{align}
\label{eq:scalarpropagator}
 i \, D(k) = \frac{i}{k^2 - m^2 + \, i \epsilon}
\end{align}
is the {\it causal Green function} or the {\it Feynman propagator}
of the scalar field. The overall factor $i$ is by convention;
the term $+ i \epsilon$ in the denominator with an infinitesimal 
$ \epsilon > 0$ is a prescription  of how to treat the pole in the 
integral~(\ref{eq:FourierGreen}); it 
corresponds to the special boundary condition of
causality for $D(x-y)$ in Minkowski space, 
which means (see Appendix~\ref{sec:AppB})
\begin{itemize}
\item  propagation of a particle from $y$ to $x$ if $x^0 > y^0$, 
\item propagation of an antiparticle from $x$ to $y$ if $y^0 > x^0$. 
\end{itemize}
In a Feynman diagram, the propagator occurs as an internal line,
whereas wave functions (resp.\ their Fourier transformed in momentum space)
are always associated with external lines representing the physical particles  
in a given process.
We introduce the following graphical symbol for the scalar propagator;
the momentum $k$ always points into the direction of the arrow which
denotes the flow of the charge of the {\it particle}
(for neutral fields the arrrow is irrelevant). 
\begin{center}
\begin{tabular}{l c r}
   $i\, D(k)\quad$  & $\bullet$- - ->- - -$\bullet$ & \\
                &       $k$             & 
\end{tabular}
\end{center}

\subsubsection{Vector fields}

A vector field $A_\mu(x)$
describes particles with spin 1. Their states $|k \lambda \!>$
 can be classified 
by momentum~$k$ and helicity $\lambda = \pm 1, 0$ for massive particles,
and  $\lambda = \pm 1$ for particles with mass zero.

\smallskip {\noindent}
{\bf Massive case.} 
For a given particle mass $m$, the Lagrangian for the free system
(`massive photon'),
\begin{align}
{\cal L} & = -\frac{1}{4}\, F_{\mu\nu} F^{\mu\nu} 
             - \frac{m^2}{2}\, A_\mu A^\mu  \quad {\rm with} \quad
  F_{\mu\nu} = \partial_\mu A_\nu - \partial_\nu A_\mu \, ,
\end{align}
yields from~(\ref{eq:LagrangeEq})
(with $\phi \to A_\nu)$  
the field equation, known as the {\it Proca equation},
\begin{align}
\big[(\Box + m^2)\, g^{\mu\nu} - \partial^\mu \partial^\nu \big] 
\, A_\nu     & = 0 \, . 
\end{align}
Special solutions are plane waves 
\begin{equation}
\label{eq:planewaves}
       \epsilon^{(\lambda)}_\mu \, e^{\pm ikx}
\end{equation} 
with three linearly independent polarization vectors 
$\epsilon^{(\lambda)}_\mu$, which are transverse and can be chosen as
orthogonal and normalized,
\begin{align}
 \epsilon^{(\lambda)}\cdot k & = 0 \, , \quad
 \epsilon^{(\lambda)*} \cdot \epsilon^{(\lambda')} = - \delta_{\lambda\lambda'}
 \, ,
\end{align}
and which fulfil the polarization sum
\begin{align}
\sum_{\lambda=1}^3 \, \epsilon^{(\lambda)*}_\mu  \epsilon^{(\lambda)}_\nu       
 & = - g_{\mu\nu} + \frac{k_\mu k_\nu}{m^2} \, .
\end{align}
The solutions~(\ref{eq:planewaves}) form a  complete set, and the field
$A_\mu$ can be written as a Fourier expansion,
\begin{align}
\label{eq:vectorfieldFourier}
 A_\mu(x) &  = \frac{1}{(2\pi)^{3/2}} \, \sum_\lambda \, 
  \int \frac{{\rm d}^3 k}{2 k^0} \,
 \big[ a_\lambda(k) \, \epsilon^{(\lambda)}_\mu(k)\,  e^{-ikx} \,
 +\, a_\lambda^\dagger(k)\, \epsilon^{(\lambda)}_\mu(k)^* \,  e^{ikx} \,  
    \big] .
\end{align}
The coefficients are the annihilation and creation operators of particle
states,
\begin{eqnarray}
 a_\lambda^\dagger(k) \,  |0\!>  & = &  |k \lambda \!>   \nn \\
 a_\lambda(k)\, |k' \lambda' \!>  & =  &
 2 k^0 \, \delta^3(\vec{k} - \vec{k}\,')\, \delta_{\lambda\lambda'}\, |0\!> \, . 
\end {eqnarray}
As in the scalar case,
the wave functions of one-particle states are given by the amplitudes
of the field operator  between the one-particle states and the vacuum,
\begin{align}
\label{eq:photonwf}
 <\! 0\, | A_\mu(x) |\, k \lambda> & =  \frac{1}{(2\pi)^{3/2}}\, 
                            \epsilon^{(\lambda)}_\mu(k)\,
                            e^{-ikx} \, , \quad
 <\! k \lambda\, | A_\mu(x) |\, 0>  =  \frac{1}{(2\pi)^{3/2}}\, 
                         \epsilon^{(\lambda)}_\mu(k)^* \,
                           e^{ikx} \, ,
\end{align}
corresponding to incoming and outgoing states.
In momentum space, the wave functions are just the polarization vectors.

The {\it Feynman propagator} of the vector field,  
$D_{\mu\nu}(x-y)$, 
is the solution of the inhomogeneous field equation with point-like source,
\begin{align}
\big[(\Box + m^2)\, g^{\mu\rho} - \partial^\mu \partial^\rho \big] \,
  D_{\rho\nu} (x-y) & = g^\mu_{\;\, \nu} \, \delta^4(x-y) \, .
\end{align}
By Fourier transformation,
\begin{align}
\label{eq:vecpropFourier}
 D_{\rho\nu} (x-y) & =
 \int \, \frac{{\rm d}^4k}{(2\pi)^4} \, D_{\rho\nu}(k) \, e^{-ik(x-y)} \, ,
\end{align}
one obtains an algebraic equation for $D_{\rho\nu}(k)$,
\begin{align}
\label{eq:GreenVectorMassive}
\big[(-k^2 + m^2)\, g^{\mu\rho} + k^\mu k^\rho \big] \, D_{\rho\nu}(k)
 & = g^\mu_{\;\, \nu} \, .
\end{align}
The solution is the Feynman propagator of a massive vector field,
\begin{align}
\label{eq:vectorpropagator}
i\, D_{\rho\nu}(k) & = \frac{i}{k^2-m^2 +i  \epsilon} 
 \left( -g_{\nu\rho} + \frac{k_\nu k_\rho}{m^2} \right) .
\end{align}
As for the scalar propagator in~(\ref{eq:scalarpropagator}), the factor $i$
is by convention, and the $+i\epsilon$ term in the denominator
corresponds to the causal boundary condition.

\smallskip
\noindent
{\bf Massless case.} 
For particles with $m=0$, like photons, the field $A_\mu$ corresponds to the
4-potential 
and the Lagrangian is that of the free electromagnetic field,
\begin{align}  
{\cal L} & = -\frac{1}{4}\, F_{\mu\nu} F^{\mu\nu} 
    \quad {\rm with} \quad
  F_{\mu\nu} = \partial_\mu A_\nu - \partial_\nu A_\mu \, .
\end{align}
The field equations are Maxwell's equations for the vector potential,
\begin{align}
\label{eq:MaxwellEqs}
\big(\Box \, g^{\mu\nu} - \partial^\mu \partial^\nu \big) 
\, A_\nu     & = 0 \, . 
\end{align}
There are two physical polarization vectors $\epsilon_\mu^{(1,2)}$
for the transverse polarization, with 
$\vec{\epsilon}^{\,(1,2)}\cdot \vec{k}= 0$.
The third solution of~(\ref{eq:MaxwellEqs}) 
with a longitudinal polarization vector $\epsilon_\mu \sim k_\mu$
is unphysical; it can be removed by a gauge transformation
\begin{align}
 A_\mu'(x) & = A_\mu(x) + \partial_\mu \chi(x) \equiv 0 \quad
 {\rm with} \quad  \chi(x) = \pm i e^{\pm ikx} \, .
\end{align}

\medskip \noindent
The equation for the propagator of the massless vector field follows
 from~(\ref{eq:GreenVectorMassive}) setting $m = 0$:
\begin{align}
\label{eq:GreenVectorMassless}
\big( -k^2 \, g^{\mu\rho} + k^\mu k^\rho \big) \, D_{\rho\nu}(k)
 & \equiv K^{\mu\rho}  \, D_{\rho\nu}(k) 
   = g^\mu_{\;\, \nu} \, .
\end{align}
This equation, however, has no solution since 
$K^{\mu\rho} k_\rho = 0$, i.e., 
$k_\rho$ is an eigenvector  of $K^{\mu\rho}$ with eigenvalue $0$, 
which means that the determinant of 
$K^{\mu\rho}$ vanishes.
It is therefore not straightforward to define a propagator for a
massless vector field. Since the basic reason is gauge invariance,
the common strategy to overcome this problem is to  break the gauge
symmetry by adding to ${\cal L}$ a gauge-fixing term
(which in classical Maxwell theory corresponds to choosing a specific gauge).
Such a term, widely used for practical calculations and corresponding to
the classical Lorentz gauge, has the following form,
\begin{align}
\label{eq:gaugefixing}
{\cal L}_{\rm fix} & = - \frac{1}{2\xi}\, 
           \big(\partial_\mu A^\mu\big)^2 \, ,
\end{align}
where $\xi$ is an arbitrary real parameter, called a
gauge-fixing parameter
(the choice $\xi=1$ defines the {\it Feynman gauge}). 
The accordingly extended Lagrangian
 \begin{align}
 {\cal L} & = -\frac{1}{4}\, F_{\mu\nu} F^{\mu\nu} 
          - \frac{1}{2\xi}\,  \big(\partial_\mu A^\mu\big)^2 
\end{align}
modifies the operator $K^{\mu\rho}$ in momentum space
as follows,
\begin{align}
K^{\mu\rho} & \to  K^{\mu\rho} - \frac{1}{\xi} \,k^\mu k^\rho  \, ,
\end{align}
and~(\ref{eq:GreenVectorMassless}) is replaced by the equation,
\begin{align}
\label{eq:GreenVectorFeyn}
\big[ -k^2 \, g^{\mu\rho} + \big(1-\frac{1}{\xi}\big) k^\mu k^\rho \big] 
\, D_{\rho\nu}(k)
 &  = g^\mu_{\;\, \nu} \, ,
\end{align}
which now has a solution for the massless propagator, namely
\begin{align}
\label{eq:photonpropagator}
i\, D_{\rho\nu}(k) & = \frac{i}{k^2 + i  \epsilon} 
 \left[ -g_{\nu\rho} + (1-\xi) \,\frac{k_\nu k_\rho}{k^2} \right] .
\end{align}
It becomes particularly simple in the Feynman gauge for $\xi=1$.
Note that adding ${\cal L}_{\rm fix}$ to the Lagrangian does not
have a physical impact since the induced extra terms in the propagator
are $\sim k_\nu$ and vanish in  amplitudes for physical processes:
photons always couple to the electromagnetic current $j^\nu$,
which is a conserved current 
with $\partial_\nu j^\nu$, or equivalently 
$k_\nu j^\nu = 0$  in momentum space.

The graphical symbol for the vector-field propagator (for both massive 
and massless) is a wavy line which carries the momentum $k$
and two Lorentz indices:

\begin{center}
\begin{tabular}{l c}
$i\, D_{\rho\nu}(k) $ &
 \includegraphics[width=2cm,clip=]{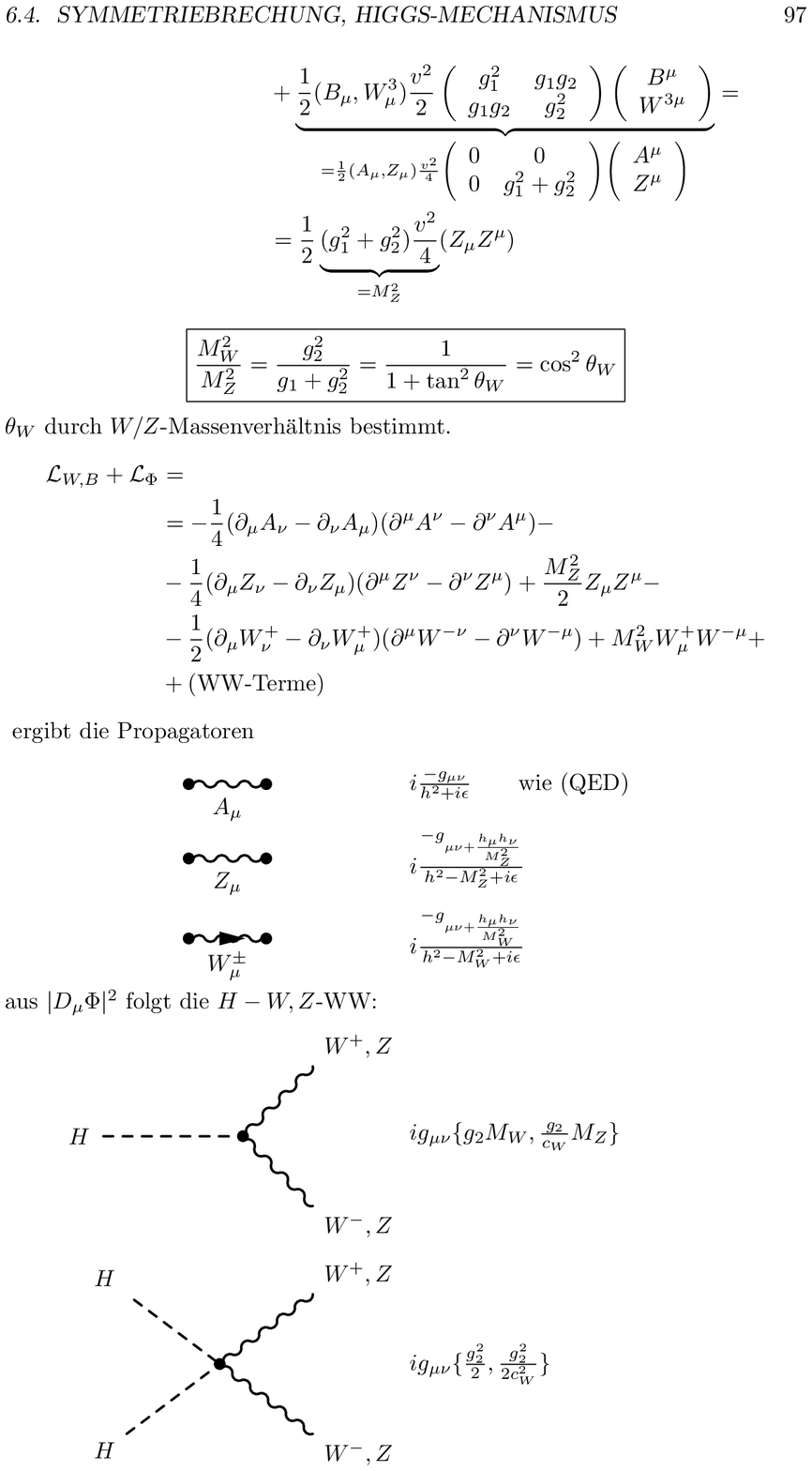} \\
               & $^\rho \qquad k\qquad ^\nu $ 
\end{tabular}
\end{center}

\subsubsection{Dirac fields}

Spin-$\frac{1}{2}$ particles, like electrons and positrons,
with mass $m$ are desribed by 
4-component spinor fields,\begin{equation}
   \psi(x) \, = \,
    \left( 
   \begin{array}{c}
    \psi_1(x) \\   \psi_2(x) \\  \psi_3(x) \\  \psi_4(x) 
    \end{array}  \right) .
\end{equation}
The dynamics of the free field is contained in the Dirac Lagrangian,
\begin{align}
\label{eq:DiracL}
{\cal L} & = \adpsi \,(i \gamma^\mu \partial_\mu - m ) \, \psi \, ,
\end{align}
involving the adjoint spinor 
\begin{align}
\label{eq:adjoint}
\adpsi & = \psi^\dagger\, \gamma^0 = 
           (\psi_1^*, \psi_2^*, - \psi_3^*, - \psi_4^*) \, . 
\end{align}
The Dirac matrices $\gamma^\mu$ {\small{($\mu=0,1,2,3$)}} 
are $4\times 4$ matrices
which can be written with the help of the Pauli matrices  $\sigma_{1,2,3}$
in the following way (the Dirac representation),
\begin{align}
\gamma^0 & =  \left(  \begin{array}{r r}
   {\bf 1} & 0 \\
     0     & - {\bf 1} 
   \end{array} \right), \quad
\gamma^k  =  \left(  \begin{array}{c c}
             0   & \sigma_k \\
    - \sigma_k   &  0
   \end{array} \right) .
\end{align}
They fulfil the anti-commutator relations
\begin{align}
 \{ \gamma^\mu, \gamma^\nu\} 
     &\equiv \gamma^\mu \gamma^\nu + \gamma^\nu \gamma^\mu
    =  2 g^{\mu\nu} \, .
\end{align}
The Lagrangian~(\ref{eq:DiracL}) yields the {\it Dirac equation} as 
the equation of motion,
\begin{align}
\label{eq:DiracE}
( i \gamma^\mu \partial_\mu - m ) \, \psi & = 0 \, .
\end{align}
There are two types of solutions, corresponding to particle
and anti-particle wave functions,
\begin{align}
\label{eq:DiracSolutions}
 u(p)\, e^{-ipx} & \quad  {\rm and}  \quad 
v(p)\, e^{ipx} 
\end{align}
where the spinors $u$ and $v$ fulfil the algebraic equations
\begin{align}
\label{eq:Diracuv}
  \big( \not{\!p} - m) u(p) &  = 0\, , \quad\quad
   \big( \not{\!p} + m) v(p)  = 0 \, . 
\end{align}
Thereby, the notation 
$\not{\!a}  = \gamma^\mu a_\mu$
applying to any 4-vector $a_\mu$ has been used. 
The solutions~(\ref{eq:Diracuv})
correspond to momentum eigenstates with eigenvalue $p^\mu$.
They can further be classified as helicity states
with helicity $\sigma=\pm 1/2$ 
by the requirement
\begin{align}
\frac{1}{2}\, \big(\vec{\Sigma}\cdot\vec{n}\big) \,  u_\sigma(p)
          & = \sigma\, u_\sigma(p) \, , \quad\quad
- \frac{1}{2}\, \big(\vec{\Sigma}\cdot\vec{n}\big) \,  v_\sigma(p) 
             = \sigma\, v_\sigma(p) 
\end{align}
with 
\begin{equation}
\vec{\Sigma}  = \left(
   \begin{array}{l l}
   \vec{\sigma} & 0 \\
            0   & \vec{\sigma} 
   \end{array}    \right)   \quad {\rm and} \quad
   \vec{n} = \frac{\vec{p}}{|\vec{p}|} \, .
\end{equation}
The normalization of the spinors is given by
\begin{align}
\label{uvnorm}
\overline{u}_\sigma \, u_{\sigma'} & = 2 m\, \delta_{\sigma\sigma'} \, , \qquad
\overline{v}_\sigma \, v_{\sigma'} = - 2 m\, \delta_{\sigma\sigma'} \, .
\end{align}
Other useful relations are
\begin{align}
\label{uvcomplete}
  \sum_\sigma\, u_{\sigma} \,\overline{u}_\sigma
   & = \, \not{\! p} + m \, , \qquad
  \sum_\sigma\, v_{\sigma} \,\overline{v}_\sigma
    = \, \not{\! p} - m \, .
\end{align}
Having determined a
complete set of solutions of the Dirac equation~(\ref{eq:DiracE}),
we can
now write the Dirac quantum field
as an expansion in terms of these solutions,
\begin{align}
\label{eq:DiracfieldFourier}
 \psi(x) &  = \frac{1}{(2\pi)^{3/2}} \, \sum_\sigma \, 
  \int \frac{{\rm d}^3 k}{2 k^0} \,
 \big[ c_\sigma(k) \, u_{\sigma}(k)\,  e^{-ikx} \,
 +\, d^{\,\dagger}_\sigma(k)\, v_{\sigma}(k) \,  e^{ikx} \,  
    \big] ,
\end{align}
where the coefficients are annihilation operators
$c_\sigma$ for particles and $d_\sigma$ for anti-particles, as well
as creation operators $c_\sigma^\dagger$ and $d_\sigma^{\dagger}$
for particles and antiparticles, respectively.
In QED, 
electrons $e^-$ are by convention the particles and positrons 
the antiparticles. Choosing the $e^\pm$ field as a concrete example,
we thus have
\begin{equation}
\begin{array}{l l}
 c_\sigma^\dagger(k) \,  |0\!> \,  = \, |e^-, k \sigma \!> \, , &
 d_\sigma^{\,\dagger}(k) \,  |0\!> \,   = \, |e^+, k \sigma \!>   \\
 c_\sigma(k)\, |e^-, k' \sigma'\!> \, = 
                 \,  2 k^0 \, \delta^3(\vec{k} - \vec{k}\,') 
                \delta_{\sigma\sigma'} \, |0\!> 
                      \, ,\qquad & 
 d_\sigma(k)\, |e^+, k \sigma' \!> \, 
          = \, 2 k^0 \, \delta^3(\vec{k} - \vec{k}\,') \, 
              \delta_{\sigma\sigma'} \, |0\!>\, .
\end{array}
\end {equation}
There are four types of wave functions, for incoming and outgoing 
particles and antiparticles, 
\begin{align}
\label{eq:Diracwf}
 <\! 0 | \psi(x) |e^-, k\sigma> & =  \frac{1}{(2\pi)^{3/2}}\, 
                      u_\sigma(k)\, e^{-ikx}   \, , \quad
 <\! e^+,k\sigma | \psi(x) | 0>  =  \frac{1}{(2\pi)^{3/2}}\, 
                     v_\sigma(k) \,e^{ikx} \, , \nn \\
 <\! 0 | \adpsi(x) |e^+, k\sigma> & =  \frac{1}{(2\pi)^{3/2}}\, 
                      \overline{v}_\sigma(k)\, e^{-ikx} \, , \quad
 <\! e^-,k\sigma | \adpsi(x) | 0>  =  \frac{1}{(2\pi)^{3/2}}\, 
                     \overline{u}_\sigma(k) \,e^{ikx} \, .
\end{align}
In momentum space, dropping the $(2\pi)^{-3/2}$ factors 
and the helicity indices,
we describe the situations as follows using a graphical notation
($k$ always denotes the physical momentum
flowing towards an interaction point for incoming 
and off an interaction point for outgoing states),

\begin{center}
\begin{tabular}{l l l}
incoming particle  & $\qquad u(k)$    &   --->---$\bullet$    \\
incoming antiparticle  & $\qquad \overline{v}(k)$  & ---<---$\bullet$   \\
outgoing antiparticle  & $\qquad v(k)$  &   $\bullet$---<---     \\
outgoing  particle  & $\qquad \overline{u}(k)$  &  $\bullet$--->---
\end{tabular}
\end{center}

\noindent
The arrows indicate the flow of the {\it particle} charge.
Note that for antiparticles the direction of the momentum is opposite
to the arrow at the line.

\medskip
We still have to determine the propagator of the Dirac field, which is the
solution of the inhomogeneous Dirac equation with point-like source,
\begin{align}
\label{eq:Diracprop}
( i \gamma^\mu \partial_\mu - m ) \, S(x-y) & =  {\bf 1}\, \delta^4(x-y) \, .
\end{align}
A Fourier transformation to $S(k)$,  
\begin{align}
\label{eq:DiracpropFourier}
  S(x-y) & =
 \int \, \frac{{\rm d}^4k}{(2\pi)^4} \, S(k) \, e^{-ik(x-y)} \, ,
\end{align}
transforms the condition~(\ref{eq:Diracprop}) into 
a condition for $S(k)$
in momentum space,
\begin{align}
( \not{\! k} - m ) \, S(k) & =  {\bf 1} \, .
\end{align}
The solution is a $4\times 4$ matrix,
\begin{align}
\label{eq:DiracPropagator}
i\, S(k) & = \frac{i}{ \not{\! k} - m + i \epsilon} 
           = \frac{i\, (\not{\! k} + m)}{k^2-m^2+ i \epsilon} \, ,
\end{align}
where the $+i\epsilon$ prescription is the causal boundary condition,
as for the scalar and vector field propagators.
We introduce a graphical symbol for the propagator,

\begin{center}
\begin{tabular}{l c r}
   $i\, S(k)\quad$  & $\bullet$--->---$\bullet$ & \\
                &       $k$             & 
\end{tabular}
\end{center}

\noindent
The arrow at the line denotes the flow of the {\it particle} charge;
the assigned momentum $k$ always points into the direction of this arrow.
The propagator appears as an internal line in Feynman diagrams.

\subsection{Interacting fields}

So far we considered only free fields, which are described by Lagrangians
that are quadratic in the field variables and yield linear equations
of motion. 
Interaction terms contain higher monomials in the fields, and a full 
Lagrangian with interaction has the form
\begin{align}
{\cal L} & = {\cal L}_0 + {\cal L}_{\rm int} \, ,
\end{align}
where ${\cal L}_0$ is the free field part
and ${\cal L}_{\rm int}$ describes the interaction.
In general, the resulting non-linear field equations cannot be solved 
in an exact way. The conventional strategy is perturbation theory
with the free fields as starting point, treating the interaction as a small
perturbation. This is justified as long as the interaction is
sufficiently weak.  

A powerful method for obtaining the perturbative 
amplitudes for physical processes is the expansion in terms of
Feynman diagrams.
As a concrete and practically useful example, we consider Quantum 
Electrodynamics (QED), the theory of electron/positron and photon
interactions. The QED Lagrangian is given by
\begin{align}
\label{eq:QEDLangrangian}
{\cal L}_{\rm QED} & = \adpsi (i \gamma^\mu \partial_\mu -m ) \psi
    - \frac{1}{4} F_{\mu\nu} F^{\mu\nu} + {\cal L}_{\rm fix} 
    \, +\, e\, \adpsi \gamma^\mu\psi \, A_\mu \, ,
\end{align}
where the interaction term
\begin{align}
\label{eq:emcurrent}
 {\cal L}_{\rm int} & = j^\mu A_\mu \quad {\rm with} \quad
    j^\mu = e\, \adpsi \gamma^\mu \psi 
\end{align}
describes the coupling of the electromagnetic current
$j^\mu = e\, \adpsi \gamma^\mu \psi $ to the photon field $A_\mu$.
The new element is an interaction point, a {\it vertex}, 
which connects the three fields in ${\cal L}_{\rm int}$ and which is 
obtained by stripping off the field operators, yielding
$e \gamma^\mu$. Also for the vertex, a graphical symbol is introduced
with lines connected to a point:

\begin{center}
 \includegraphics[width=0.24\linewidth,clip=]{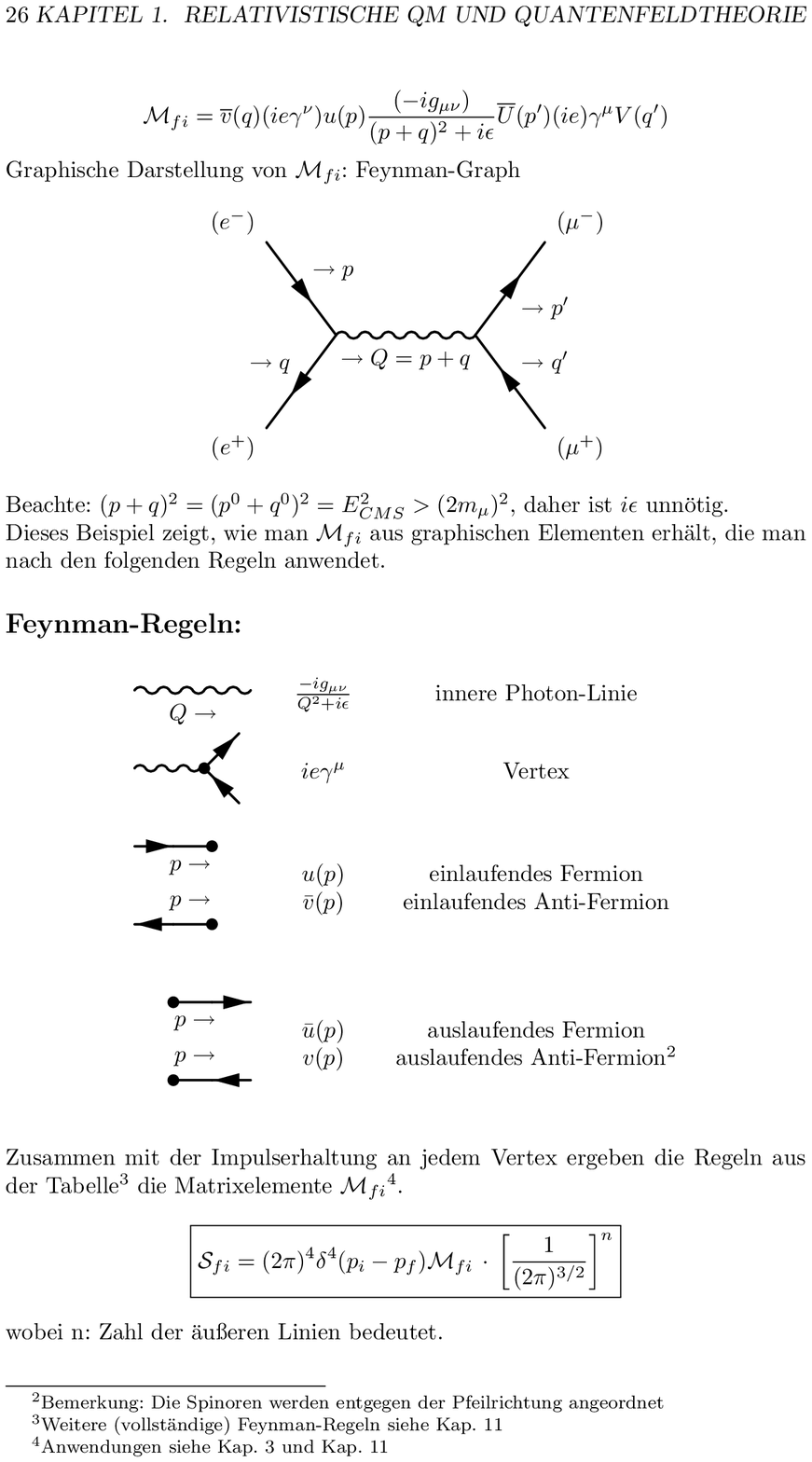}
\end{center}

\noindent
Note that the factor $i$ is a convention.
The lines can be either propagators (internal) or wave functions
(external) in momentum space. They carry momenta which have to
 fulfil momentum conservation. Formally, momentum conservation 
follows via Fourier transformation from
the exponentials in the wave functions~(\ref{eq:photonwf},\ref{eq:Diracwf})
and the propagators~(\ref{eq:vecpropFourier},\ref{eq:DiracpropFourier}) 
when going to momentum space.

Collecting all the information, we give the complete
list of Feynman rules for QED, with the photon propagator
in the Feynman gauge. For fermions different from $e$
(or $\mu, \tau$), an extra factor for the 
different charge appears in the vertex,
as indicated in the brackets. Helicity indices are suppressed for the
wave functions. 

\begin{minipage}{.5\linewidth}
\raggedright
 \includegraphics[width=0.6\linewidth,clip=]{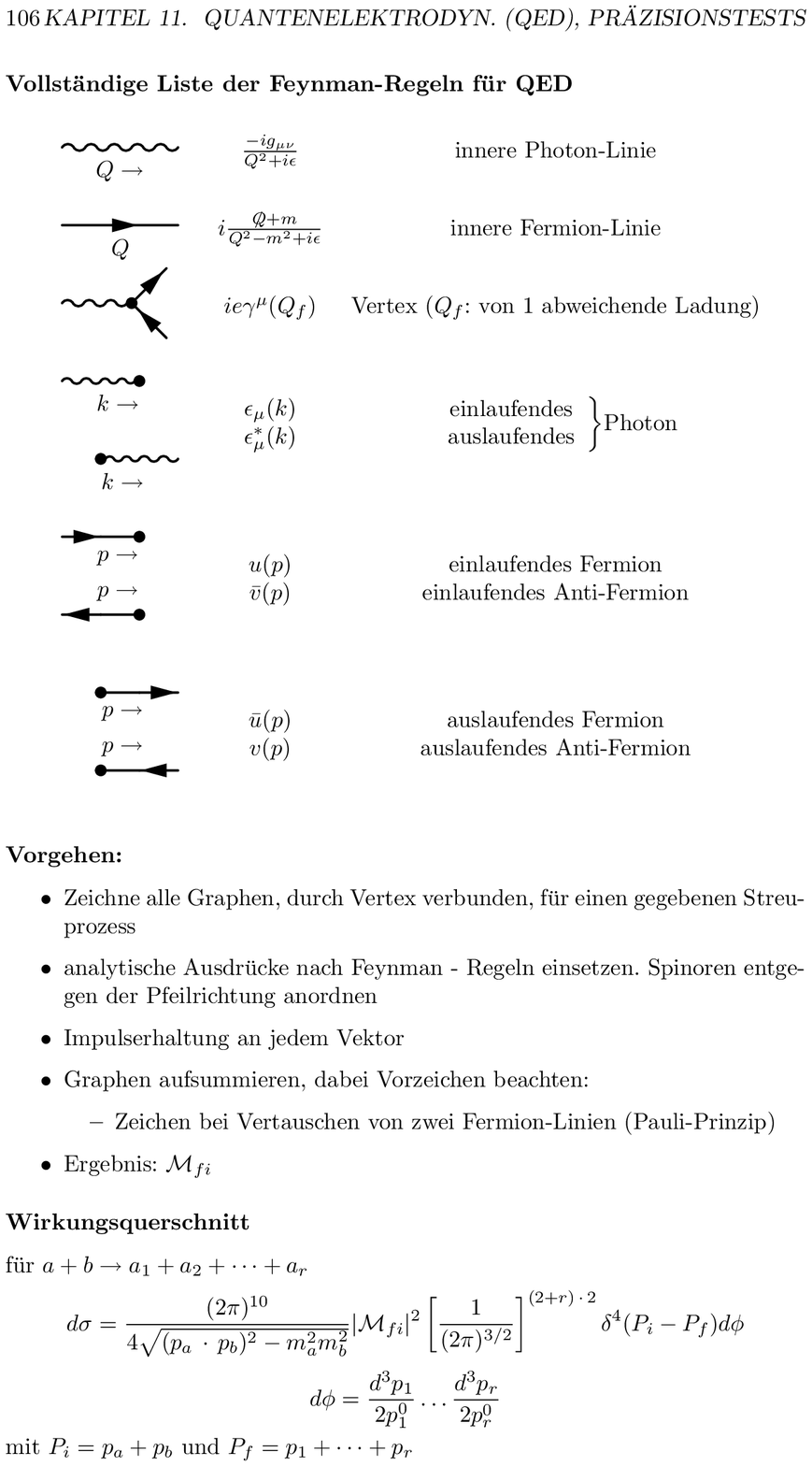} 
\end{minipage}\hfill
\begin{minipage}{.5\linewidth}

\vspace*{-0.1cm}
photon propagator ($\xi=1$)\\[0.8cm]
fermion propagator\\[0.8cm]
electron--photon vertex \\[1.1cm]
incoming photon\\[0.1cm]
outgoing photon\\[1.5cm]
incoming fermion\\[0.1cm]
incoming anti-fermion\\[1.4cm]
outgoing fermion\\[0.1cm]
outgoing anti-fermion
\end{minipage}

\begin{figure}[htb]
\begin{center}
 \includegraphics[width=0.35\linewidth,clip=]{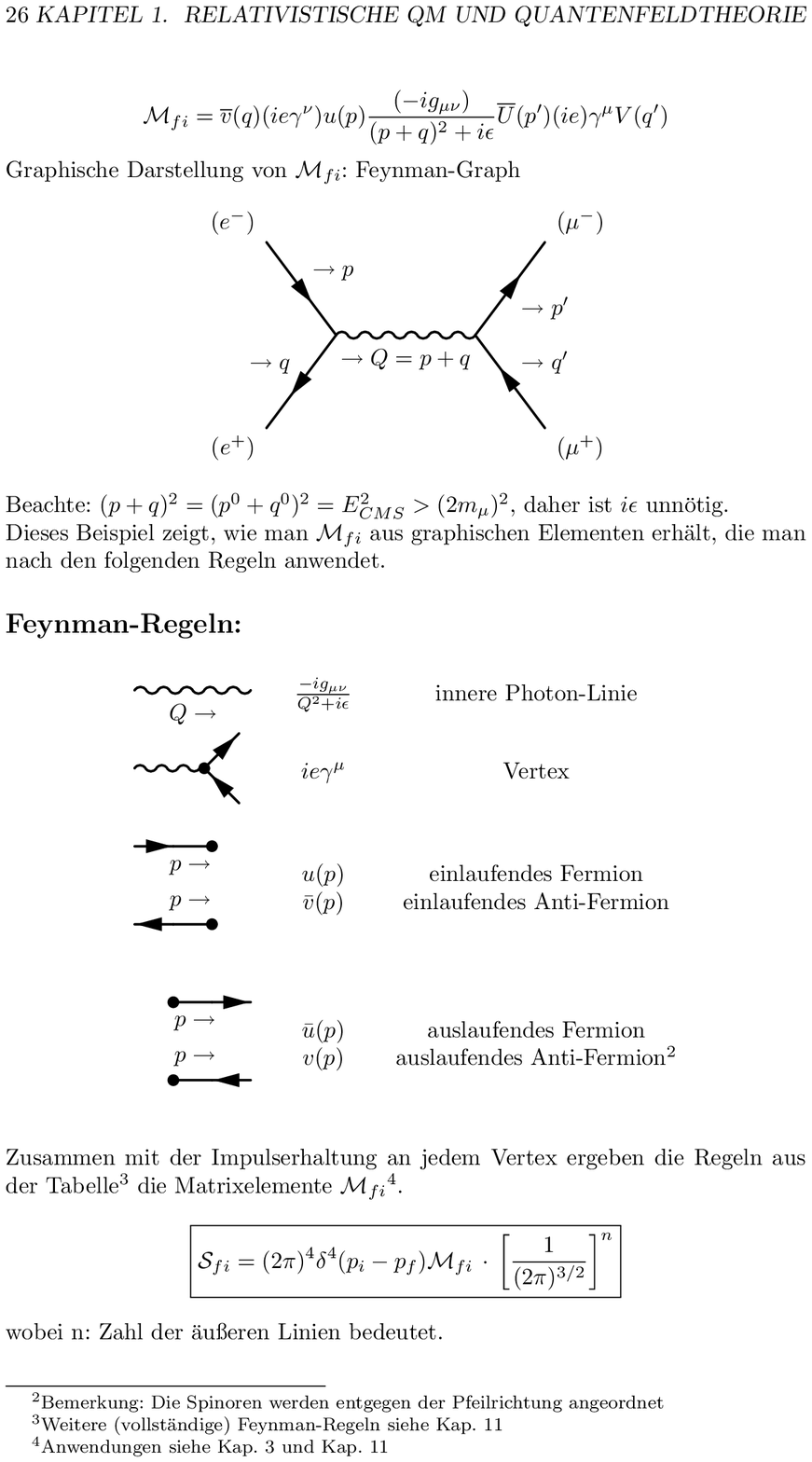}
\end{center}
\caption{Lowest-order Feynman graph for $e^+ e^- \to \mu^+\mu^-$.
The  momenta with directions are indicated at each line.}
\label{Fig:eemumu}
\end{figure}

\noindent
To obtain the transition matrix element,
the amplitude ${\cal M}_{fi}$
for a physical process $|i\!> \to |f\!>$ 
(see Section~\ref{sec:xsec}), one has the following
recipe. 
\begin{itemize}
\item[$\bullet$]
For a process with given external particles 
draw all diagrams connecting the external lines by vertices and propagators.
The lowest order corresponds to diagrams involving 
the minimum number of vertices,
which determines the power of the coupling constant $e$
in the matrix element.  
\item[$\bullet$]
Insert the analytical expressions for the wave functions,
propagators and vertices from the Feynman rules. The arrangement of 
spinors is thereby opposite to the arrow at a fermion line.
\item[$\bullet$]
 Impose momentum conservation at each vertex.
\item[$\bullet$] Sum over all diagrams, paying attention
to the relative sign which occurs when two fermion lines 
are interchanged (according to Pauli's principle). 
\end{itemize}
Note that the factors $(2\pi)^{-3/2}$ from each wave function
are omitted so far. They are collected globally and reappear
in the $S$-matrix element and the cross section, respectively
(Section~\ref{sec:scattering})
We demonstrate the method for the process of
electron--positron annihilation into muon pairs,
$e^+ e^- \to \mu^+\mu^-$.
There is only one Feynman diagram in lowest order,
displayed in Fig.~\ref{Fig:eemumu}.
The analytical expression for the amplitude according to this diagram
is given by
\begin{align}
{\cal M}_{fi} & = \overline{v}(q) i e \gamma^\mu u(p) \,
 \left( \frac{-i g_{\mu\nu} }{Q^2 + i\epsilon} \right) \,
  \overline{u}(p') i e \gamma^\nu v(q') \, = \, i\, 
  \frac{e^2}{Q^2} \; \overline{v}(q) \gamma^\mu u(p) \;
                     \overline{u}(p') \gamma^\nu v(q') \, .
\end{align}
Since $Q^2 = (p+q)^2 >  4 m_\mu^2$, the $i\epsilon$ term in the 
photon propagator is irrelevant and can be dropped.

\begin{figure}[htb]
\begin{center}
 \includegraphics[width=0.8\linewidth,clip=]{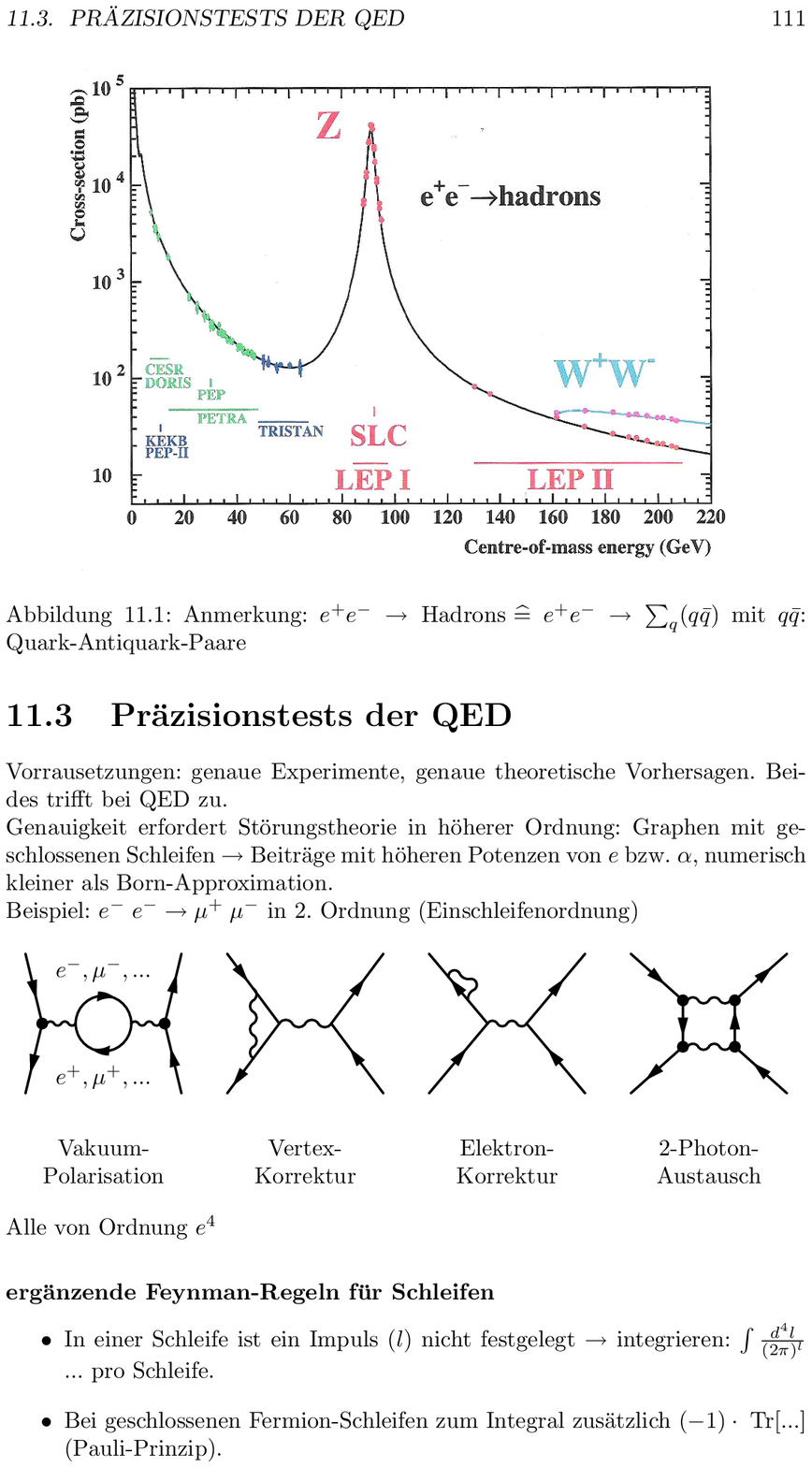}
\end{center}
\caption{One-loop order Feynman graphs for $e^+ e^- \to \mu^+\mu^-$
 (examples)}
\label{Fig:eemumuloop}
\end{figure}

The next-order contribution to ${\cal M}_{fi}$, which is $\sim e^4$,
contains diagrams with closed loops. 
Examples are displayed in Fig.~\ref{Fig:eemumuloop}.
Since inside a loop one momentum is free, 
not fixed by momentum conservation,
loop diagrams involve a 4-dimensional integration over the free momentum
(Section~\ref{subsec:loopcalc}).

\section{Cross sections and decay rates}
\label{sec:xsec}

This section provides the kinematical relations necessary for
getting from the matrix elements for physical processes to observable
quantities, like cross sections and decay rates.

\subsection{Scattering processes}
\label{sec:scattering}

For a given scattering process 
$\quad a + b  \rightarrow b_1 +b_2 + \cdots + b_n $
the $S$-matrix element $S_{fi} = <f|S|i>$  
is the probability amplitude for the transition   
from an initial state $ |a(p_a), b(p_b)\!>\, = |i\!>$ to a final state 
$|b_1(p_1), \cdots b_n(p_n)\! >\, = |f\!> $ of free particles.
For  $ |i\!> \neq |f\!> $, one can write
\begin{align}
 S_{fi} & =  (2\pi)^4 \; \delta^4(P_i-P_f)\; {\cal M}_{fi}  \; (2\pi)^{-3(n+2)/2}
\end{align}
with the $\delta$-function from momentum conservation,
\begin{align}
P_i = p_a+p_b  & =   P_f = p_1 + \cdots +p_n \, , 
\end{align}
the $(2\pi)^{-3/2}$ factors from the normalization of the 
external wave functions,
and with the genuine matrix element ${\cal M}_{fi}$
derived from the Feynman graphs for the scattering process.   
The differential cross section for scattering into the  
Lorentz-invariant phase space element  

\begin{align}
{\rm d}\Phi & = \frac{{\rm d}^3 p_1}{2p_1^0}\, \cdots  
     \frac{{\rm d}^3p_n}{2p_n^0}
\end{align}
is given by
\begin{align}
{\rm d}\sigma & = 
 \frac{(2\pi)^{4}}{4\sqrt{(p_a\cdot p_b)^2-m_a^2 m_b^2} }\;\;
 |{\cal M}_{fi}|^2 \; (2\pi)^{-3n}\; 
\delta^4(P_i-P_f) \; {\rm d}\Phi \, .
\end{align}
The expression in the denominator is the relativistically-invariant version
of the incoming flux-normal\-ization factor.
As a special example of practical importance, we give the cross section
for a two-particle final state 
$a+b \rightarrow  b_1 + b_2$ 
in the centre-of-mass system (CMS), where
$\vec{p}_a+\vec{p}_b = 0 = \vec{p}_1+\vec{p}_2$:
\begin{align}
\frac{{\rm d}\sigma}{{\rm d} \Omega}  & = 
\frac{1}{64 \pi^2 s}\,  \frac{|\vec{p}_1|}{|\vec{p}_a|} \;
 |{\cal M}_{fi}|^2 \,
\end{align}
with $s = (p_a +p_b)^2 = (p_a^0 + p_b^0)^2 $ and the solid angle
${\rm d} \Omega = \sin\!\theta\, {\rm d}\theta\, {\rm d}\varphi$
involving the scattering angle $\theta = \langle \vec{p}_a,\vec{p}_1 \rangle$,
and the azimuth $\varphi$ with respect to the polar axis 
given by $\vec{p}_a$.
For high energies, when the particle masses are negligible,
one has the further simplification $|\vec{p}_1| = |\vec{p}_a|$.

\subsection{Particle decays}
\label{sec:decays}

For a decay process 
$a \rightarrow b_1 +b_2 + \cdots + b_n$
where 
$ |a(p_a)\! >\, = |i\!>, \;\; |b_1(p_1), \cdots b_n(p_n)\! >\,  = |f\!> $,
the (differential) decay width into the phase space element ${\rm d}\Phi$
is given by
\begin{align}
\label{eq:diffwidth}
{\rm d}\Gamma & = 
 \frac{(2\pi)^4}{2\, m_a }\;\;
 |{\cal M}_{fi}|^2 \; (2\pi)^{-3n}\; 
\delta^4(p_a-P_f) \; {\rm d}\Phi\, .
\end{align}
In the special case of a two-particle decay with final-state masses
$m_1=m_2 =m$ one has the simple expression 
\begin{align}
\label{eq:diffwidthmm}
\frac{{\rm d}\Gamma}{{\rm d} \Omega} & = 
\frac{1}{64 \pi^2\,  m_a}\, \sqrt{1-\frac{4 m^2}{m_a^2} }\;
        |{\cal M}_{fi}|^2\, .
\end{align}

\section{Gauge theories}
\label{sec:gaugeinvariance}

The powerful principle of gauge invariance dictates the structure of the
interactions between fermions and vector bosons as well as the 
vector boson self-interactions. It is the generalization of the
Abelian gauge symmetry
found in Quantum Electrodynamics (QED) to the non-Abelian case.

\subsection{Abelian gauge theories --- QED}
\label{subsec:QED}

QED can be derived by the requirement that the global $U(1)$ symmetry
of the Lagrangian for the free charged fermion field $\psi$, i.e.,\ 
the symmetry of 
\begin{align}
\label{eq:free}
{\cal L}_0 & =  \overline{\psi} \, (\gamma^\mu \partial_\mu - m)\, \psi
\end{align} 
under the phase transformation
\begin{align}
 \psi(x) & \rightarrow\, \psi'(x) = e^{i \alpha}\, \psi(x)
\end{align}
for arbitrary real numbers $\alpha$,
can be extended to a symmetry under local transformations
where $\alpha \to \alpha(x)$ is now an arbitrary real function.
This necessitates the presence of a vector field $A_\mu$ 
and the {\it minimal substitution}
of the derivative in ${\cal L}_0$ by the
{\it covariant derivative}  
\begin{align}
\partial_\mu & \to D_\mu = \partial_\mu - i e A_\mu \, ,
\end{align}
yielding a Lagrangian that is invariant under the local gauge transformations
\begin{align}
\label{eq:gaugetrafo}
 \psi(x)\, & \rightarrow\, \psi'(x)\;  =\,  e^{i \alpha(x)} \, \psi(x) \, ,
                                                     \nn\\
 A_\mu(x)\, & \rightarrow\,  A_\mu'(x) =
             A_\mu(x) + \frac{1}{e} \, 
            \partial_\mu \alpha(x)  \, ,
\end{align}
which form the electromagnetic gauge group $U(1)$.
As an immediate consequence, the invariant Lagrangian describes an
interaction of the vector field with the electromagnetic 
current~(\ref{eq:emcurrent}),
\begin{align}
  {\cal L} &  = \overline{\psi} \, ( i \gamma^\mu D_\mu - m) \, \psi \,
=\,  {\cal L}_0 \, +\, e \;  \overline{\psi} \gamma^\mu  \psi \; A_\mu \,
=\, {\cal L}_0 + {\cal L}_{int} \, . 
\end{align}
The vector field  $A_\mu$ itself is not yet a dynamical field since a 
kinetic term is still missing. Such a term can easily be added invoking the
expression well known from classical electrodynamics,
\begin{align}
\label{eq:Akineticterm}
{\cal L}_A & = - \frac{1}{4} \, F_{\mu\nu} F^{\mu\nu} \quad
 {\rm with\; the\; field\; strengths} \quad  
 F_{\mu\nu} = \partial_\mu A_\nu -  \partial_\nu A_\mu \, ,
\end{align}
which is invariant under the local gauge transformation~(\ref{eq:gaugetrafo}). 
$A_\mu$ thus becomes the photon field obeying Maxwell's equations.

\subsection{Non-Abelian gauge theories}
\label{subsec:non-abelian}

The three basic steps yielding QED as the gauge theory of the
electromagnetic interaction:

(i)
identifying the global symmetry of the free Lagrangian,

(ii)
replacing $\partial_\mu$ via minimal substitution by the
covariant derivative $D_\mu$ with a vector field,

(iii)
adding a kinetic term for the vector field, \\[.15cm]
can now be extended to the case of non-Abelian symmetries
as follows.

\bigskip
(i) The given non-interacting system is described by
a multiplet of fermion fields with mass $m$, 
$\Psi = (\psi_1,\psi_2,\dots \psi_n)^{\rm T}$, and the free
dynamics by the Lagrangian
\begin{align}
\label{eq:freeLagrangian}
{\cal L}_0 & = \overline{\Psi}\, (\gamma^\mu \partial_\mu - m)\, \Psi
\quad  {\rm with} \quad  
\overline{\Psi} = (\overline{\psi}_1,\dots \overline{\psi}_n)\, .
\end{align}
${\cal L}_0$ is invariant under global transformations
\begin{align} 
\label{eq:globalU}
\Psi(x) & \rightarrow \, U(\alpha^1,\dots \alpha^N) \Psi(x) \, ,
\end{align}
with unitary matrices $U$ from an $n$-dimensional representation
of a non-Abelian Lie group $G$ of rank $N$, depending on $N$ real
parameters $\alpha^1, \dots \alpha^N$. 
Physically relevant cases are
$G=SU(2)$ and $G=SU(3)$, where the fermion fields $\psi_1,\dots \psi_n$
form the fundamental representations 
with $n=2$ and $n=3$, respectively.     

The matrices $U$ can be written as follows,
\begin{align} 
\label{eq:exp}
U(\alpha^1,\dots\alpha^N) & = 
e^{i(\alpha^1 T_1 + \dots + \alpha^N T_N)} \, , 
\end{align}
with the generators of the Lie group, $T_1, \dots T_N$.
These Hermitian matrices form the Lie algebra
\begin{align}
\label{eq:LieAlgebra} 
[T_a,T_b] = i\, f_{abc} \, T_c
\end{align}
with the structure constants $f_{abc}$ as real numbers 
characteristic for the group. Conventionally, the generators are normalized
according to
\begin{align}
\label{eq:Tnormalization}
 {\rm Tr}\,  (T_a T_b) & = \frac{1}{2} \, \delta_{ab} \, .
\end{align}

\smallskip
(ii) The global symmetry can now be extended to a local symmetry
by converting the constants $\alpha^a$ in~(\ref{eq:exp}) to real functions
$\alpha^a(x)$, $a=1,\dots N$, and simultaneously introducing a covariant 
derivative in~(\ref{eq:freeLagrangian}), via
\begin{align}
\label{eq:covderiv} 
\partial_\mu & \rightarrow D_\mu \, = \, \partial_\mu
   - i g \, {\bf W}_\mu \, , 
\end{align}
involving a vector field ${\bf W}_\mu$, together with a coupling constant $g$
(the analogue of $e$ in QED).
Since $D_\mu$ acts on the $n$-dimensional column $\Psi$, the vector field is 
a $n\times n$ matrix and can be expanded in terms of the generators,
\begin{align}
\label{eq:Wfieldmatrix}
{\bf W}_\mu(x) & = T_a\, W_\mu^a(x) \quad
({\rm summation\; over} \; a = 1, \dots N) \, .
\end{align}
In this way, a set of $N$ fields $W_\mu^a(x)$, the gauge fields, 
enters the Lagrangian~(\ref{eq:freeLagrangian})
and induces an interaction term,
\begin{align}
{\cal L}_0 &\to {\cal L} =
{\cal L}_0 + {\cal L}_{\rm int} \quad {\rm with} \quad 
{\cal L}_{\rm int}  = g \, \overline{\Psi} \gamma^\mu {\bf W}_\mu \Psi \;
     = g\,  \overline{\Psi} \gamma^\mu T_a  \Psi \; W_\mu^a \, ,
\end{align}
which contains the interaction of $N$ currents 
$j^\mu_a = g \overline{\Psi} \gamma^\mu T_a \Psi$
with the gauge fields $W_\mu^a$. 

The local gauge transformation that leaves ${\cal L}$ invariant, 
involves the matrix $U \equiv U(\alpha^1(x), \dots )$
and reads as follows,
\begin{align}
\label{eq:nonabeliangaugetrafo}
\Psi & \rightarrow \, \Psi' = U\, \Psi \, ,  \nn \\
{\bf W}_\mu  & \rightarrow \, {\bf W}'_\mu = 
 U\, {\bf W}_\mu\, U^{-1}
- \frac{i}{g} (\partial_\mu U) U^{-1}  \, . 
\end{align}
The gauge transformation for the vector field looks more familiar when
written for the components and expanded for infinitesimal $\alpha^a(x)$: 
\begin{align}
\label{eq:infinitesimal}
W_\mu^a & \to {W'}_\mu^{\,a} =  
   W_\mu^a + \frac{1}{g} \, \partial_\mu \alpha^a 
  + f_{abc} \, W_\mu^b\, \alpha^c \, .
\end{align}
The derivative term corresponds to~(\ref{eq:gaugetrafo}) in the Abelian case,
the last term is of pure non-Abelian origin.

{\small  
Note: The construction works in the same way for a multiplet of scalar
fields $\Phi = (\phi_1, \dots \phi_n)^T$, with
\begin{align}
\label{eq:scalarLfree}
{\cal L}_0 & = (\partial_\mu \Phi)^\dagger  (\partial^\mu \Phi)
               - m^2 \, \Phi^\dagger \Phi 
  \quad \to \quad  {\cal L} =
 (D_\mu \Phi)^\dagger  (D^\mu \Phi) - m^2 \, \Phi^\dagger \Phi \, .
\end{align} 
}

\smallskip
(iii) The kinetic term for the $W$ fields can be obtained from
a generalization of the electromagnetic field strength tensor $F_{\mu\nu}$
in~(\ref{eq:Akineticterm}),
\begin{align}
{\bf F}_{\mu\nu} & = T_a  F^a_{\mu\nu} =
 \partial_\mu  {\bf W}_\nu -  \partial_\nu  {\bf W}_\mu 
 - i\, g \, [ {\bf W}_\mu, {\bf W}_\nu] \, ,
\end{align}
with the $N$ components
\begin{align}
 F^a_{\mu\nu} & =
 \partial_\mu W^a_\nu -  \partial_\nu W^a_\mu 
 +  g f_{abc} \, W^b_\mu\,  W^c_\nu \, . 
\end{align}
Under the gauge transformation~(\ref{eq:nonabeliangaugetrafo}) the field
strength is transformed according to
\begin{align}
{\bf F}_{\mu\nu} & \to {\bf F}'_{\mu\nu} = U {\bf F}_{\mu\nu} U^{-1} \, .
\end{align}
As a consequence, the trace ${\rm Tr}({\bf F}_{\mu\nu}  {\bf F}^{\mu\nu})$
is gauge invariant,
\begin{align}
{\rm Tr} ({\bf F'}_{\mu\nu}  {\bf F'}^{\mu\nu}) & =
{\rm Tr} (U {\bf F}_{\mu\nu} U^{-1} \, U  {\bf F}^{\mu\nu} U^{-1}) =
{\rm Tr} (U^{-1} U {\bf F}_{\mu\nu} U^{-1} \, U  {\bf F}^{\mu\nu}) =
{\rm Tr} ({\bf F}_{\mu\nu}  {\bf F}^{\mu\nu})  \, ,
\end{align}
and provides the non-Abelian analogue of~(\ref{eq:Akineticterm}) for
the kinetic term of the gauge fields $W_\mu^a$,
\begin{align}   
\label{Wkinetic}
{\cal L}_W &  = - \frac{1}{2} \, {\rm Tr} ({\bf F}_{\mu\nu}  {\bf F}^{\mu\nu})
 = - \frac{1}{4} \, F^a_{\mu\nu}\, F^{a,\mu\nu} \, .
\end{align}
The quadratic part of ${\cal L}_W$  
describes the free propagation of the $W$ fields, but there are also
cubic and quartic terms describing self-interactions of the vector fields
that are determined exclusively through the gauge symmetry:
\begin{align}   
\label{Wkineticexplicit}
{\cal L}_W \,  = & - \frac{1}{4} \,
      (\partial_\mu W_\nu^a - \partial_\nu W_\mu^a) \,
      (\partial^\mu W^{a,\nu} - \partial^\nu W^{a,\mu} ) \nn \\
    & -\frac{g}{2} \, f_{abc} \,  
      (\partial_\mu W_\nu^a - \partial_\nu W_\mu^a)\,  W^{b,\mu}\, W^{c,\nu} \nn \\
    & -\frac{g^2}{4} \, f_{abc} f_{ade} \, 
                       W_\mu^b\,  W_\nu^c\, W^{d,\mu}\, W^{e,\nu} \, .
\end{align} 
In the gauge field Lagrangians ${\cal L}_W$  and ${\cal L}_A$,
the vector fields are strictly massless. Mass terms 
$\frac{m^2}{2} W_\mu^a W^{a,\mu}$ are not invariant under
gauge transformations and thus would break the gauge symmetry.

\section{Formulation of QCD}
\label{sec:physics:sm:formulationQCD}

\begin{figure}[htb]
\begin{center}
 \includegraphics[width=0.67\linewidth,clip=]{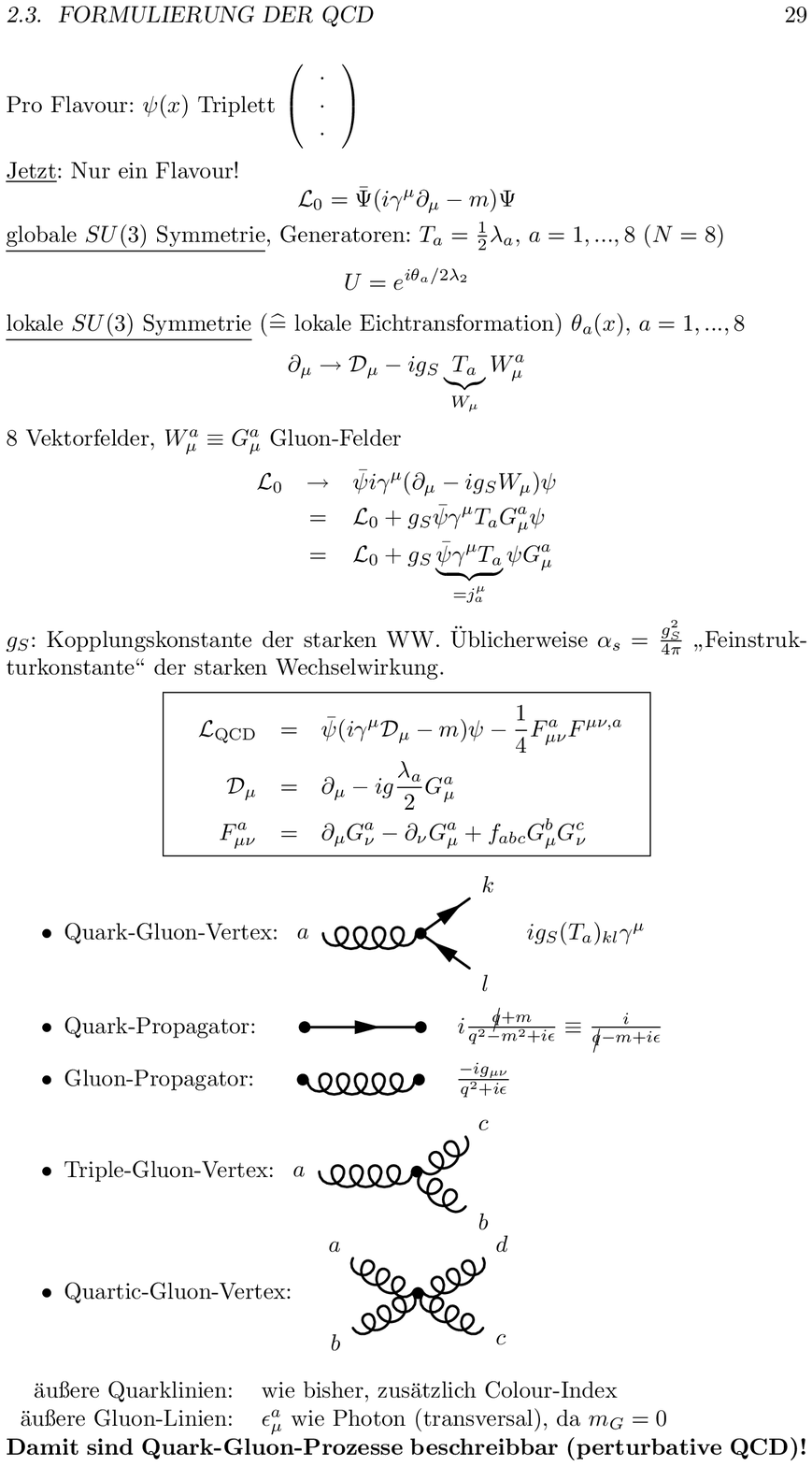} 
\end{center}
\caption{Propagators and interactions in QCD}
\label{Fig:QCDgraphs}
\end{figure}

Quantum Chromodynamics (QCD), the gauge theory of the strong interaction,
is formulated following the principle of the previous section for the
specific case of the symmetry group $G=SU(3)$. 
The basic fermions are quarks in three different colour states, forming 
the fundamental representation of the group.
They are described by triplets of fermion fields $\Psi = (q_1,q_2,q_3)^T$ 
for each  quark flavor $u,\, d, \dots$. 
The colour group
$SU(3)$ has eight generators $T_a$, which in the triplet representation
\begin{align}
 T_a & = \frac{1}{2} \, \lambda_a \, , \quad
         a = 1, \dots 8 \, ,
\end{align}
are expressed in terms of eight $3\times3$ matrices,
the Gell-Mann matrices $\lambda_a$.
The covariant derivative, acting on the quark triplets $\Psi$,
\begin{align}\
D_\mu & =  \partial_\mu - i g_s\, \frac{\lambda_a}{2}\, G_\mu^a \, , 
\end{align}
and the field strengths 
\begin{align}
G_{\mu\nu}^a & = \partial_\mu G_\nu^a - \partial_\nu G_\mu^a
      + g_s\, f_{abc}\, G_\mu^b G_\nu^c \, ,
\end{align}
involve eight gauge fields, the gluon fields $G_\mu^a$, and
the coupling constant of QCD, the strong coupling constant
$g_s$, which is commonly expressed in terms of the 
finestructure constant of the strong interaction,
\begin{align}
 \alpha_s & = \frac{g_s^2}{4 \pi} \, .
\end{align}
The Lagrangian of QCD (for a given species of quarks) can then
easily be written down according to the rules of 
Section~\ref{sec:gaugeinvariance}
(see also Ref.~\cite{Gavan}),
\begin{align}
{\cal L}_{\rm QCD} 
 & =  \overline{\Psi} \, (i \gamma^\mu D_\mu - m) \Psi 
             \, +\, {\cal L}_G  \nn \\
 & =  \overline{\Psi} \, (i \gamma^\mu \partial_\mu - m) \Psi
 \, +\, g_s \, \overline{\Psi} \gamma^\mu \frac{\lambda_a}{2} \Psi \, G^a_{\mu}
   \, -\, \frac{1}{4} \,  G^a_{\mu\nu} G^{a,\mu\nu} \, .
\end{align}
It involves the interaction of the quark currents with the gluon fields
as well as the triple and quartic gluon self interactions as specified 
in~(\ref{Wkineticexplicit}), graphically displayed as 
Feynman rules for QCD in Fig.~\ref{Fig:QCDgraphs}.
There is also a gauge-fixing term in the Lagrangian for each gluon field
(not explicitly written here),
which can be chosen in the same way as for the photon field 
in~(\ref{eq:gaugefixing}) yielding the same form for the
gluon propagators as for the photon propagaor 
in~(\ref{eq:photonpropagator}).

The quark mass $m$ appears in QCD as a free parameter for a given
colour triplet. It is different for different quark flavours;
its origin is of electroweak nature and will be discussed 
in the subsequent section.   

Note that the Lagrangian above considers only a single species 
of flavour. For the realistic physical situation of six flavours, 
one has to introduce a colour triplet for each flavour 
$q=u,d,\dots t$
and to perform
a summation over $q$, with individual masses $m_q$.

\section{Formulation of the electroweak Standard Model}
\label{sec:formulationEW}

The fundamental  fermions,
as families of leptons and quarks with left-handed doublets and
right-handed singlets, 
appear as the fundamental representations of the
group $SU(2)\times U(1)$,
$$ \left( \barr{l} \nu_e \\ e  \earr \right)_L \, , \;\;\;
   \left( \barr{l} \nu_{\mu} \\ \mu   \earr \right)_L \, , \;\;\;
   \left( \barr{l} \nu_{\tau} \\ \tau  \earr \right)_L \, , \;\;\;
   e_R, \;\;\; \mu_R, \;\;\; \tau_R   $$
\begin{equation}
 \qquad\qquad\quad  \left( \barr{l} u \\ d \earr \right)_L, \;\;\;
    \left( \barr{l} c \\ s \earr \right)_L, \;\;\;
    \left( \barr{l} t \\ b \earr \right)_L, \;\;\;
    u_R, \;\;\; d_R, \;\;\; c_R,  \;\;\; 
    s_R, \;\;\; t_R, \;\;\; b_R  
\label{eq:families}
\end{equation}
They can be classified by the quantum numbers of the weak
isospin $I$, $I_3$, and the weak hypercharge $Y$.
Left-handed fields have $I=\frac{1}{2}$ and thus form doublets,
right-handed fields are singlets with $I=0$.
The Gell-Mann--Nishijima relation establishes the relation of these
basic quantum numbers to 
the electric charge $Q$:
\begin{align}
\label{eq:GellmannNishi}
   Q & =  I_3 \, + \, \frac{Y}{2} \, .
\end{align}
The assignment of the quantum numbers to
the fundamental lepton and quark fields is contained 
in Table~\ref{tab:charges} for the fermions of the first generation
(identical for the second and third generation).

\begin{table}[b]
\begin{center}
\caption{Quantum numbers isospin $I_3$ and hypercharge $Y$ for the left- and
       right-handed leptons and quarks, together with the electric charge $Q$}
\label{tab:charges}
\begin{tabular}{l | c c | c | c c | c | c }
      & $\nu_L$ &  $e_L$ & $e_R$ & $u_L$ & $d_L$ & $u_R$ &  $d_R$ \\
\hline
$I_3$ & +1/2    &  -1/2  &  0    & +1/2  & -1/2 &  0    & 0       \\  
$Y$   &  -1     &  -1    &  -2   &  +1/3 & +1/3 & +4/3  & -2/3    \\ 
$Q$   &   0     &  -1    &  -1   &  +2/3 & -1/3 & +2/3  & -1/3    \\ 
\hline
\end{tabular}
\end{center}
\end{table}

This structure can be embedded in a gauge invariant
field theory
of the unified electromagnetic and weak interactions by interpreting
$SU(2)\times U(1)$
as the group of gauge transformations under which the Lagrangian
is invariant. 
The group has four generators,
\begin{align}
T_a & = I_a \; (a=1,2,3) \quad {\rm and} \quad T_4 = Y \, , 
\end{align} 
where $Y$ is the Abelian hypercharge, and $I_a$ are the isospin operators,
forming the Lie algebra 
\begin{align}
 [I_a, I_b] = i \, \epsilon_{abc} \, I_c \, , \quad 
 [I_a, Y] = 0 \, .
\end{align}
This electroweak symmetry has to be broken 
down to the electromagnetic gauge symmetry $U(1)_{\rm em}$, otherwise the
$W^{\pm},\, Z$ bosons would be massless. 
In the Standard Model, this is done by the Higgs mechanism 
in its minimal formulation requiring a
single Higgs field which is a doublet under $SU(2)$.

According to the general principles of constructing a gauge-invariant
field theory with spontaneous symmetry breaking, the gauge,
Higgs, fermion and Yukawa parts of the electroweak Lagrangian
\begin{align}
\label{eq:Lagrangian}
{\cal L}_{\rm EW} & = {\cal L}_G+{\cal L}_H+{\cal L}_F +{\cal L}_Y
\end{align}
are specified in the following way.

\smallskip \noi
{\bf Gauge fields.} 
$SU(2)\times U(1)$
is a non-Abelian  group with generators $I_a, Y$, where 
$I_a\, (a=1,2,3)$ are the isospin operators 
and $Y$ is the hypercharge.
Each of these generalized charges 
is associated with a vector field: a triplet
of vector fields
$W_{\mu}^{1,2,3}$ with $I_{1,2,3}$, and a singlet field $B_{\mu}$
with $Y$.
The isotriplet $W_{\mu}^a$ and the isosinglet  $B_{\mu}$
lead to the field strength tensors
\begin{align}
\label{eq:fieldstrength} 
 W_{\mu\nu}^a & = \partial _{\mu}W_{\nu}^a- \partial_{\nu}W_{\mu}^a
   +g_2 \, \epsilon_{abc} \, W_{\mu}^bW_{\nu}^c , \nonumber \\
   B_{\mu\nu} & = \partial_{\mu}B_{\nu}-\partial_{\nu}B_{\mu}.
\end{align}
Since the gauge group is semi-simple and contains two factors,
there are two independent gauge coupling constants, denoted by
$g_2$ for the non-Abelian factor $SU(2)$ and by
$g_1$ for the Abelian factor $U(1)$. From the field
tensors~(\ref{eq:fieldstrength}) the pure gauge field Lagrangian
\begin{align}
\label{eq:gaugepart}
 {\cal L}_G & = -\frac{1}{4} \, W_{\mu\nu}^aW^{\mu\nu,a}-
     \frac{1}{4}\, B_{\mu\nu}B^{\mu\nu}
\end{align}
is constructed, which is invariant under gauge transformations 
composed of~(\ref{eq:nonabeliangaugetrafo})
and~(\ref{eq:gaugetrafo}).
Explicit mass terms for the gauge fields are forbidden because
they violate gauge invariance. Masses for the vector bosons of the
weak interaction will be introduced  
in a second step below by breaking the electroweak symmetry 
spontaneously with the help of the Higgs mechanism.

\smallskip \noi
{\bf Fermion fields and  fermion--gauge interactions.}  
Since the representations of the gauge group are different for 
fermions with different chirality, we have to distinguish
between the left- and right-handed fields. 
We use the generic notation for the chiral fields, 
\begin{align}
\psi_L & = \frac{1-\gamma_5}{2}\, \psi \, , \quad
\psi_R   = \frac{1+\gamma_5}{2}\, \psi \, .
\end{align}
The left-handed fermion fields of each lepton and quark family
with generation index $j$ are grouped into $SU(2)$ doublets 
and  the right-handed fields into singlets,
 \begin{align}
\psi^j_L & = \left ( \begin{array}{c}
              \psi_{L+}^j \\
              \psi_{L-}^j
              \end{array} \right ) , \quad \psi_{R\sigma}^{j}
\end{align}
with the component index $\sigma = \pm$ denoting $u$-type fermions ($+$)
and $d$-type fermions ($-$).   
Each left- and right-handed multiplet is an eigenstate of the weak
hypercharge $Y$ such that the relation~(\ref{eq:GellmannNishi}) 
is fulfilled 
(see Table~\ref{tab:charges}). 
The covariant derivative
\begin{align}
 D_{\mu}^{L,R} & = \partial_{\mu}\, -\, i\, g_2\, I_a^{L,R} W_{\mu}^a\,
         +\, i\,g_1\, \frac{Y}{2}\, B_{\mu} \, \quad
 {\rm with} \quad I_a^L = \frac{1}{2} \sigma_a \, , 
  \; \;      I_a^R = 0
\label{eq:covderivative}
\end{align}
induces the fermion--gauge field interaction via the minimal
substitution rule,
\begin{align}
\label{eq:fermiongauge}
{\cal L}_F & = \sum_j\,  
  \overline{\psi}^{\,j}_L\, i\gamma^{\mu}D_{\mu}^L\psi_L^j\,
     +\, \sum_{j,\sigma}  \,
  \overline{\psi}_{R\sigma}^{\,j}\, i \gamma^{\mu}D_{\mu}^R \psi_{R\sigma}^{j}
  \, ,
\end{align}
where the index $j$ runs over the three lepton and quark 
generations~(\ref{eq:families}).
Note that the covariant derivatives are different for the $L$ and $R$
fields.

Mass terms are avoided at this stage. They would mix left- and 
right-handed fields as, for example, in 
$m_e (\overline{e}_L e_R + \overline{e}_R e_L)$ and  hence would
explicitly break gauge invariance. They will be introduced later with the help
of gauge-invariant Yukawa interactions of the fermions 
with the Higgs field.
Note that in the genuine Standard Model neutrinos are considered as massless
and there are no right-handed neutrino fields.

\smallskip   \noi
{\bf Higgs field and Higgs interactions.}
Here we describe how
spontaneous breaking of the $SU(2)\times U(1)$ symmetry 
can be obtained, leaving
the electromagnetic gauge subgroup $U(1)_{\rm em}$ unbroken.
For this aim, a single isospin 
doublet of complex scalar fields  
with hypercharge $Y=1$,
\begin{align}
 \Phi(x) & = \left ( \begin{array}{c}
              \phi^+(x) \\ \phi^0(x)
          \end{array} \right ) ,
\label{eq:Higgsfield}
\end{align}
is introduced and 
coupled to the gauge fields via minimal substitution
as indicated in~(\ref{eq:scalarLfree}),
\begin{align}
 {\cal L}_H & = (D_{\mu}\Phi)^\dagger (D^{\mu}\Phi) - V(\Phi) \, ,
\label{eq:HiggsLagrange}
\end{align}
with the covariant derivative for $I=\frac{1}{2}$ and $Y=1$ given by
\begin{align}
 D_{\mu} & = \partial_{\mu}\, -\, i\, g_2\, \frac{\sigma_a}{2}\, W_{\mu}^a\,
         +\,i\,  \frac{g_1}{2}\, B_{\mu}  \, .
\end{align}
The Higgs field self-interaction enters through the Higgs potential
with constants $\mu^2$ and $\lambda $,
\begin{align}
\label{eq:potential}
 V(\Phi) & = -\mu^2\, \Phi^\dagger \Phi + \frac{\lambda}{4}\, 
                     (\Phi^\dagger \Phi)^2 \, .
\end{align}
In the ground state, the vacuum, the potential has a minimum.
For  $\mu^2, \lambda > 0$, the minimum does not occur for
$\Phi=0$; instead, $V$ is minimized by all non-vanishing
field configurations with $\Phi^\dagger \Phi = 2\mu^2/\lambda$.
Selecting the one which is real and electrically neutral, $Q\Phi = 0$,
with 
\begin{align}
\label{Qgenerator}
Q & = I_3 + \frac{Y}{2} = 
      \left( 
        \begin{array}{c c}
       1 & 0 \\
       0 & 0 
         \end{array}  \right) ,
\end{align}
one gets the {\it vacuum expectation value} 
\begin{align}
\label{eq:vacuum}
 <\!\Phi\! > & =  \frac{1}{\sqrt{2}}
             \left( \begin{array}{c}
              0 \\  v  \end{array}  \right) \, \quad  
 {\mbox{with}} \quad        
v=\frac{2\mu}{\sqrt{\lambda}} \; .
\end{align}
Although the Lagrangian is symmetric under gauge transformations
of the full $SU(2)\times U(1)$ group, the vacuum configuration 
$<\!\Phi\! >$ does not have this symmetry: the symmetry has been
{\it spontaneously broken}.
$<\!\Phi\! >$ is still symmetric under transformations of the
electromagnetic subgroup $U(1)_{\rm em}$, which is generated 
by the charge $Q$, thus preserving the electromagnetic gauge symmetry.

The field~(\ref{eq:Higgsfield}) can be written in the following way,
\begin{align}
\label{eq:Higgscomponents}
 \Phi(x) & = \left ( \begin{array}{c}
          \phi^+(x) \\
          \big(v+H(x)+i\chi(x)\big)/\sqrt{2}
          \end{array} \right ) \, ,
\end{align}
where the components $\phi^+$, $H$, $\chi$ have vacuum
expectation values zero.
Expanding the potential~(\ref{eq:potential}) 
around the vacuum configuration in terms of the components 
yields a mass term for $H$, whereas $\phi^+$,
and $\chi$ are massless. 
Exploiting the invariance of the Lagrangian,
the components $\phi^+,\,\chi$ can be eliminated by a suitable gauge
transformation; 
this means that they are unphysical degrees of freedom 
(called Higgs ghosts or would-be Goldstone bosons). 
Choosing this particular gauge where $\phi^+=\chi=0$, 
denoted as the unitary gauge, the Higgs doublet field has
the simple form
\begin{align}
\label{eq:higgsunitarygauge}
\Phi(x) & = \frac{1}{\sqrt{2}} \left(
\begin{array}{c}
  0 \\
  v+H(x)
\end{array} \right) \, ,
\end{align}
and the  potential~(\ref{eq:potential}) reads
\begin{align}
\label{eq:potentialunitary}
V & =  \mu^2 H^2 \, +\, \frac{\mu^2}{v} H^3 \, + \, 
       \frac{\mu^2}{4v^2} H^4 
   \; = \; 
    \frac{M_H^2}{2} H^2 \, +\, \frac{M_H^2}{2v} H^3 \, +
      \frac{M_H^2}{8v^2} H^4 \, .
\end{align}
The real field $H(x)$ thus describes physical neutral scalar
particles, the Higgs bosons, with mass
\begin{align}
 M_H & = \mu\sqrt{2} \, ,
\end{align}
as well as triple and quartic self interactions 
with couplings proportional to $M_H^2$.
The couplings to the gauge fields
follow from the kinetic term of~(\ref{eq:HiggsLagrange})
and give rise to trilinear $HWW,\, HZZ$ and quadrilinear
$HHWW,\, HHZZ$ vertices.

In order to solve the mass problem for the fermions, 
Yukawa interactions between the Higgs field and 
the fermion fields are introduced in addition 
to get the charged fermions massive.
The gauge-invariant Yukawa term in the Lagrangian,
for one family of leptons and quarks, 
is a compact expression in terms of the doublets 
$L_L=(\nu_L,l_L)^T, \, Q_L = (u_L,d_L)^T$ and the Higgs field~$\Phi$ 
and its charge-conjugate 
$\Phi^c = i \sigma_2 \Phi = ( \phi^{0*}, - \phi^-)^T$
with $\phi^-$ as the adjoint of $\phi^+$,
\begin{align}
\label{eq:Yukawa}
\lyu & = - G_l\, \, \overline{L}_L \Phi \, l_R
   - G_d\, \, \overline{Q}_L \Phi \, d_R
   - G_u \, \overline{Q}_L \Phi^c \, u_R   \, + h.c.
\end{align}
It reads explicitly in terms of the Higgs field 
components~(\ref{eq:Higgscomponents}) 
\begin{align}
\label{eq:YukawaOneGeneration}
  \lyu      = & -G_l\,(\adnu_L\,\plus\,l_R\,+\,\adl_R\,\mis\,\nu_L
 \,+\, \adl_L\,\nul\,  l_R\, +\, \adl_R\, \phi^{0*}\,  l_L ) \nn \\
              & -\, G_d \,
 (\adu_L\,\plus\,d_R\,+\,\add_R\,\mis\,u_L\,+\,\add_L\,\nul\,d_R
                    \,+\,\add_R\,\phi^{0*}\,d_L )     \nn  \\
               &    -\,G_u \,
 (-\adu_R\,\phi^+\,d_L\,-\,\add_L\,\phi^-\,u_R\,+\,
  \adu_R\,\phi^0\,u_L\,+\,\adu_L\,\phi^{0*}\,u_R ) \, .
\end{align}
The fermion mass terms follow from the $v$ part of $\phi^0$
in~(\ref{eq:higgsunitarygauge}), relating 
the individual Yukawa coupling constants $G_{l,d,u}$ to the
masses of the charged fermions by 
\begin{align}
 m_f & = G_f\, \frac{v}{\sqrt{2}} \, . 
\end{align}
In the unitary gauge~(\ref{eq:higgsunitarygauge})
the Yukawa Lagrangian becomes particularly simple:
\begin{align}
\label{YukawaUnitaryGauge}
\lyu & = \,
-\sum_f \,m_f\,\overline{\psi}_f \psi_f - \sum_f \frac{m_f}{v}\,
 \overline{\psi}_f\psi_f \, H \, .
\end{align}
As a remnant of this mechanism,
Yukawa interactions between the massive
fermions and the physical Higgs field occur with coupling
constants proportional to the fermion masses.


In the realistic case of three generations, one has to take into account 
flavour mixing in the quark sector (in the lepton sector, lepton number is
conserved and flavour mixing is absent in the minimal model).
Quark-family mixing is induced by Yukawa interactions with the Higgs field
as before, but the Yukawa couplings are now matrices in generation space
with complex entries, $G_u=(G^u_{ij}),\,  G_d = (G^d_{ij})$, and the 
generalization of~(\ref{eq:YukawaOneGeneration}) for the quark sector
reads as follows, with the notation $Q_L^i = (u_L^i,d_L^i)^T$ for
the three left-handed doublets
[$u^i= u,c,t$ and $d^i=d,s,b$]:

\begin{align}
\label{eq:YukawaThreeGenerations}
  \lyu^{\rm quarks} = &  
  - G^d_{ij} \, \overline{Q}_L^{\,i} \Phi \, d_R^j 
  - G^u_{ij} \, \overline{Q}_L^{\,i} \Phi^c \, u_R^j   \, + h.c.
 \end{align}
The mass term is obtained from  replacing  $\Phi$ by its vacuum 
configuration, $\Phi \to <\!\Phi\!>$ from~(\ref{eq:vacuum}),
\begin{align}
\label{eq:quarkmassterm}
 &  - \frac{v}{\sqrt{2}}\,  G^d_{ij} \, \add_L^{\,i} d_R^j \,
   -\, \frac{v}{\sqrt{2}} 
    G^u_{ij} \, \adu_L^{\,i} u_R^j   \, + h.c.
 \end{align}
This bilinear term in the quark fields can be diagonalized with the 
help of four unitary matrices  $V_{L,R}^q$ ($q=u,d$),
yielding the mass eigenstates 
\begin{align}
\label{eq:quarktransformation}
 \tilde{u}_{L,R}^i & = (V^u_{L,R})_{ik}\,  u_{L,R}^k, \quad
 \tilde{d}_{L,R}^i =  (V^d_{L,R})_{ik}\,  d_{L,R}^k \, ,
\end{align}
as well as the $u$- and $d$-type quark masses as diagonal 
mass matrices,
\begin{align}
{\rm diag} (m_q) 
& = \frac{v}{\sqrt{2}}\, 
                  V^q_L\, G_q\,  V_R^{q \, \dagger} \, , \quad q =u,d \, .
\end{align}
Introducing the mass eigenstates in the fermion--gauge Lagrangian~(\ref{eq:fermiongauge}) 
does not change the flavour-diagonal terms,
i.e., the kinetic term and the interaction terms with the neutral
gauge bosons, because of the unitarity of the 
transformations~(\ref{eq:quarktransformation}).
Also the Yukawa interaction of the physical Higgs field 
with the quarks, when expressed in terms 
of the quark masses and the mass eigenstates,
retains its
structure as given in~(\ref{YukawaUnitaryGauge}). 
The only modification occurs in the flavour-changing quark
interaction with the charged vector bosons in~(\ref{eq:fermiongauge})
where the insertion of the 
mass eigenstates for the left-handed quark fields 
introduces the unitary CKM matrix, 
\begin{align}
\label{eq:CKM}
 V_L^u \, V_L^{d \,\dagger} & \equiv \, V_{\rm CKM} \, . 
\end{align} 
Given the constraints  from unitarity,
 $V_{\rm CKM}$ has four independent physical parameters, 
three real angles 
and one complex phase.

For neutrino masses zero, no generation mixing in the lepton sector
occurs. 
It is, however, possible to augment the Standard Model by
introducing also right-handed neutrinos and neutrino
mass terms in analogy
to those of the $u$-type quark sector allowing 
for lepton-flavour mixing as well.
The general treatment of lepton masses and mixing would, however, go beyond
the scope of these lectures
(for a discussion of neutrino masses see Ref.~\cite{Lindner}).

\smallskip  \noi
{\bf Physical fields and parameters.}   
The gauge invariant Higgs--gauge field interaction in the kinetic part
of~(\ref{eq:HiggsLagrange}) gives rise to mass terms for the vector bosons 
in the non-diagonal form
\begin{equation}
\label{eq:massterm}
 \frac{1}{2}\, \left ( \frac{g_2}{2}v \right )^2\,
                     (W_1^2+W_2^2)
 +\frac{1}{2} \left(\frac{v}{2}\right)^2 \, 
  \left ( W_{\mu}^3,B_{\mu} \right )
  \left ( \begin{array}{cc}
       g_2^2  & g_1g_2 \\
       g_1g_2 & g_1^2
       \end{array} \right )
  \left ( \begin{array}{c}
          W^{3,\mu} \\
          B^{\mu}
          \end{array} \right ) \; .
\end{equation}
The physical content becomes transparent by performing a
transformation from the fields $W_{\mu}^a$, $B_{\mu}$ (in terms of
which the symmetry is manifest) to the physical fields
\begin{align}
 W_{\mu}^{\pm} & = \frac{1}{\sqrt{2}}\, (W_{\mu}^1\mp i W_{\mu}^2)
\end{align}
and
\begin{align}
\label{eq:rotation}
 \left(   
 \begin{array}{c}
 Z_{\mu}  \\  A_{\mu}  \end{array}  
 \right)
  &  = 
 \left( 
 \begin{array}{r r}  
 \cos\theta_W &  \quad \sin\theta_W \\
  -\sin\theta_W  & \quad \cos\theta_W 
 \end{array} 
 \right) 
 \left( \begin{array}{c}
  W_{\mu}^3 \\ B_{\mu}   \end{array} \right) \, . 
\end{align}
In these fields the mass term~(\ref{eq:massterm}) is diagonal and has the form
\begin{equation}
 M_W^2\, W_{\mu}^+W^{- \mu}\, +\,
  \frac{1}{2}\, (A_{\mu},Z_{\mu})
  \left ( \begin{array}{cc}
        0 & \quad 0 \\
        0 & \quad M_Z^2
        \end{array} \right )
  \left ( \begin{array}{c}
         A^{\mu} \\
         Z^{\mu}
         \end{array} \right )
\end{equation}
with
\begin{align}
 M_W & =  \frac{1}{2}\, g_2 v \, , \quad
 M_Z  =  \frac{1}{2}\sqrt{g_1^2+g_2^2}\, v \, . 
\end{align}

\noindent
The  mixing angle in the rotation~(\ref{eq:rotation})
is determined by 
\begin{align}
\label{eq:WZmassratio}
 \cos\theta_W & =
 \frac{g_2}{\sqrt{g_1^2+g_2^2}} =\frac{M_W}{M_Z} \, .
\end{align}
Inserting the rotation~(\ref{eq:rotation}) into the interaction part
of ${\cal L}_F$ in~(\ref{eq:fermiongauge}) and
identifying $A_{\mu}$ with the photon field which couples via  the
electric charge $e$ to the electron, $e$ can be
expressed in terms of the gauge couplings in the following way:
\begin{align}
 e & = \frac{g_1g_2}{\sqrt{g_1^2+g_2^2}},  \quad \mbox{or} \quad
 g_2 = \frac{e}{\sin\theta_W},\;
 g_1 = \frac{e}{\cos\theta_W} .
\end{align}
The relations above
allow us to replace the original set of parameters
$
 g_2,\,     g_1,\,  \lambda,\,  \mu^2, \, G_f 
$
by the equivalent set of more physical parameters
$
 e,\,     M_W,\,     M_Z,\,     M_H,\,  m_f,  \, V_{\rm CKM} ,
$
where each of them can
(in principle)  be measured directly in a suitable experiment.
At present, all parameters are empirically known with the exception of 
the mass of the Higgs boson, $M_H$.

\smallskip  \noi
{\bf Gauge interactions.}
The fermion--gauge interactions are part of the fermion--gauge
Lagrangian~(\ref{eq:fermiongauge});
expressed in the physical field and parameters, they 
appear as interactions of the electromagnetic current $J^\mu_{\rm em}$,
the weak neutral current  $J^\mu_{\rm NC}$,
and the weak charged current $J^\mu_{\rm CC}$
with the corresponding vector fields, 
\begin{align}
{\cal L}_{\rm FG} & = J^\mu_{\rm em}\, A_\mu 
                +  J^\mu_{\rm NC}\, Z_\mu 
                + J^\mu_{\rm CC}\, W_\mu^+ 
                + {J^\mu_{\rm CC}}^\dagger\, W_\mu^-   \, ,
\end{align}
with the currents
\begin{align}
J^\mu_{\rm em} & = - e\, \sum_{f=l,q} \, Q_f \,
                \overline{\psi}_f \gamma^\mu \psi_f \, , \nn \\
J^\mu_{\rm NC} & = \frac{g_2}{2\cosw} \, \sum_{f=l,q} \,
  \overline{\psi}_f (v_f \gamma^\mu
                     - a_f \gamma^\mu \gamma_5)  \psi_f \, , \nn \\ 
J^\mu_{\rm CC} & = \frac{g_2}{\sqrt{2}}\, 
                  \left ( \sum_{i=1,2,3} \,
                  \adnu^i \gamma^\mu \frac{1-\gamma_5}{2} e^i 
                +  \sum_{i,j=1,2,3} \,
              \adu^i \gamma^\mu \frac{1-\gamma_5}{2} V_{ij} d^j \right) . 
\label{eq:currents}
\end{align}
In analogy to the notation for the quark fields in~(\ref{eq:YukawaThreeGenerations}),
the lepton families are labelled by $e^i = e,\mu,\tau$
for the charged leptons and $\nu^i = \nu_e,\nu_\mu,\nu_\tau$ for the
corresponding neutrinos.
The neutral current coupling constants in~(\ref{eq:currents}) 
are determined by the charge $Q_f$  and isospin $I_3^f$
of $f_L$,
 \begin{align}
\label{eq:NCcouplingstree}
  v_f & =        I_3^f-2Q_f\,\sin^2\theta_W \, ,   \nn \\
  a_f & =        I_3^f      \, .
 \end{align}
The quantities $V_{ij}$ in the charged current
are the elements of the CKM matrix~(\ref{eq:CKM}),
which describes family mixing in the quark sector.
Owing to the unitarity of $V_{\rm CKM}$, the electromagnetic and the weak
neutral current interaction are flavour-diagonal. Hence,
flavour-changing processes resulting from neutral current interactions
can only occur at higher order; they are mediated by loop contributions
and are consequently suppressed by additional powers of the fine-structure
constant $\alpha$.

Besides the fermion--gauge interactions, the non-Abelian structure
of the gauge group induces self-interactions between the vector bosons.
These gauge self-interactions are contained in the pure gauge-field part~(\ref{eq:gaugepart}) 
of the Lagrangian. Expressing the fields $W^a_\mu$ and
$B_\mu$ in~(\ref{eq:fieldstrength}) resp.~(\ref{eq:gaugepart}) 
by the physical fields $A_\mu$, $Z_\mu$, and
$W^\pm_\mu$  yields a self-interaction term with
triple and quartic couplings, which by use of the notation
$ F_{\mu\nu} = \partial_\mu A_\nu - \partial_\nu A_\mu , \, 
  Z_{\mu\nu} = \partial_\mu Z_\nu - \partial_\nu Z_\mu 
$
can be written in the following way, 
\begin{align}
\label{eq:selfgauge}
 {\cal L}_{\rm G,self}  =\, & e \left[
      ( \partial_\mu W^+_\nu - \partial_\nu W^+_\mu)\, W^{-\mu} A^\nu 
       \,  + \, W^+_\mu W^-_\nu\, F^{\mu\nu} \, +\, h.c. \right] \nn \\
    +\, & e \cot\theta_W  \left[
  (\partial_\mu W^+_\nu - \partial_\nu W^+_\mu)\, W^{-\mu} Z^\nu  
   \, + \, W^+_\mu W^-_\nu\, Z^{\mu\nu}  \, +\,  h.c. \right] \nn \\
 & - e^2/(4\siw)\, 
   [ (W^-_\mu W^+_\nu - W^-_\nu W^+_\mu) W^+_\mu W^-_\nu   + h.c.] \nn \\
 &   - e^2/4 \;  (W^+_\mu A_\nu - W^+_\nu A_\mu) 
     (W^{-\mu} A^\nu - W^{-\nu} A^\mu)                   \nn \\
 &   - e^2/4 \; \cot^2\theta_W \;
     (W^+_\mu Z_\nu - W^+_\nu Z_\mu) 
     (W^{-\mu} Z^\nu - W^{-\nu} Z^\mu)                 \nn \\
&   + e^2/2 \; \cot\theta_W \;
     (W^+_\mu A_\nu - W^+_\nu A_\mu) 
     (W^{-\mu} Z^\nu - W^{-\nu} Z^\mu)   + h.c.                
\end{align}
In the Standard Model the coefficients 
of the self-couplings are exclusively determined by the gauge
symmetry.
Deviations from these values could only be of non-standard origin,
e.g., as remnants from new physics at some higher mass scale.

\section{Electroweak parameters and precision observables}
\label{sec:physics:sm:EWobservables}

Before predictions can be made from the electroweak theory,
the input parameters have to be determined from experiments.
As specified in the previous section, a convenient choice is 
the set of physical parameters given by the particle masses
and the electromagnetic coupling $e$, which is commonly expressed
in terms of the fine-structure constant $\alpha=e^2/4\pi$,
a very precisely known low-energy parameter.
Apart from the flavour sector with the fermion masses and mixing
angles, only three independent quantities are required for 
fixing the input for the gauge sector and the fermion--gauge interactions.
Conveniently, the vector-boson masses $M_{W,Z}$ and $\alpha$ 
are selected (equivalent to $g_1$, $g_2$, $v$).

\subsection{Lowest-order relations}

In the unitary gauge~(\ref{eq:higgsunitarygauge}),
the propagators of the $W$ and $Z$ have the form as given
in~(\ref{eq:vectorpropagator}) for massive vector fields,
but with a finite width $\Gamma$ according to a Breit--Wigner shape 
for unstable particles,
\begin{align}
i\, D_{\rho\nu}(k) & = \frac{i}{k^2-M_{W,Z}^2 + i\, M_{W,Z} \Gamma_{W,Z} }
 \left( -g_{\nu\rho} + \frac{k_\nu k_\rho}{M_{W,Z}^2} \right) .
\end{align}
In processes with light fermions as external
particles, the $k_\rho k_\nu$ terms are negligible since they are
suppressed by powers of $m_f/M_{W,Z}$.
The widths become important around the poles,
i.e., when the vector bosons can be produced on-shell, like in
$e^+ e^-$ annihilation or in Drell--Yan processes in hadron--hadron
collisions. 

A very precisely measured low-energy parameter is the Fermi constant
$\Gmu$, which is the effective 4-fermion coupling constant in
the Fermi model, obtained from the muon lifetime to be~\cite{PDG08}
$\Gmu  = 1.16637(1)\cdot 10^{-5}\, {\rm GeV}^{-2}$.

\begin{figure}[htb]
\begin{center}
 \includegraphics[width=0.4\linewidth,clip=]{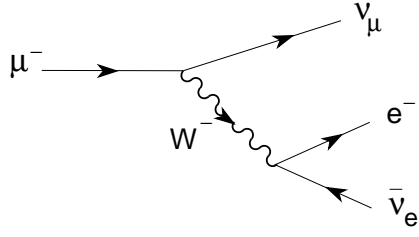}
\end{center}
\caption{Muon decay lowest-order amplitude in the Standard Model}
\label{Fig:mudecay}
\end{figure}

Muon decay is described in the Standard Model in lowest order by 
exchange of a $W$ boson between the fermionic charged currents,
as shown in Fig.~\ref{Fig:mudecay}.
Consistency of the Standard Model  at the muon mass scale
much smaller than
$M_W$, where the momentum in the $W$ propagator can be neglected, 
with the Fermi model requires the identification
\begin{align}
\label{eq:Gfermitree}
\frac{\Gmu}{\sqrt{2}} & = \frac{g_2^2}{8 \mw} =  
      \frac{e^2}{8\siw M_W^2}  =
       \frac{e^2}{8\siw\cow M_Z^2} \, ,
\end{align}
which allows us to relate  the vector
boson masses to the parameters $\alpha,\, \Gmu$, $\siw$ 
and to establish also the $M_W$--$M_Z$ interdependence
in terms of precise low-energy parameters,
\begin{align}
\label{WZcorrelation}
 \mw\left(1-\frac{\mw}{\mz}\right) & =
 \frac{\pi\alpha}{\sqrt{2}\Gmu}\,  \equiv A^2 \, , \quad
 A = 37.2805 \, {\rm GeV} \, .
\end{align}
Moreover, it yields the vacuum expectation value expressed in terms
of the Fermi constant, also denoted as the Fermi scale,
\begin{align}
\label{eq:Fermiscale}
v & = \big( \sqrt{2} G_F \big)^{-\frac{1}{2}} = 246\; {\rm GeV} \, .
\end{align}
The relation~(\ref{eq:Gfermitree}) can be further exploited to  express the 
normalization of the NC couplings in~(\ref{eq:currents})
in terms of the Fermi constant,
\begin{align}
\label{eq:NCnormalization}
 \frac{g_2}{2\cosw} = \big( \sqrt{2} \Gmu \mz\big)^{\frac{1}{2}} .  
\end{align}
In this way, the NC vector and axial vector coupling constants of
each fermion species to the $Z$ are determined and can be used to calculate
the variety of observables at the $Z$ resonance, like $Z$ width and partial
widths, 
\begin{align}
\Gamma_Z & = \sum_f \Gamma(Z\rightarrow f \bar{f}),  \qquad
 \Gamma(Z\rightarrow f \bar{f})  =
   \frac{M_Z}{12\pi} \, (v_f^2 + a_f^2)
\end{align}
and a series of asymmetries, such as forward--backward asymmetries
from the cross sections integrated over the forward ($\sigma_F$)
and the backward ($\sigma_B$) hemisphere,
\begin{align}
 A_{FB} & = \frac{\sigma_F-\sigma_B}{\sigma_F+\sigma_B} 
          =  \frac{3}{4} \, A_e\, A_f \, ,
\end{align}
and the left--right asymmetry from the cross sections $\sigma_{L,R}$
for left- and right-handed polarized electrons,
\begin{align}
 A_{LR} & = \frac{\sigma_L-\sigma_R}{\sigma_L+\sigma_R}  = A_e \, ,
\end{align}
all of them being determined by the ratios
\begin{align}
A_f = \frac{2 v_f a_f }{v_f^2+a_f^2}
\end{align}
with the coupling constants $v_f, a_f$ given 
in~(\ref{eq:NCcouplingstree}). The asymmetries are particularly
sensitive to the electroweak mixing angle $\siw$.


\subsection{Higher-order contributions}

\subsubsection{Loop calculations}
\label{subsec:loopcalc}

These lowest-order relations given above,
however, turn out to be significantly 
insufficient when confronted with the experimental data, which have
been measured with extraordinary accuracy during the LEP and Tevatron era
and require the inclusion 
of terms beyond the lowest order in pertubation theory.
The high experimental precision 
makes the observables sensitive to the quantum structure of the theory
which appears in terms of higher-order contributions 
involving diagrams with closed loops in the Feynman-graph expansion.  
These loop diagrams contain, in general, integrals
that diverge for large integration momenta,  
for example in the self-energy diagrams for a propagator, typically

\vspace*{-0.2cm}
\begin{center}
 \includegraphics[width=0.2\linewidth,clip=]{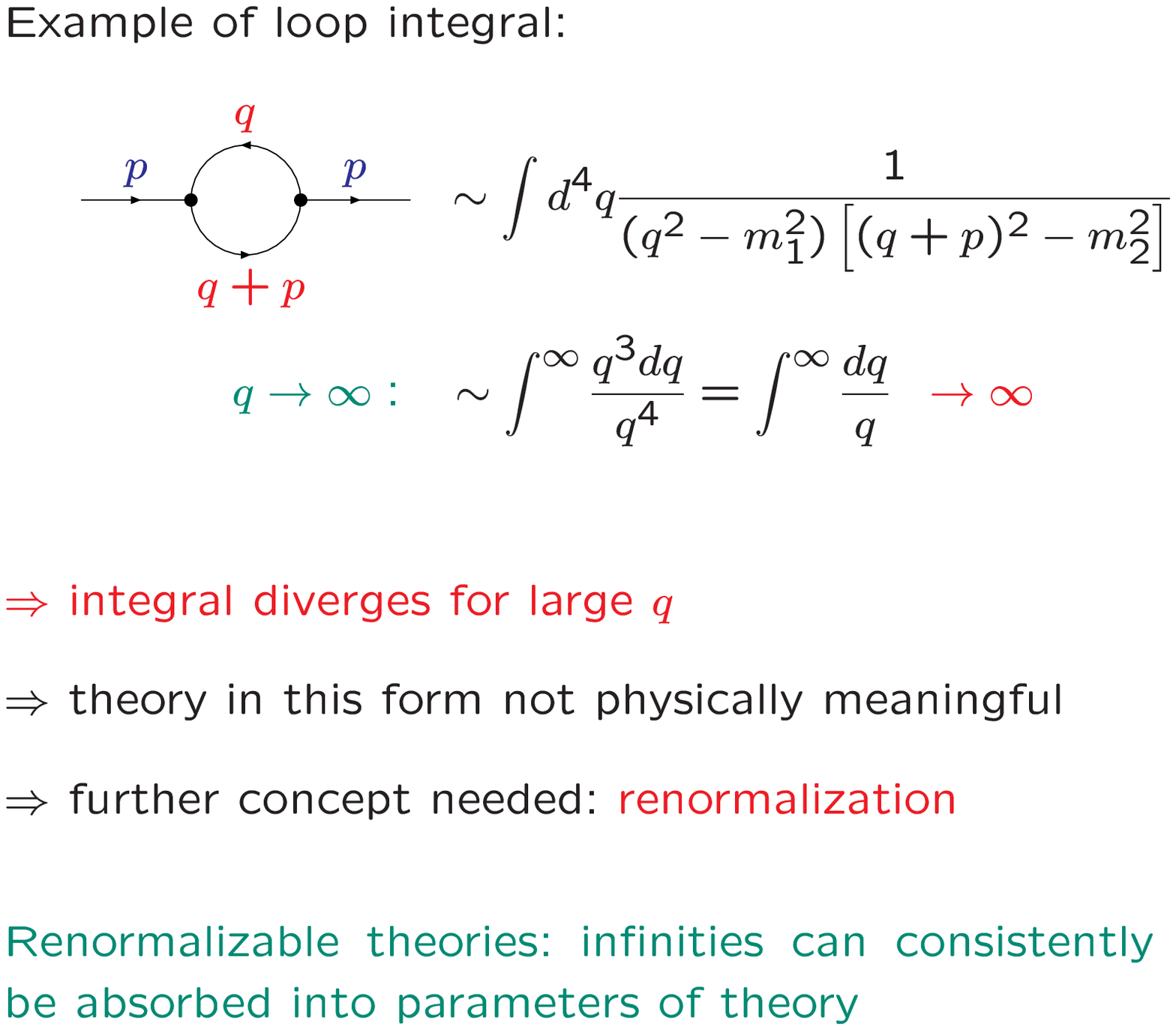}
\end{center}

\vspace*{-0.8cm}
\begin{align}
\label{eq:twopointint} 
 \int\, {\rm d}^4 q \; \frac{1}{(q^2-m_1^2)\,  [(q+p)^2-m_2^2]}\,  \sim\,
  \int\, \frac{{\rm d}^4 q}{q^4} \, \to \, \infty \, .
 \end{align}
Nevertheless, the relations between physical observables
result as finite and testable predictions, owing to the virtue of 
renormalizability.
The possibility to perform such higher-order calculations is based on the 
formulation of the Standard Model as a renormalizable quantum field theory
preserving its predictive power also beyond the tree level.
Renormalizability is thereby guaranteed by local gauge invariance
of the basic Lagrangian.

The first step  to deal with the divergent integrals is a method for
regularization, which is a procedure to redefine the integrals
in such a way that they become finite and mathematically well-defined
objects.
The widely used regularization procedure for gauge theories is that
of dimensional regularization which is Lorentz and gauge invariant:
replace the dimension 4 by a lower
dimension $D$ where the integrals are convergent 
(see Appendix~\ref{sec:AppC}),
\begin{equation}
\label{eq:dimreg}
 \int \, {\rm d}^4 q \quad \to \quad \mu^{4-D} \, \int \, {\rm d}^D q\, . 
\end{equation}
Thereby, an (arbitrary) mass parameter $\mu$ is introduced to
maintain the mass dimensions of the integrals.

The divergences manifest themselves in terms of poles in the dimension
$\sim 1/(4-D)$. In renormalizable theories these divergences
can be absorbed in the basic parameters of the Lagrangian,
like masses and coupling constants.
Formally this procedure, called renormalization, 
is done by introducing a counter term for each parameter
[for example $m^2 \to m^2 + \delta m^2$
for a mass parameter $m$] 
which cancels the singularities; 
the finite part of the
counter terms, however, is not a priori fixed and has to be
defined by a renormalization scheme. The selection of a renormalization 
scheme defines the physical meaning of each parameter and its
relation to measurable quantities. These relations are then
independent of $D$ and thus one can set $D\to 4$.

In pure QCD, considering quarks as massless, 
the only basic parameter is the strong coupling constant $\alpha_s$.
Since there is no intrinsic mass scale, the frequently used scheme is
the $\overline{MS}$ scheme~\cite{MSbar}, 
where the counter term for $\alpha_s$ 
consists only of the singular pole part (together with a  universal
numerical constant). The coupling is then defined  
for the chosen mass scale $\mu$ in~(\ref{eq:dimreg}), the 
renormalization scale, and thus becomes a scale-dependent quantity,
the running coupling constant $\alpha_s(\mu)$
(see Ref.~\cite{Gavan}).

The Lagrangian of the electroweak Standard Model
involves quite a few free parameters which are not
fixed by the theory but have to be taken from experiment. 
In QED and in the electroweak theory, 
classical Thomson scattering and the particle masses
set natural scales where the parameters can be defined. 
A distinguished choice for the basic parameters 
is thus given by the fundamental charge $e$ 
and the masses of the particles, 
$M_Z,M_W,M_H,m_f$,  
and a common choice for the renormalization is the on-shell scheme:
the mass parameters 
coincide with the poles of corresponding propagators (pole masses), 
and the charge $e$ is
defined in the classical limit. 
The on-shell scheme
hence defines the counter terms in the following way
(see, e.g., Ref.~\cite{denneretal} for details):
\begin{itemize}

\item[$\bullet$]
The mass counter term $\delta m^2$, for 
any free mass parameter $m$, is determined by the condition
\begin{align}
\label{eq:massren}
  \delta m^2 = \Sigma (m^2) \, ,
\end{align}
where $\Sigma$ is the  self-energy of the corresponding particle, 
schematically depicted in~(\ref{eq:twopointint}) and yielding 
a dressed propagator
\begin{align}
  \frac{i}{p^2 - (m^2+\delta m^2) + \Sigma(p^2) } \; ,
\end{align}
which by mass renormalization
now includes also the mass counterterm. 
The condition~(\ref{eq:massren}) ensures that $m^2$ still
remains the pole of the propapator.
\footnote{In the $\overline{MS}$ scheme, $\delta m^2$ only absorbs the
divergent part of $\Sigma(m^2)$. 
The remaining finite part depends on the
renormalization scale $\mu$, and in that scheme the mass becomes a
$\mu$-dependent parameter, the running mass $m(\mu)$, which is different
from the pole mass.}

\item[$\bullet$]
The counter term $\delta e$ for the electric charge,
$e \to e +\delta e$,
is determined by the requirement that $e$ be the electron--photon
coupling in the classical limit, i.e., for the electron--photon vertex
for real photons, $k^2=0$, and
for low photon energy,
%
\begin{center}
 \includegraphics[width=0.125\linewidth,angle=270,clip=]{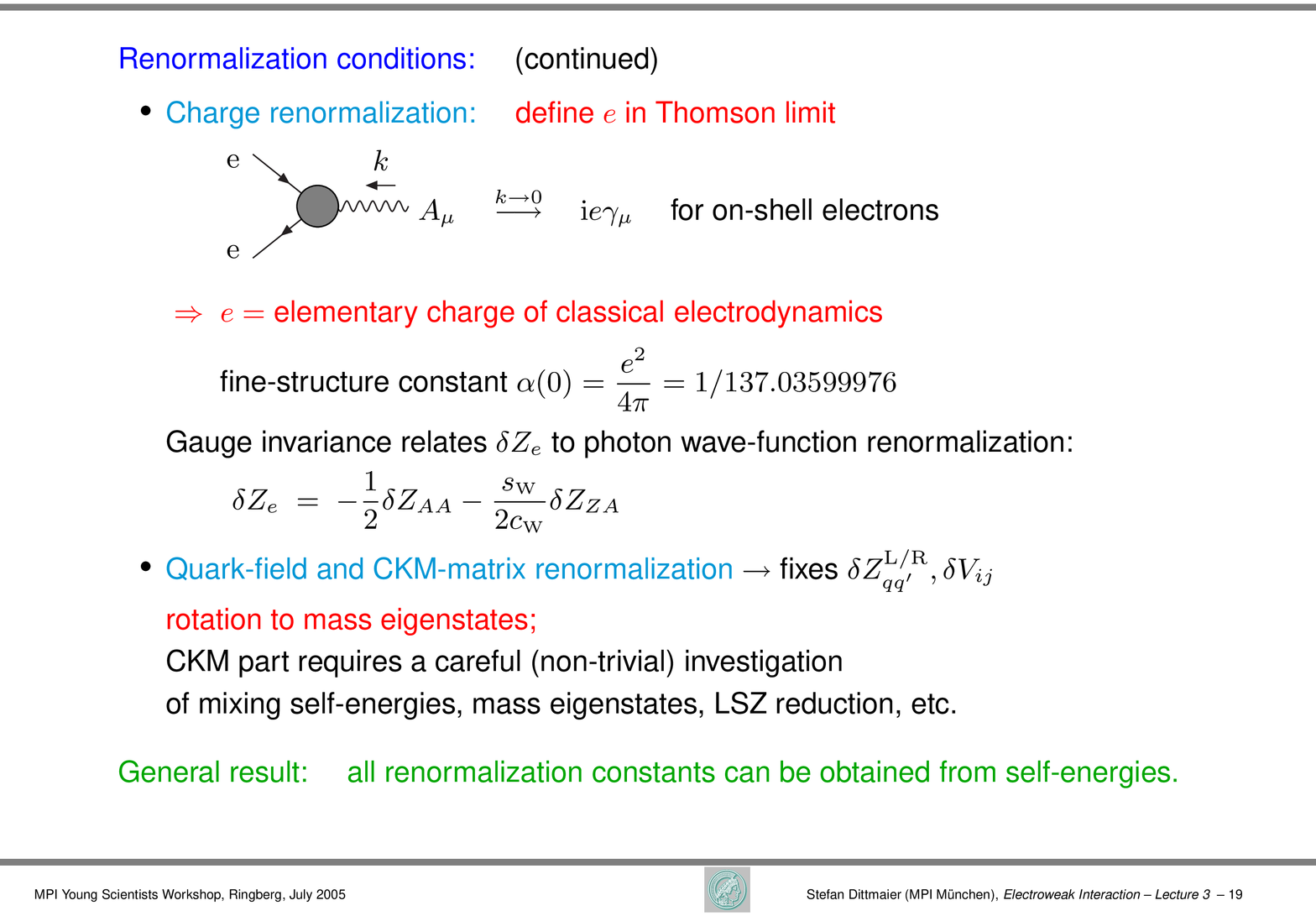}
\end{center}
%

$\delta e$ is essentially given by the
charged-light-fermion contribution to  
the photon vacuum polarization at zero momentum, $\Pi^\gamma(0)$, 
\begin{center}
 \includegraphics[width=0.2\linewidth,clip=]{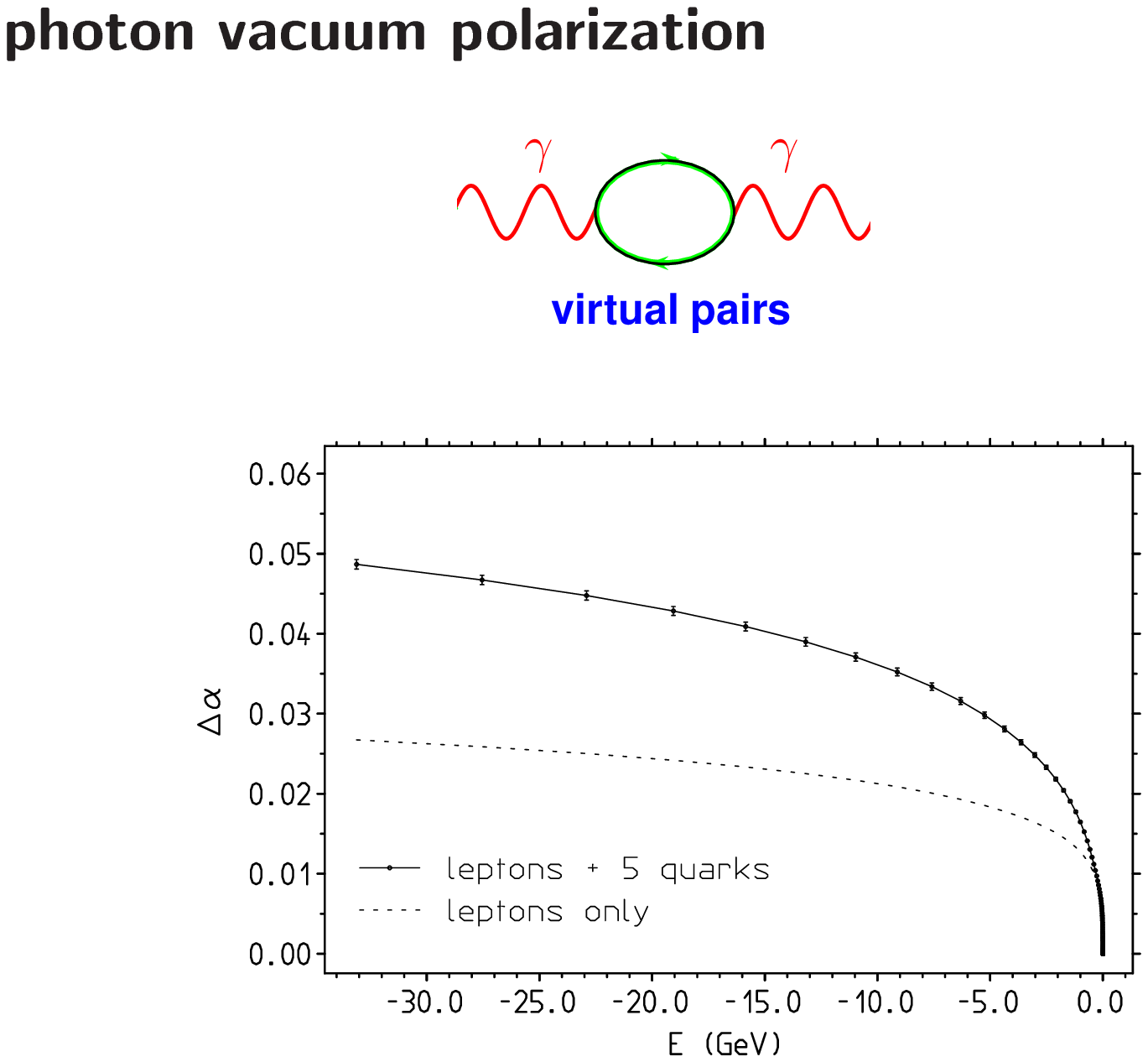}
\end{center}%
which has a finite part 
$\Delta\alpha = \Pi^\gamma(0) - \Pi^\gamma(\mz)$
yielding a shift  
of $\Delta\alpha  \simeq 0.06$ in the electromagnetic fine-structure
constant $\alpha \to \alpha (1+\Delta\alpha)$. 
$\Delta\alpha$ can be resummed according to the renormalization group,
accommodating all the leading logarithms of the type
  $ \alpha^n\log^n(M_Z/m_f)$ from the light fermions.
The result is an effective fine-structure
constant at the $Z$ mass scale
\begin{align}
\label{alphaeff}
   \alpha(\mz) &  = \frac{\alpha}{1-\Delta\alpha} 
                 \simeq \frac{1}{129} \, .
\end{align}
It corresponds to a resummation of the iterated 1-loop vacuum
polarization from the light fermions  to all orders.
$\Delta\alpha$ is an input of crucial importance because of its universality
and remarkable numerical size~\cite{DeltaAlpha,teubneretal}.

\end{itemize}

%

\noi
The loop contributions to the electroweak observables contain all
particles of the Standard Model spectrum, in particular also the Higgs
boson, as, for example, in the vector-boson self-energies
\begin{center}
 \includegraphics[width=0.12\linewidth,angle=270,clip=]{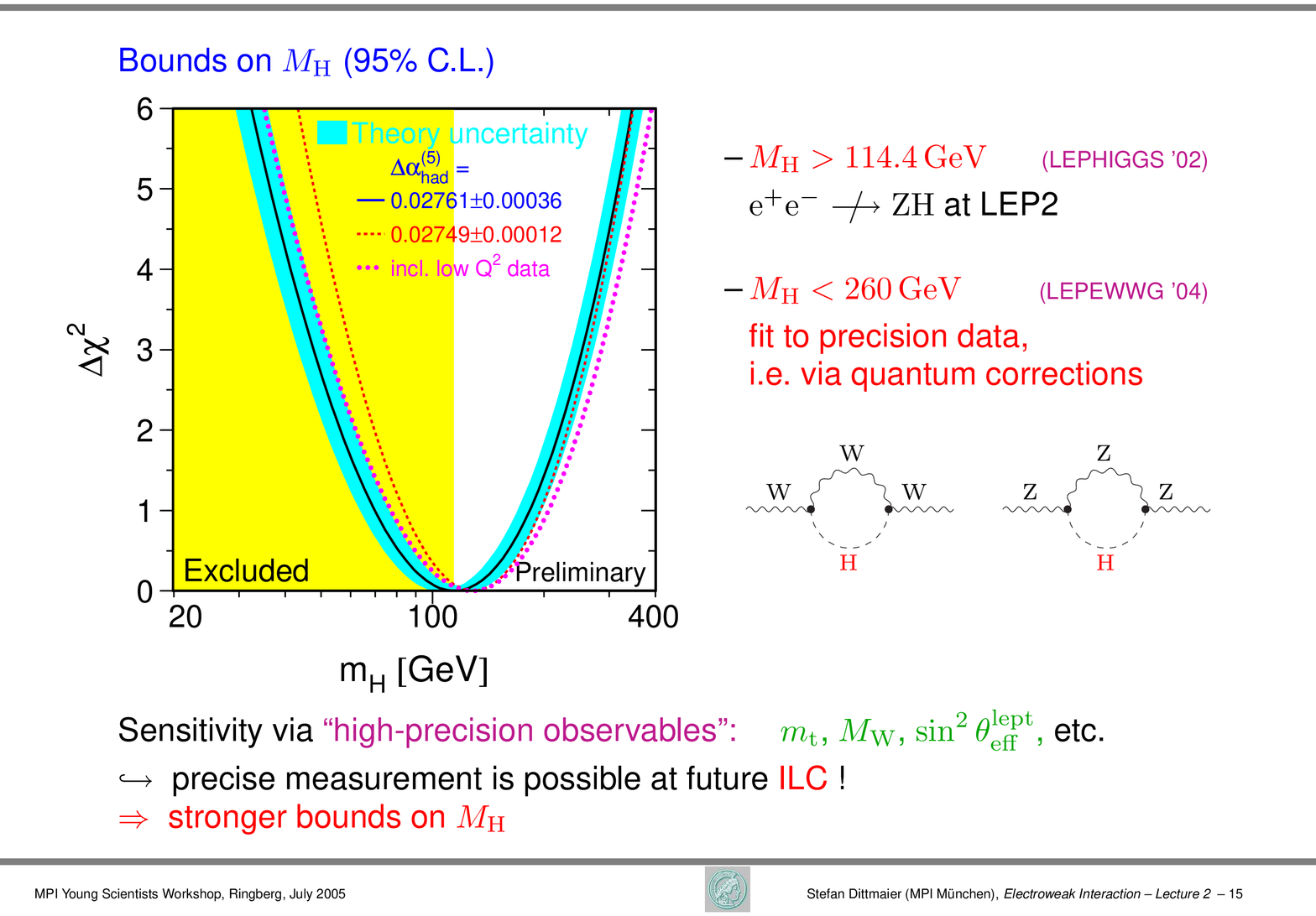}
\end{center}
The higher-order terms thus induce a dependence of the observables
on the Higgs-boson mass $M_H$, which by means of precision measurements
becomes indirectly accessible, although still unknown from direct searches.   
For more details see Ref.~\cite{ewgr} and references therein.

\subsubsection{Vector boson masses and Fermi constant}

\noi
The implementation of higher-order terms can be done in a compact way 
for the $W$--$Z$ mass correlation, 
\begin{align}
\label{eq:mw}
 \mw\left(1-\frac{\mw}{\mz}\right) & = \frac{A^2}{1-\Delta r} \, .
\end{align}

\begin{figure}[hbt]
\begin{center}
 \includegraphics[width=0.15\linewidth,angle=270,clip=]{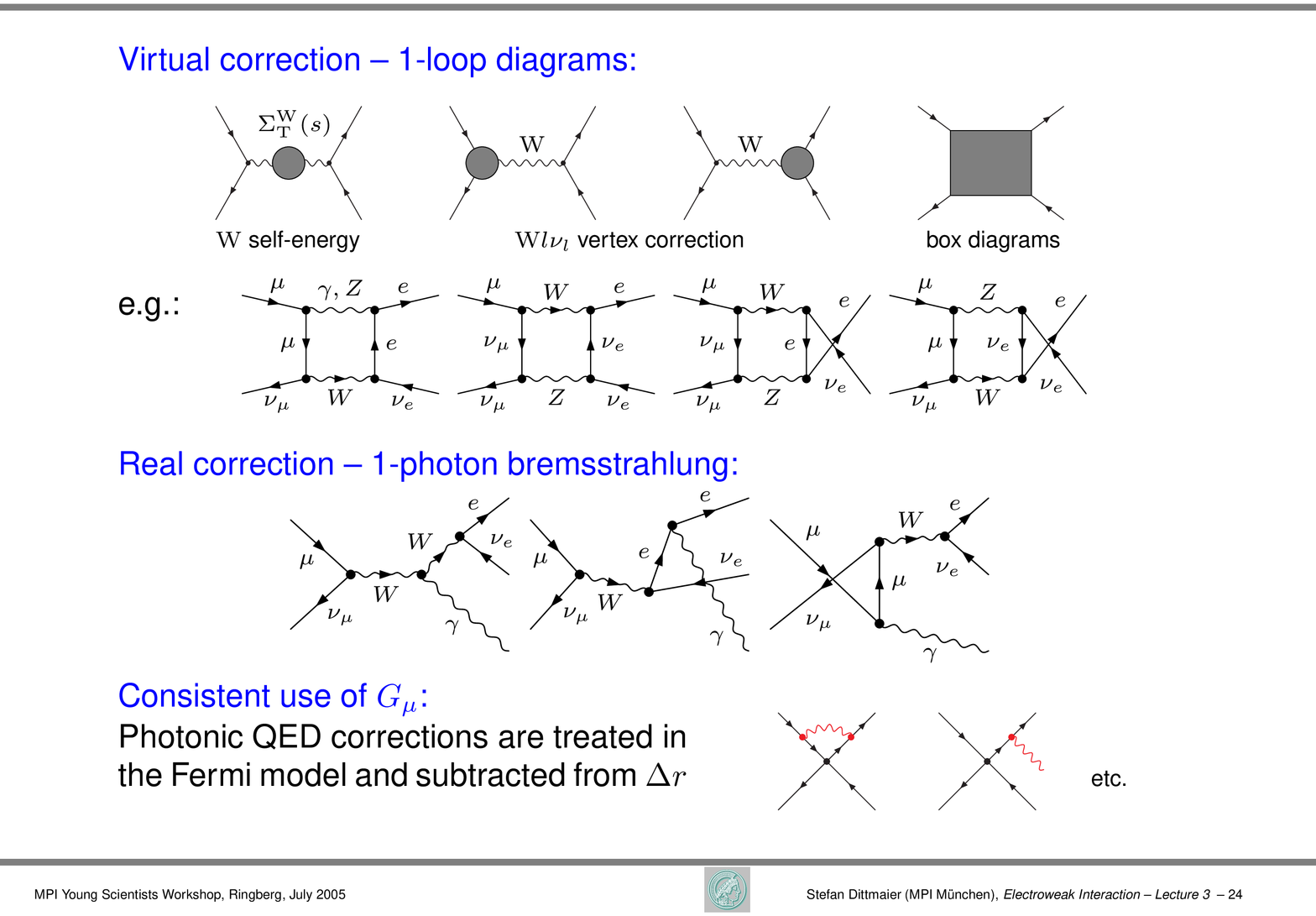}
\end{center}
\caption{Loop contributions to the muon decay amplitude}
\label{Fig:muonloops}
\end{figure}

\begin{figure}[htb]
\begin{center}
 \includegraphics[width=0.55\linewidth,clip=]{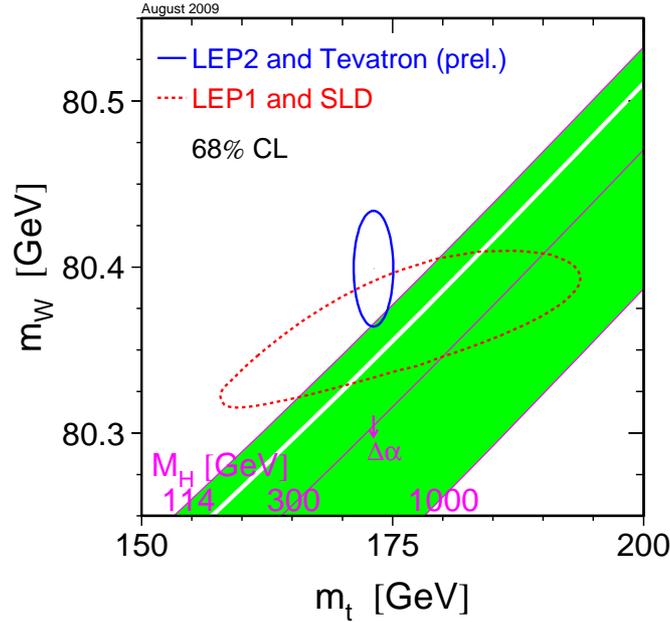}
\end{center}

\vspace*{-1cm}
\caption{Standard Model predictions for the dependence of $M_W$ on the
          masses of the top quark and Higgs boson} 
\label{Fig:mwmt09}
\end{figure}

\noi
Therein, the contributions from the 
loop diagrams to the muon decay amplitude,
schematically depicted in Fig.~\ref{Fig:muonloops}, 
are summarized 
by the quantity $\Delta r = \Delta r(m_t,M_H)$, which 
at one-loop order
depends logarithmically on
the Higgs-boson mass and quadratically on
the top-quark mass. 
The calculation of $\Delta r$ is complete at the 
two-loop level~\cite{HOdeltar}
and comprises the leading terms also at the three- and 
four-loop level~\cite{beyondtwoloop}.
The prediction of $M_W$ from~(\ref{eq:mw}) is shown in
Fig.~\ref{Fig:mwmt09}~\cite{lepewwg}.

\subsubsection{Observables at the Z resonance}

\begin{figure}[htb]
\begin{center}
 \includegraphics[width=0.45\linewidth,clip=]{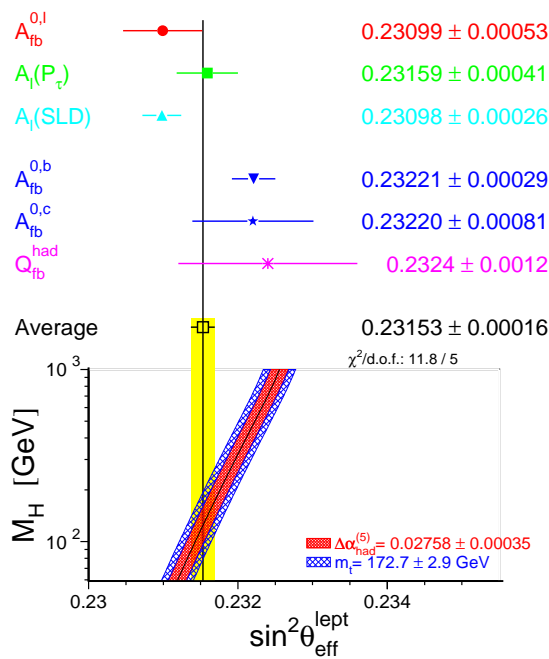}
\end{center}
%
\caption{Standard Model predictions for the dependence of 
          $\sin^2\theta^{\rm lept}_{\rm eff}$
          on the mass of the  Higgs boson and the experimental 
          $1\sigma$-range from averaged measurements done at LEP and SLC} 
\label{Fig:sineffvsMH}
\end{figure}

\noi
The NC couplings dressed by higher-order terms can also be written in 
a compact way, 
replacing the lowest-order couplings~(\ref{eq:NCcouplingstree})
by effective couplings~\cite{ewgr},
\begin{align}
  g_V^f & =   \sqrt{\rho_f} \, 
             (I_3^f-2Q_f\,\sin^2\theta^f_{\rm eff} ) \, , \quad
  g_A^f  =   \sqrt{\rho_f}\,  I_3^f      \, ,
\end{align}
which comprise the higher-order contributions in terms of the 
form factor $\rho_f(m_t,M_H)$ 
and the effective mixing angle 
$\sin^2\theta^f_{\rm eff}(m_t,M_H)$, 
being now a fermion-type dependent quantity.
Again, their dependence on $m_t$ is quadratic, whereas they depend on
$M_H$ only logarithmically. Nevertheless, the leptonic effective
mixing angle
is one of the most constraining observables for the
mass of the Higgs boson, as shown in
Fig.~\ref{Fig:sineffvsMH}~\cite{lepewwg}. 
Like for $\Delta r$, the calculation 
is complete at the two-loop level~\cite{HOsineff} 
and supplemented by 3- and 4-loop leading terms~\cite{beyondtwoloop}.

\begin{figure}[hbt]
\centerline{
 \includegraphics[width=0.45\linewidth,clip=]{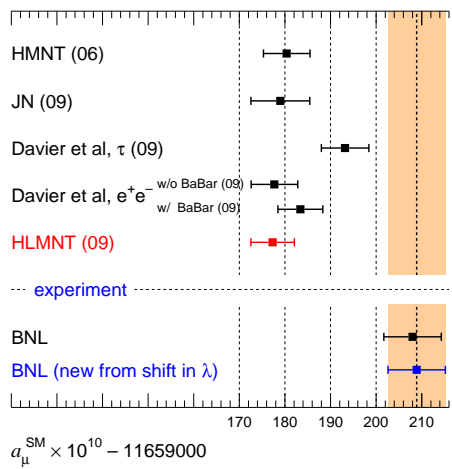} }
\caption{Measurements and Standard Model predictions
for $a_\mu = (g_\mu-2)/2$}
\label{amuon}
\end{figure}

\clearpage

\subsubsection{Muon magnetic moment}

The anomalous magnetic moment of the muon
\begin{align}
   a_{\mu} & = \frac{g_{\mu}-2}{2}
\end {align}
provides a precision test at low energies.
The experimental result of E 821 at Brookhaven National
Laboratory~\cite{BNL} has reached a substantial improvement in accuracy.
It shows a deviation from the Standard Model prediction by 3--4 
standard deviations depending on the evaluation of the hadronic
vacuum polarization from data based on $e^+e^-$ annihilation 
as shown in Fig.~\ref{amuon}~\cite{teubneretal}.
For a recent review see Ref.~\cite{JN}.

\subsection{The vector-boson self-interaction}

The success of the Standard Model in the correct description of the
electroweak precision observables is simultaneously an indirect
confirmation of the Yang--Mills structure of the gauge boson
self-interaction. For conclusive confirmations direct 
experimental investigation is required. 
At LEP 2 (and higher energies), pair production of on-shell
$W$ bosons allows direct experimental tests
of the trilinear vector boson self-couplings and precise
$M_W$ measurements. 
From LEP 2, an error of 33 MeV in $M_W$ has been reached. 
Further improvements have been obtained from the
Tevatron with currently 31 MeV uncertainty,
yielding the world average for the $W$ mass 
$M_W=80.399 \pm 0.023$ GeV~\cite{lepewwg}.

\begin{figure}[htb]
\centerline{
  \includegraphics[width=0.6\linewidth,clip=]{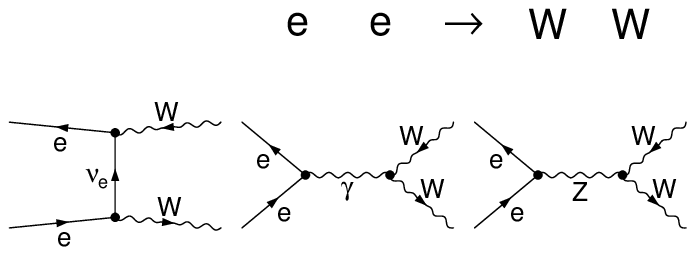} }
\caption{Feynman graphs for $e^+e^- \to W^+ W^-$ in lowest order}
\label{Fig:eeww}
\end{figure}

Pair production of $W$ bosons in the  Standard Model is described 
by the amplitude based on the Feynman graphs in
Fig.~\ref{Fig:eeww} (in Born approximation) and higher-order 
contributions~\cite{lepreports}.

\begin{figure}[htb]
\centerline{
  \includegraphics[width=0.4\linewidth,clip=]{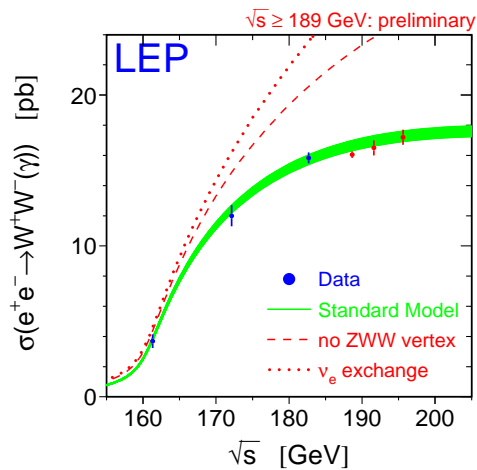} } 
\caption{Cross-section for $e^+e^- \to W^+W^-$, measured at LEP,
         and the Standard Model prediction}
\label{Fig:wwsec}
\end{figure}

\noindent
Besides the $t$-channel $\nu$-exchange diagram, which involves
only the $W$--fermion coupling, the $s$-channel diagrams 
contain the triple gauge interaction between the vector bosons.
The gauge self-interactions of the vector bosons, as specified
in~(\ref{eq:selfgauge}) are essential
for the high-energy behaviour of the production cross-section
in accordance with the principle of unitarity. 
Deviations from these values spoil the high-energy behaviour
of the cross-sections and would be visible at energies
sufficiently above the production threshold.
Measurements of the cross section for $e^+e^-\to WW$ at LEP 
have confirmed the prediction of the Standard Model, as
visualized in Fig.~\ref{Fig:wwsec}~\cite{lepewwg}.

\subsection{Global fits and Higgs boson mass bound}
\label{sec:globalfit}

\begin{figure}[htb]
\begin{center}
 \includegraphics[width=0.4\linewidth,clip=]{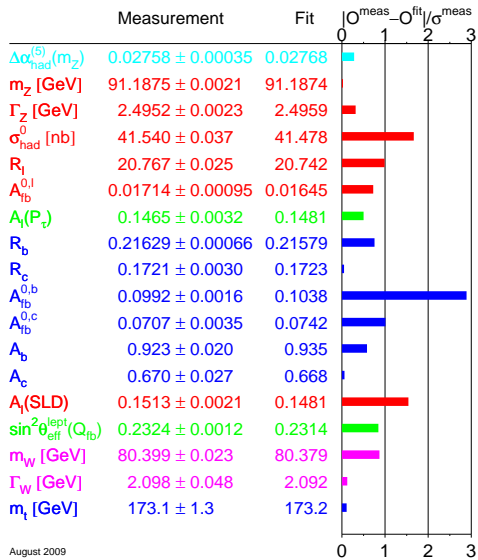}
\end{center}
%
\caption{Experimental measurements versus best-fit Standard Model values}
\label{Fig:pull}
\end{figure}

\begin{figure}[htb]
\begin{center}
 \includegraphics[width=0.55\linewidth,clip=]{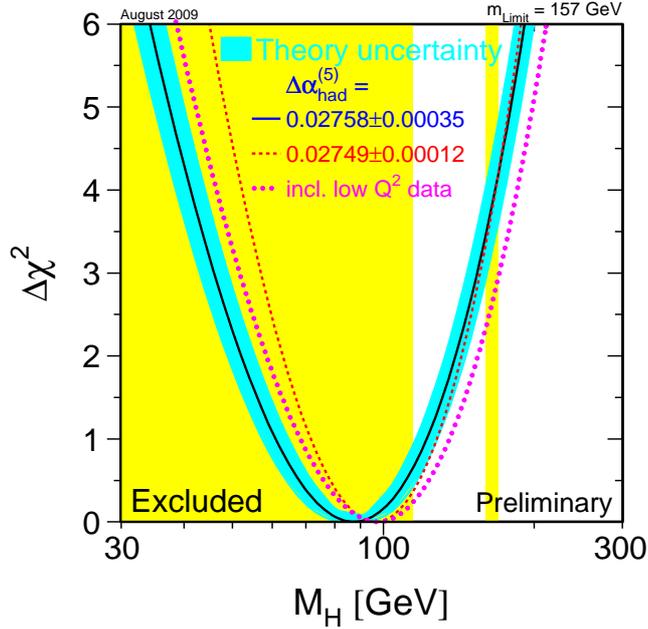}
\end{center}
%
\caption{$\chi^2$ distribution from a global electroweak fit to $M_H$} 
\label{Fig:blueband}
\end{figure}

The $Z$-boson observables from LEP 1 and SLC
together with $M_W$ and the top-quark mass from LEP~2 
and the Tevatron, constitute the set of high-energy quantities
entering a global precision analysis. 
Global fits within the Standard Model to the electroweak precision
data contain $M_H$ as the only free parameter, 
yielding the results~\cite{lepewwg} shown in Fig.~\ref{Fig:pull} and 
an upper limit to the Higgs mass at the 95\% C.L. of 
$M_H < 157$ GeV, 
including the present theoretical
uncertainties of the  Standard Model
predictions visualized as the blue band~\cite{lepewwg} 
in Fig.~\ref{Fig:blueband}.
Taking into account the lower exclusion bound of 114 GeV for $M_H$
from the direct searches via renormalizing the probability shifts
the  95\% C.L. upper bound to 186 GeV~\cite{lepewwg}.
For similar analyses see Ref.~\cite{GFITTER}.

The anomalous magnetic moment of the muon is practically independent
of the Higgs boson mass; hence its inclusion in the fit does not change the 
bound on $M_H$, but it reduces the goodness of the overall fit.

\subsection{Perspectives for the LHC and the ILC}

In the LHC era, further improved measurements of the 
electroweak parameters are expected, especially on the $W$ mass and the mass
of the top quark, as indicated in Table~\ref{Tab:errors}.
The accuracy on the effective mixing angle, measureable from
forward--backward asymmetries, will not exceed the one already obtained in
$e^+ e^-$ collisions~\cite{LHCworkshop}.
The detection of a Higgs boson would go along with a determination
of its mass with an uncertainty of about 100 MeV.

\begin{table}[b]
\begin{center}
\caption{Present experimental accuracies and expectations for future colliders}
\label{Tab:errors}
\begin{tabular}{l c  c c c c }
\hline
Error for       &      &  Now 
                       & {Tevatron/LHC} 
                       & {LC} 
                       & {GigaZ}                               \\  
\hline 
 {$M_W$ [MeV]}  &  & 23  & 15    & 10 &   7                      \\ 
 {$\sin^2\theta_{\rm eff}$} 
                 &     & 0.00016  & 0.00021 &  & 0.000013       \\ 
 {$m_{\rm top}$ [GeV]}  &
                      & 1.3 &  1.0    & 0.2 &  0.13                  \\ 
 {$M_{\rm Higgs}$ [GeV]}  &  
                      & -- &  0.1 & 0.05 &  0.05                 \\  \hline 
\end{tabular}
\end{center}
\end{table}

At a future electron--positron collider, the International Linear Collider
(ILC), the accuracy on $M_W$ can be substantially improved via the scanning
of the $e^+ e^-\to W^+ W^-$ threshold region~\cite{TESLA}.
The GigaZ option, a high-luminosity
$Z$ factory, can provide in addition a significant reduction of the
errors in the $Z$ boson observables, in particular  
for  the leptonic effective mixing angle, denoted by
$\sin^2\theta_{\rm eff}$,  
with an error being an order of magnitude smaller than the present one. 
Moreover, the top-quark mass
accuracy can also be considerably improved.
The numbers are collected in Table~\ref{Tab:errors}.

\begin{figure}[hbt]
\begin{center}
\includegraphics[width=0.4\linewidth,clip=]{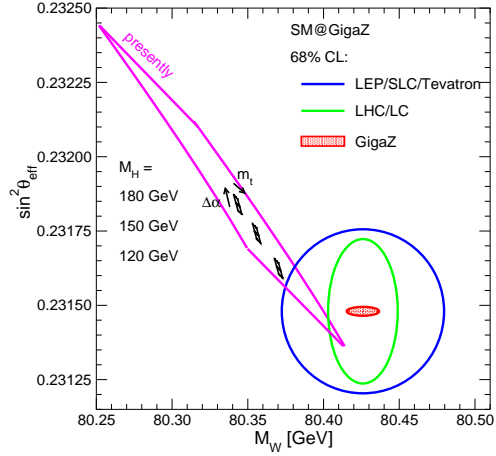}
\end{center}
\caption{Perspectives for Standard Model 
 precision tests at future colliders}
\label{Fig:gigaZplot}
\end{figure}

An ultimate precision test of the Standard Model  
that would be possible in
the future scenario with GigaZ~\cite{gigaZ} 
is illustrated in Fig.~\ref{Fig:gigaZplot}.
The figure displays the 68\% C.L. regions for $M_W$ and 
$\sin^2\theta_{\rm eff}$
expected from
the LHC and ILC/GigaZ measurements; the small quadrangles 
denote the Standard Model 
predictions for a possible, experimentally determined,
Higgs boson mass with the sides reflecting the parametric uncertainties
from $\Delta \alpha$  and the top-quark mass (for  $\Delta \alpha$,
a projected uncertainty of $\delta\Delta\alpha =5\cdot 10^{-5}$ is assumed).
If the Standard Model 
is correct, the two areas with the theory prediction and 
the future experimental results have to overlap. 
The central values chosen in Fig.~\ref{Fig:gigaZplot}
are just examples; the main message is the development
of the uncertainties.

\section{Higgs bosons}
\label{sec:Higgsbosons}

The minimal model with a single scalar doublet is the simplest way
to implement the electroweak symmetry breaking. 
The Higgs potential of the Standard Model given 
in~(\ref{eq:potential}) 
involves two independent parameters $\mu$ and
$\lambda$, which can equivalently be replaced by
the vacuum expectation value $v$ and
the Higgs boson mass $M_H$, as done in~(\ref{eq:potentialunitary}).
The vacuum expectation value $v$
is determined  by the gauge sector, 
as explained in~(\ref{eq:massterm}) and~(\ref{eq:Fermiscale});
$M_H$ is independent and cannot be predicted but has to be taken from
experiment.
Thus in the Standard Model
the mass $M_H$ of the Higgs boson appears as the only free parameter
that is still undetermined as yet. 
Expressed in terms of $M_H$,
the Higgs part of the electroweak Lagrangian in the unitary gauge
reads as follows:
\begin{align}
\label{LHiggsinunitarygauge}
{\cal L}_{\rm H} \, = & \, \frac{1}{2} 
  \big( \partial_\mu H \big) \big(\partial^\mu H \big) \,  
        - \frac{M_H^2}{2}\, H^2
        - \frac{M_H^2}{2v} \, H^3
        - \frac{M_H^2}{8 v^2}\, H^4 \nonumber \\[0.1cm]
  & +\, 
 \left( M_W^2\, W^+_\mu W^{- \mu} + \frac{M_Z^2}{2}\, Z_\mu Z^\mu \right) 
  \left( 1 + \frac{H}{v} \right)^2 \,
      -\,  \sum_f \, m_f\,  \adpsi_f \psi_f \left(1+ \frac{H}{v} \right) , 
\end{align}
involving interactions of the Higgs field with the massive fermions and
gauge bosons, as well as Higgs self interactions 
proportional to $M_H^2$.

\subsection{Empirical bounds}

\begin{figure}[htb]
\begin{center}
 \includegraphics[width=0.15\linewidth,angle=270,clip=]{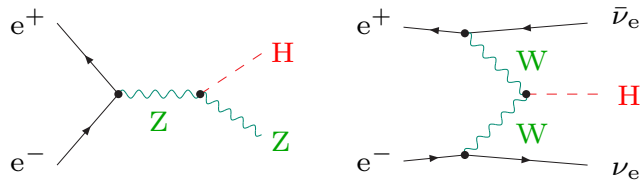}  
\end{center}
\caption{Processes for Higgs boson production in $e^+e^-$ collisions}
\label{Fig:HiggsatLEP}
\end{figure}

The existence of the Yukawa couplings 
and the couplings to the vector bosons $W$ and $Z$
is  the basis  for the experimental searches 
that have been performed until now at LEP 
and the Tevatron. 
At $e^+ e^-$ colliders, Higgs bosons can be produced by
Higgs-strahlung from $Z$ bosons and by vector boson fusion 
(mainly $WW$) as displayed in Fig.~\ref{Fig:HiggsatLEP}. 

\begin{figure}[htb]
\begin{center}
 \includegraphics[width=0.15\linewidth,angle=270,clip=]{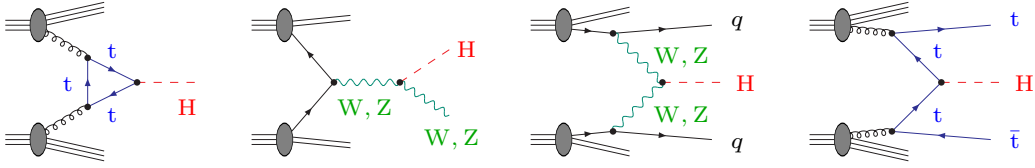}  
\end{center}
\caption{Processes for Higgs boson production at hadron colliders}
\label{Fig:HiggsatHAD}
\end{figure}

At LEP energies, Higgs-strahlung is the relevant process.
The lower limit at 95\% C.L.\ resulting from the search at
LEP is 114.4~GeV~\cite{PDG08}.
From searches at the Tevatron~\cite{TevHiggsWG}
(see Fig.~\ref{Fig:HiggsatHAD} for various mechanisms)
the mass range from 162~GeV to 166~GeV 
has been excluded (95\% C.L.).

Indirect determinations of $M_H$ from precision data 
yield an upper limit and have
already been discussed in Section~\ref{sec:globalfit}.
As a general feature,
it appears that the data prefer a light Higgs boson.

\subsection{Theoretical bounds}

\begin{figure}[htb]
\centerline{
\includegraphics[height=0.45\linewidth,angle=90]{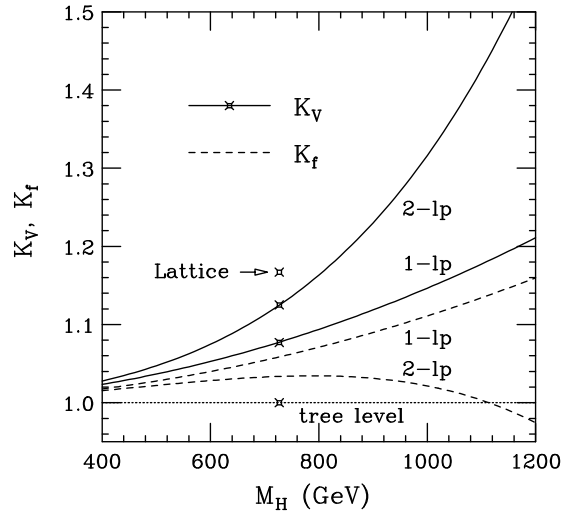}}
\caption{Correction factors $K_V, K_f$ from higher orders for the Higgs decay
         widths $H\to VV\; (V=W,Z)$ and 
         $H \to f \bar{f}$ in 1- and 2-loop order} 
\label{kfactors}
\end{figure}

There are also theoretical constraints on the Higgs mass
from vacuum stability and absence of a Landau pole~\cite{lindner,higgsbounds,hambye},
and from lattice calculations~\cite{lattice,decaylattice}. 
Explicit perturbative calculations
of the decay width for $H\to W^+W^-,ZZ$  in the large-$M_H$ limit,
$\Gamma(H\to VV)=  K_V\cdot \Gamma^{(0)}(H\to VV)$
up to 2-loop order~\cite{ghinculov} have  shown that the 2-loop
contribution exceeds the 1-loop term in size (same sign) for
 $M_H > 930$~GeV (Fig.~\ref{kfactors}~\cite{riess}).
This result  is confirmed by the calculation of the next-to-leading
order
correction  in the $1/N$ expansion, where the Higgs sector is treated
as an $O(N)$ symmetric 
$\sigma$-model~\cite{binoth}. 
A similar increase of the 2-loop perturbative contribution 
with $M_H$
is observed for the fermionic 
decay width~\cite{Hff}, 
$\Gamma(H\to f\bar{f})) =  K_f\cdot \Gamma^{(0)}(H\to f\bar{f}))$, 
but with opposite sign 
leading to a cancellation of the 1-loop correction 
for $M_H\simeq 1100$ GeV (Fig.~\ref{kfactors}).
The lattice result~\cite{decaylattice} for the bosonic Higgs decay
in Fig.~\ref{kfactors} for $M_H=727$ GeV is not far from
the perturbative 2-loop result;
the difference may at least partially be interpreted as missing 
higher-order terms.

The behaviour of the quartic Higgs self-coupling $\lambda$,
as a function of a rising energy scale~$Q$, follows
from the renormalization group equation 
\begin{align}
\label{HiggsRGE}
 \frac{{\rm d} \lambda}{{\rm d} t} 
  &  = \frac{1}{16\pi^2}\, (12 \lambda^2 + 6\, \lambda\,g_t^2
                    - 3 \, g_t^4 + \cdots ), \quad
  t=\log \frac{Q^2}{v^2} \, ,
\end{align}
with the $\beta$-function
dominated by the contributions from $\lambda$ and
the top-quark Yukawa coupling $g_t$ in the loop contributions to
the quartic interactions,

\centerline{
\includegraphics[height=0.17\linewidth,clip=]{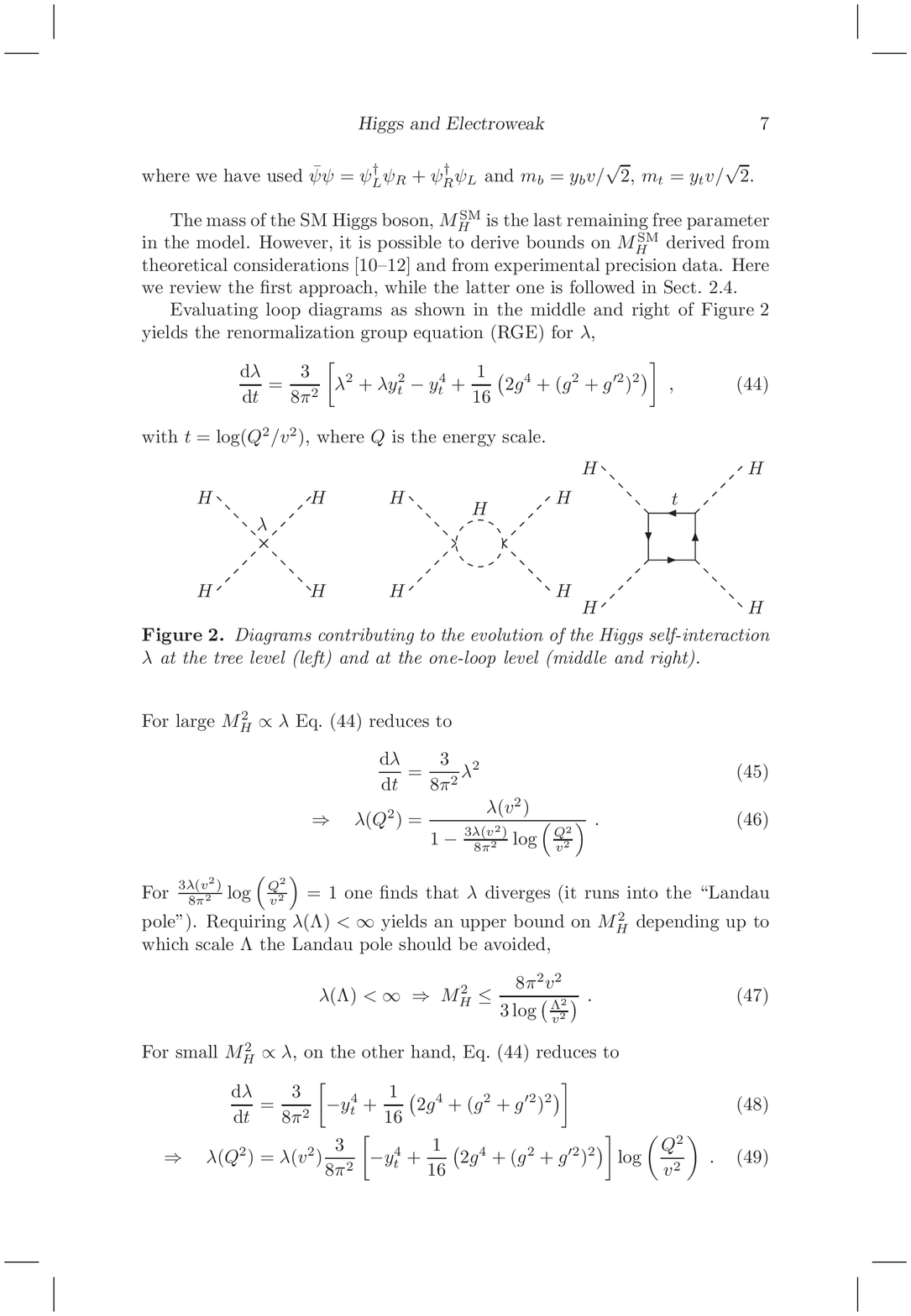}}

\begin{figure}[htb]
\centerline{
\includegraphics[height=0.45\linewidth,clip=]{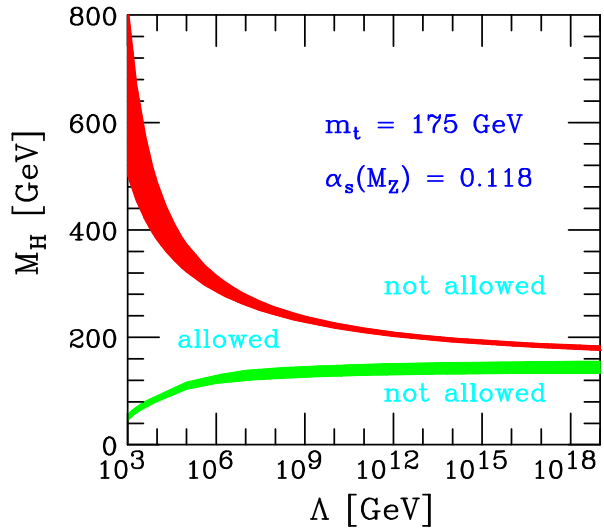}}
\caption{Theoretical limits on the Higgs boson mass from
         the absence of a Landau pole and from vacuum stability}
\label{higgslimits}
\end{figure}

\noindent
Owing to the second diagram, the first term in~(\ref{HiggsRGE}),
$\lambda(Q)$ increases with $Q$ and diverges at a critical scale,
the Landau pole, which moves towards lower values for 
increasing mass $M_H$.
The requirement of a perturbative, small coupling $\lambda(Q)$ 
 up to a scale $\Lambda$ thus yields
 an upper bound for $M_H$. 
In order to avoid unphysical negative quartic couplings
from the negative top-loop contribution,
a lower
bound on the Higgs mass is derived. In combination, 
the requirement that the Higgs coupling remain finite and
positive up to a scale $\Lambda$ yields constraints
on the Higgs mass $M_H$, which have been evaluated at the 
2-loop level~\cite{higgsbounds,hambye}.  
These bounds on $M_H$ are shown in Fig.~\ref{higgslimits}~\cite{hambye}  
as a function of the cut-off scale $\Lambda$ up to which the
standard Higgs sector can be extrapolated.
The allowed 
region is the area between the lower and the upper curves.
The bands indicate the theoretical uncertainties associated
with the solution of the renormalization group equations~\cite{hambye}. It is interesting to note that the
indirect determination of the Higgs mass range from
electroweak precision data via radiative corrections 
is compatible with a value of $M_H$ where $\Lambda$ can be
extended up to the Planck scale.

\subsection{Future searches}

\begin{figure}[ht]
\begin{center}
 \includegraphics[width=0.45\linewidth,angle=270,clip=]{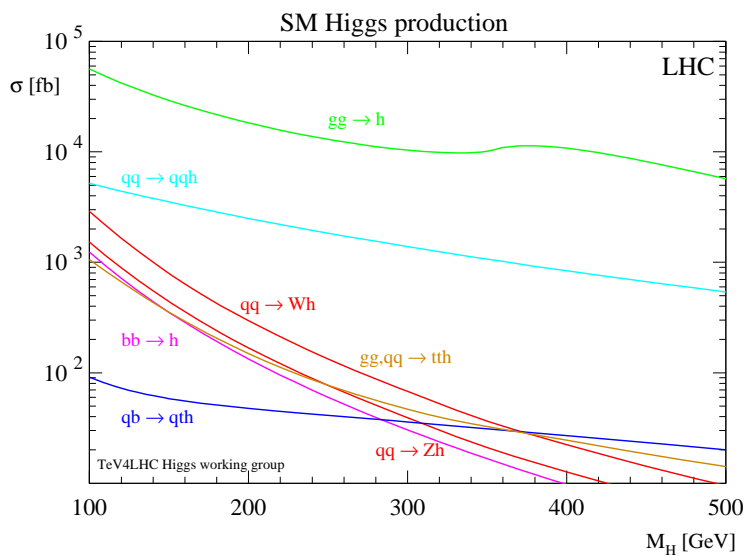}  
\end{center}
\caption{Cross sections for Higgs boson production at the LHC}
\label{Fig:LHCxsec}
\end{figure}

\begin{figure}[hb]
\begin{center}
 \includegraphics[width=0.55\linewidth,clip=]{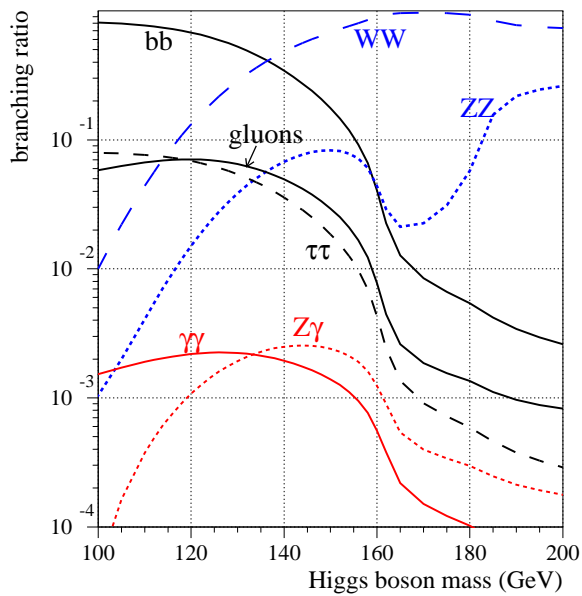}  
\end{center}
\caption{Branching ratios for Higgs boson decays}
\label{Fig:HiggsBR}
\end{figure}

For the coming experimental searches at the LHC, 
it is important to have precise and reliable predictions
for the production and decay rates.
Higgs bosons can be produced through various 
mechanisms at the partonic level. The main partonic
processes for Higgs boson production are
depicted in Fig.~\ref{Fig:HiggsatHAD}, and the corresponding
production cross sections are shown in 
Fig.~\ref{Fig:LHCxsec}~\cite{TeVtoLHC}.
The largest cross section arises from gluon--gluon fusion.
The experimental signal, however, 
is determined by the product 
\begin{align}
  & \sigma(AB\to H) \cdot BR(H\to X)
\end{align}
of the production cross section $\sigma(AB\to H)$
from initial-state partons $A, B$ 
and the branching ratio $BR(H\to X)$
for the decay of the Higgs boson into a specific final state $X$
(see Fig.~\ref{Fig:HiggsBR} for the branching ratios~\cite{BRs}).
A light Higgs boson, well below the $WW$ threshold,
decays predominantly into $b\bar{b}$ quarks, owing to 
the largest Yukawa couplings in the kinematically allowed 
fermionic decay channels.
This signal, however, is experimentally unaccessible
because it is covered by a huge background of QCD-generated
$b$-quark jets. Therefore, in the low mass range, the rare decay
channel $H\to \gamma \gamma$ has to be selected reducing the total
number of events considerably, in spite of the large production
cross section, and makes Higgs detection a cumbersome business.
For larger masses, $M_H \gtrsim 140$ GeV, the decay modes 
$H\to WW, ZZ \to 4 f$ make detection relatively easy.
The vector-boson fusion channel 
(third diagram of Fig.~\ref{Fig:HiggsatHAD}) with subsequent
leptonic decay $H\to \tau^+ \tau^-$ 
is a promising alternative.

\clearpage

\noindent
For completeness we list the (lowest-order) expressions for the 
dominant Higgs decay rates into fermion and vector-boson pairs,
\begin{align}
\Gamma (H\to f\bar{f}) & = \, N_C
\frac{G_F M_H\, m_f^2}{4 \pi \sqrt{2}} 
 \sqrt{1-\frac{4m_f^2}{M_H^2} }  \;\;{\rm with} \;\;
   N_C=3 \; {\rm for} \,  f=q, \;\; N_C=1 \;  
                    {\rm for} \,  f=\ell , \nn \\
\Gamma (H\to VV) & = \,
\frac{G_F M_H^3}{16 \pi \sqrt{2}} \, R_V(x_V), \quad
    x_V = \frac{M_V^2}{M_H^2},  \qquad (V= W, Z)  
\end{align}
with 
\begin{align}
R_Z & = R(x_Z) , \quad  R_W = 2 \, R(x_W), \quad
 R(x) = \sqrt{1-4x} \, (1-4x+12x^2) \, . 
\end{align}
{\small
As an exercise, these formulae can easily be derived from the
$Hff$ and $HVV$ vertices in~(\ref{LHiggsinunitarygauge}) 
with the help of the
Feynman rules of Section~\ref{sec:QFT}
and the general expression for the width 
in~(\ref{eq:diffwidthmm}). }

\subsection{Supersymmetric Higgs bosons}
\label{sec:susyhiggs}

Among the extensions of the Standard Model, the 
Minimal Supersymmetric Standard Model (MSSM)~\cite{MSSM} 
is a theoretically favoured scenario as 
the most predictive framework beyond the Standard Model.
A light Higgs boson, 
as indicated in the analysis of the electroweak precision data,
would find a natural explanation by the structure of the 
Higgs potential. For a review on MSSM Higgs bosons see Ref.~\cite{djouadi}.

\begin{figure}[hbt]
\begin{center}
 \includegraphics[width=0.45\linewidth,clip=]{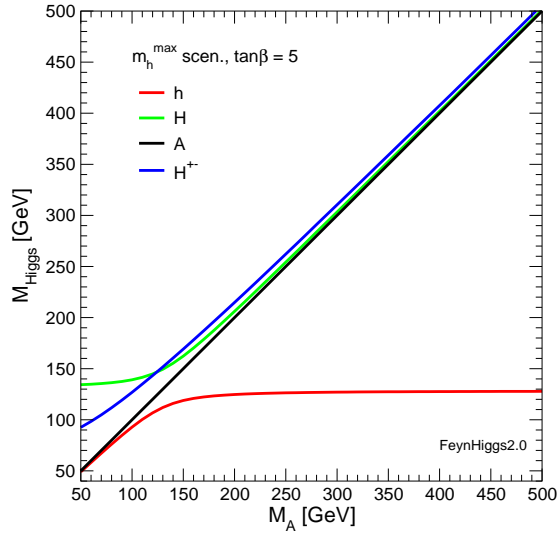}  
\end{center}
\caption{Example of the Higgs boson mass spectrum in the MSSM}
\label{Fig:spectrumHiggs}
\end{figure}

The five physical Higgs particles of the MSSM consist of two $CP$-even
neutral bosons $h^0,H^0$, a $CP$-odd $A^0$ boson, and a pair of charged
Higgs particles $H^\pm$. At tree level, their masses are determined 
by the $A^0$ boson mass, $M_A$, and the ratio of the two vacuum expectation
values, $v_2/v_1 = \tan\beta$,
\begin{align}
M_{H^+}^2    & = \, M_A^2 + M_W^2 \, , \nn \\
M_{H^0,h^0}^2 & = \, \frac{1}{2} \,  
  \left( M_A^2 + M_Z^2 \pm 
  \sqrt{\big(M_A^2 + M_Z^2\big)^2 - 4 M_Z^2 M_A^2 \cos^2 2\beta} \right) .
\end{align}
These relations are sizeably modified by higher-order contributions
to the Higgs boson vacua and propagators. 
A typical example of a spectrum is shown in Fig.~\ref{Fig:spectrumHiggs},
based on the {\sc FeynHiggs} code~\cite{feynhiggs}.
In particular the mass of the lightest Higgs boson $h^0$ is substantially 
influenced
by loop contributions; for large $M_A$, the $h^0$ particle 
behaves like the standard Higgs boson,
but its mass is dependent on basically all the
parameters of the model and hence yields another powerful
precision observable.
A definite prediction of the MSSM
is thus the existence of a light Higgs boson
with mass below $\sim 140$ GeV. 
The detection of a
light Higgs boson could be a significant hint for
supersymmetry.

The structure of the MSSM as a renormalizable quantum field theory
allows a similarly complete calculation of
the electroweak precision observables
as in the Standard Model in terms of one Higgs mass
(usually taken as $M_A$) and $\tan\beta$,
together with the set of
SUSY soft-breaking parameters fixing the chargino/neutralino and
scalar fermion sectors~\cite{mssm_ren}.
For updated discussions of precision observables in the MSSM
see Ref.~\cite{Heinemeyer:2004gx} .

\begin{figure}[hbt]
\begin{center}
 \includegraphics[width=0.45\linewidth,clip=]{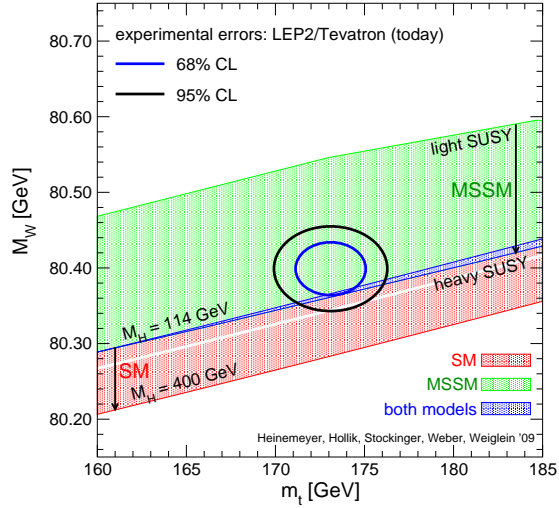}  
\end{center}
\caption{The $W$ mass range
         in the Standard Model (lower band) and in the
         MSSM (upper band) respecting bounds are from the non-observation of 
         Higgs bosons and SUSY particles} 
\label{Fig:susymw}
\end{figure}
As an example, Fig.~\ref{Fig:susymw}
displays the range of predictions for $M_W$
in the Standard Model
and in the MSSM, together with the present experimental errors
and the expectations for the  LHC measurements. 
The MSSM prediction is in slightly better agreement with the
present data for $M_W$, although not conclusive as yet.
Future increase in the experimental accuracy, however, will
become decisive for the separation between the  models.

Especially for the muonic $g-2$, the MSSM can significantly improve
the agreement between theory and experiment:
one-loop terms with relatively light scalar muons, sneutrinos,  
charginos and neutralinos, 
\vspace*{-0.2cm}
\centerline{
\includegraphics[width=0.16\linewidth,angle=270,clip=]{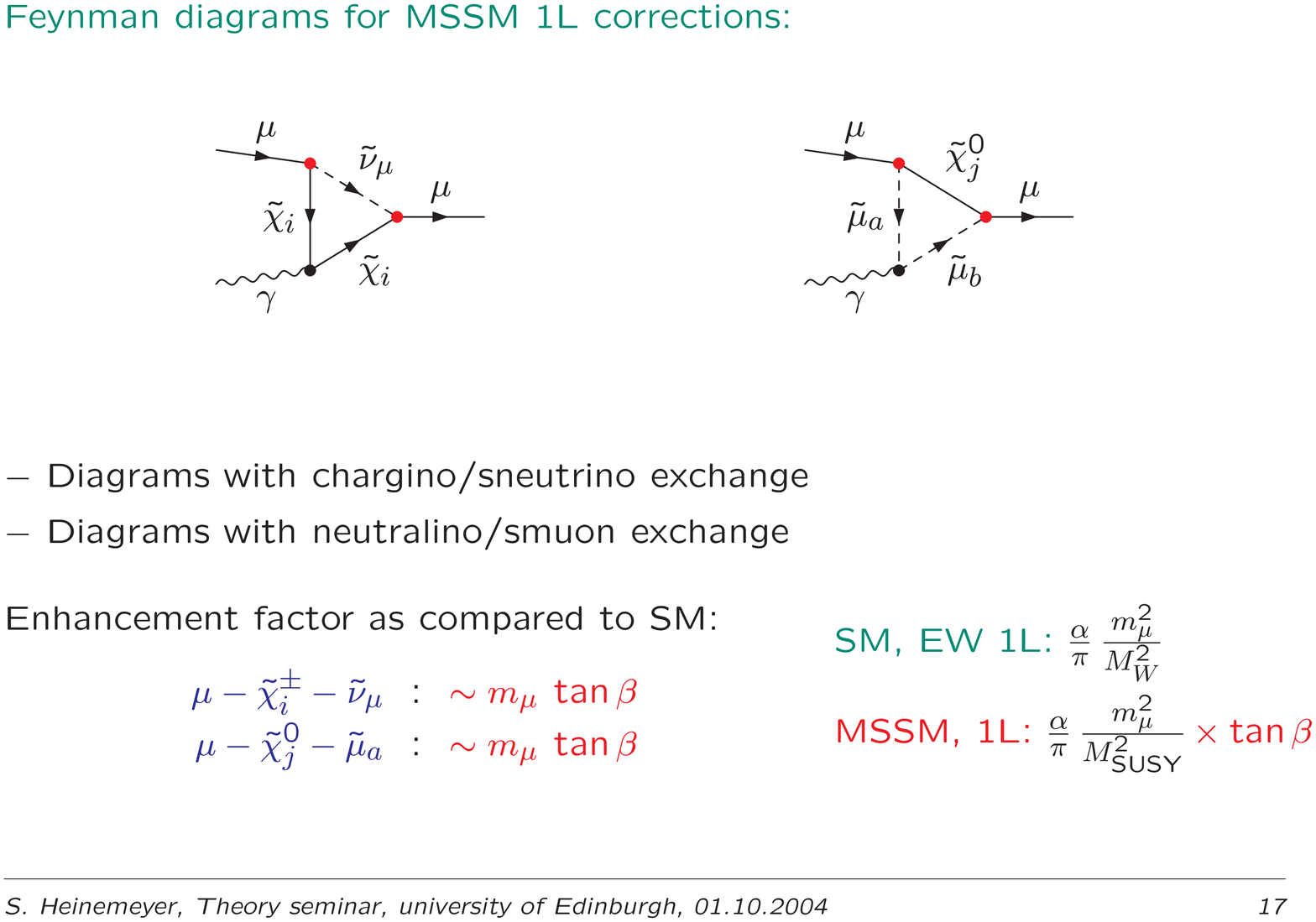} }

\vspace*{0.3cm}
\noindent
in the mass range 200--600~GeV, 
together with a large value of $\tan\beta$
can provide a positive contribution $\Delta a_\mu$, 
which can entirely
explain the difference $a_\mu^{\rm exp} - a_\mu^{\rm SM}$
(see Ref.~\cite{stoeckinger} for a review).

\begin{figure}[hbt]
\begin{center}
 \includegraphics[width=0.7\linewidth,clip=]{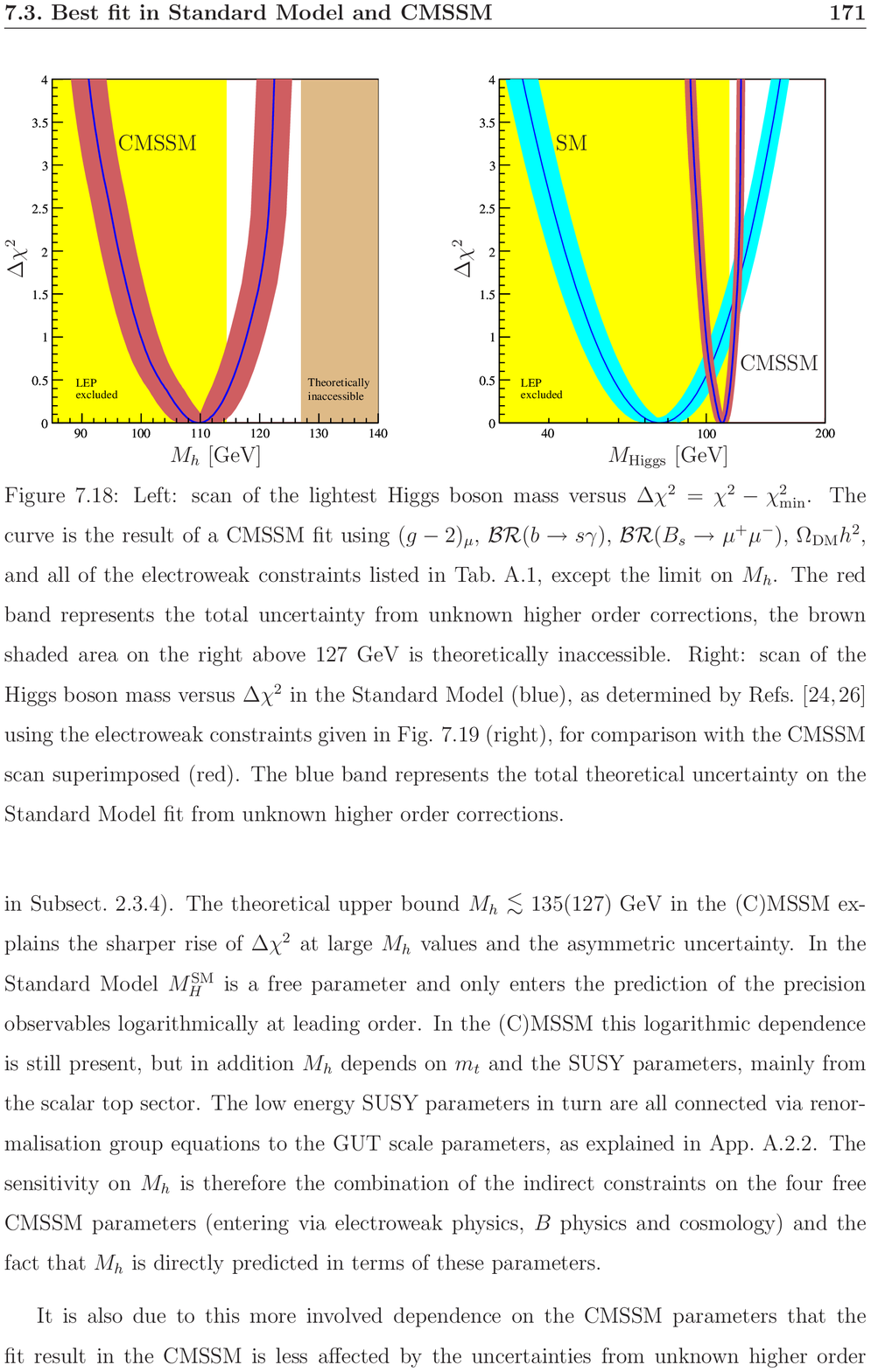}
\end{center}
\caption{$\chi^2$-distribution for cMSSM fits, expressed in terms of $M_h$} 
\label{Fig:cMSSMfit}
\end{figure}

The MSSM yields a  comprehensive
description of the precision data, in a  similar way to the 
Standard Model.
Global fits, varying the MSSM parameters, have been performed   
to all electroweak precision data~\cite{deboer}
showing that the description within the MSSM is slightly better
than in the Standard Model. 
This is mainly due to the improved
agreement for $a_\mu$. 
The fits 
have been updated recently for the constrained
MSSM (cMSSM), including also bounds from
$b\to s\gamma$ and from the cosmic relic density.
The $\chi^2$-distribution for the fit parameters can be 
shown~\cite{buchmulleretal}
as a $\chi^2$-distribution for the lightest Higgs boson mass $M_H$,
displayed in Fig.~\ref{Fig:cMSSMfit}.
The mass range $M_h = 110^{+9}_{-10}$ GeV obtained from this fit
is in much better agreement with the lower bound from the direct
search than in the case of the Standard Model.

%

\section{Outlook}
\label{sec:beyondSM}

In spite of the success of the Standard Model in describing a large
variety of phenomena, at a high level of accuracy on both the
theoretical and the experimental side, there is a list of shortcomings 
that motivate the quest for physics beyond the Standard Model. 

A rather direct augmentation is enforced by the need for accommodating
massive neutrinos. 
The Standard Model in its strictly minimal version is incomplete 
with respect to a mass term for neutrinos.
Neutrino mass terms can be added~\cite{Lindner} without 
touching on the basic architecture of the Standard Model.
Besides this rather immediate modification
one is confronted, however, with a series of basic 
conceptual problems:
\begin{itemize}
\item
the smallness of the electroweak scale 
$v \sim 1/\sqrt{G_F}$ 
compared to the Planck scale $M_{\rm Pl}\sim 1/\sqrt{G_N}$
(the {\it hierarchy problem})
and the smallness of the Higgs boson mass of ${\cal O}(v)$, which is not
protected against large quantum corrections of ${\cal O}( M_{\rm Pl})$; 
\item
the large number of free parameters 
(gauge couplings, vacuum expectation value, $M_H$, fermion
masses, CKM matrix elements), which are not predicted but
have to be taken from experiments; 
\item
the pattern that occurs in the arrangement of the fermion 
masses;
\item
the quantization of the electric charge, or the values of the 
hypercharge, respectively;
\item
the missing way to connect to gravity.
\end{itemize}
Moreover, there are also phenomenological shortcomings,
like missing answers to the questions about 
\begin{itemize}
\item
the nature of dark matter that constitutes the largest fraction
of matter in the Universe,
\item 
the origin of the  
baryon asymmetry of the Universe.
\end{itemize} 
The class of models based on supersymmetry, 
briefly addressed in the last subsection~\ref{sec:susyhiggs}, 
can at least provide partial
answers, e.g., for dark matter,  the further unification of forces
and hierarchy of mass scales, new sources of CP violation,
and can be related to string theory as a candidate for a microscopic
theory of gravity. The LHC experiments may soon shed light 
on our unanswered questions, 
or may also surprise us with answers to questions we did not ask.

\clearpage

\appendix

\section{Canonical commutation relations}
\label{sec:AppA}

The commutators between the canonically conjugate variables 
$Q_j, P_k$ in quantum mechanics,
\begin{align}
\label{eq:Heisenberg}
[Q_j, P_k] & =\, i \, \delta_{jk} , \quad
[Q_j, Q_k]  =\,  [P_j, P_k]  = 0 , 
\end{align}
are translated in quantum field theory to commutators
for a (generic) field operator $\phi(x)\equiv \phi(t,\vec{x})$
and its conjugate canonical momentum
\begin{align}
\label{eq:canonmom}
\Pi(x) & = \frac{\partial {\cal L}}
                        {\partial (\partial_o \phi)}
\end{align}
derived from the basic Lagrangian ${\cal L}$ for the system.
This procedure, known as canonical field quantization,
is specified by the equal-time 
commutation relations, where the discrete indices $j, k$ 
in~(\ref{eq:Heisenberg}) are replaced by the continuous indices $\vec{x}, \vec{x}\,'$: 
\begin{align}
\label{eq:QFTCR}
[\phi(t,\vec{x}), \Pi(t,\vec{x}\,')] & = 
    i\, \delta^3(\vec{x} - \vec{x}\,'), \quad
[\phi(t,\vec{x}), \phi(t,\vec{x}\,')] =
[\Pi(t,\vec{x}), \Pi(t,\vec{x}\,')] = 0 \, .
\end{align}
For fermionic field variables $\psi(x)$ the commutators have to be replaced
by anti-commutators.

\subsection{Scalar field}

We illustrate the method of canonical quantization   
choosing the scalar field as a specific example. 
Starting from the Lagrangian~(\ref{eq:Lscalar}) for a general, complex,
free scalar field, 
we find the canonical field momenta via~(\ref{eq:canonmom}) to be
\begin{align}
 \frac{\partial {\cal L}}{\partial (\partial_o \phi)}
    &  = \partial^0 \phi^\dagger  =  \dot{\phi}^\dagger 
      =   \Pi, \nn \\
  \frac{\partial {\cal L}}{\partial (\partial_o \phi^\dagger)}
    &  = \partial^0 \phi \;   = \, \dot{\phi}\;\;  =  \Pi^\dagger \, .
 \end{align}
Accordingly, the canonical commutation relations are given by
\begin{align}
\label{eq:CRscalar}
 & [\phi(t,\vec{x}), \dot{\phi}^\dagger(t,\vec{x}\,')] 
   = i\, \delta^3(\vec{x} - \vec{x}\,') ,  \nn \\
 & [\phi(t,\vec{x}), \phi(t,\vec{x}\,')]  = \,
[\dot{\phi}(t,\vec{x}), \dot{\phi}(t,\vec{x}\,')] = 0 \, .
\end{align}
These relations can equivalently be expressed in terms of the
annihilation and creation operators $a,b, a^\dagger, b^\dagger$
in the Fourier expansion of the  scalar field $\phi(x)$
in~(\ref{eq:scalarfieldFourier}). They
fulfil the following canonical commutation relations 
in momentum space and can be interpreted as those for 
a continuous set of quantized harmonic oscillators, 
labelled by $\vec{k}$,
with frequencies $\omega = k^0 = \sqrt{{\vec{k\,}}^{2} +m^2}$ 
and with the relativistic normalization:
\begin{align}
\label{CRscalarmomentum}
& [a(k), a(k')] = [b(k), b(k')] = 0 , \qquad
  [a^\dagger(k), a^\dagger(k')] = [b^\dagger(k), b^\dagger(k')] = 0 ,
   \nn \\
& [a(k), a^\dagger(k')] = 2 k^0 \, \delta^3(\vec{k} - \vec{k}\, ') , \qquad
[b(k), b^\dagger(k')]  = 2 k^0 \, \delta^3(\vec{k} - \vec{k}\, ') , \nn \\
& [a(k), b(k')] = [a(k), b^\dagger(k')] =  
  [a^\dagger(k), b(k')] = [a^\dagger(k), b^\dagger(k')] =  0 . 
\end{align}
Since we do not make use of the formulation of quantization in space-time,
but use instead the creation and annihilation operators, which are
closer to the physical picture of particles and particle states,
we list the commutators for the vector and spinor fields only
in momentum space.

\subsection{Vector field}

For the vector field~(\ref{eq:vectorfieldFourier}) 
the annihilation and creation operators
$a_\lambda, a_\lambda^\dagger$ carry helicity indices in additon to the 
momenta. Otherwise the commutation rules are analogous to the scalar
case:
\begin{align}
\label{CRvector}
& [a_\lambda(k), a_{\lambda'}(k')] = 
  [a_\lambda^\dagger(k), a_{\lambda'}^\dagger(k')] = 0 , 
   \nn \\
& [a_\lambda(k), a_{\lambda'}^\dagger(k')] 
   = 2 k^0\, \delta_{\lambda \lambda'} \, \delta^3(\vec{k} - \vec{k}\, ') .
\end{align}

\subsection{Dirac field}

The Dirac field~(\ref{eq:DiracfieldFourier}) 
involves fermionic annihilation and creation
operators $c_\sigma, d_\sigma, c^\dagger_\sigma,  d^\dagger_\sigma$
for each momentum $\vec{k}$ and helicity $\sigma$.
According to the antisymmetry of fermionic states, 
all commutators applying to bosonic states in the canonical
quantization above have to be replaced by anti-commutators:
\begin{align}
\label{CRDirac}
& \{c_\sigma(k), c_{\sigma'}(k')\} = 
  \{c_\sigma^\dagger(k), c_{\sigma'}^\dagger(k')\} = 0 , \qquad
  \{c_\sigma(k), c_{\sigma'}^\dagger(k')\}
   = 2 k^0\, \delta_{\sigma \sigma'} \, \delta^3(\vec{k} - \vec{k}\, ') , \nn \\
& \{d_\sigma(k), d_{\sigma'}(k')\} = 
  \{d_\sigma^\dagger(k), d_{\sigma'}^\dagger(k')\} = 0 , \qquad \!\!
  \{d_\sigma(k), d_{\sigma'}^\dagger(k')\}
   = 2 k^0\, \delta_{\sigma \sigma'} \, \delta^3(\vec{k} - \vec{k}\, ') , \nn \\
& \{c_\sigma(k), d_{\sigma'}(k')\} =
  \{c^\dagger_\sigma(k), d^\dagger_{\sigma'}(k')\} =
  \{c_\sigma(k), d^\dagger_{\sigma'}(k')\} = 
  \{c_\sigma^\dagger(k), d_{\sigma'}(k')\} = 0 . 
 \end{align}

\section{Green functions and causality}
\label{sec:AppB}

We demonstrate, for the example of the scalar field, 
how the $+i\epsilon$ prescription
in the Fourier representation of the Feynman propagator
leads to causal behaviour of particle/antiparticle propagation
in space-time.
Making use of the time-ordered product of any two field quantities 
$A(x)$ and $B(x)$,
\begin{align}
\label{eq:timeordering}
T A(x) B(y) & =\, \Theta(x^0-y^0)\, A(x) B(y) + 
                \Theta(y^0-x^0)\, B(x) A(y) \, , 
\end{align}
one can define the 2-point function for a (complex) scalar field $\phi(x)$
in the following way:
\begin{align}
\label{eq:twopoint}
<\!0| T \phi(x) \phi^\dagger (y) |0\!> & = \,
 \Theta(x^0-y^0)\, <\!\!0| \phi(x) \phi^\dagger (y) |0\!>   \nn \\
    &  +\,  \Theta(y^0-x^0)\, <\!\!0| \phi^\dagger(y) \phi(x) |0\!> \, .
\end{align}
Invoking the Fourier expansion for $\phi$ and $\phi^\dagger$
in terms of creation and 
annihilation  operators~(\ref{eq:scalarfieldFourier}),
one can see that~(\ref{eq:twopoint})  
describes particles created at time $y^0$ and
annihilated at time $x^0$ if $x^0 > y^0$,
and anti-particles  created at time $x^0$ and
annihilated at time $y^0$ if $y^0 > x^0$.

On the other hand, starting from the 
Fourier integral~(\ref{eq:FourierGreen})
and performing the $k^0$ integration by means of a contour integral
in the complex plane, one obtains the expression
\begin{eqnarray}
D(x-y) & = &  \int  \frac{{\rm d}^4k}{(2\pi)^4} \,
 \frac{e^{-ik(x-y)}}{k^2 - m^2 + \, i \epsilon} \nn \\
 & = &  \int \frac{{\rm d}^3k}{(2\pi)^3} \, e^{i \vec{k} (\vec{x}-\vec{y})}  
      \int \frac{{\rm d}k^0}{2\pi} \,
      \frac{e^{-i k^0 (x^0-y^0)} } 
           {(k^0)^2 - \vec{k}^{\, 2} - m^2 + i\epsilon} 
 \nn \\
 & =&  \,-\, \frac{i}{(2\pi)^3} \int 
      \frac{{\rm d}^3k}{2 k^0} \, 
           e^{i \vec{k} (\vec{x}-\vec{y})- i k^0 (x^0-y^0)} 
           \,|\,_{k^0 = \sqrt{\vec{k\,}^{2} + m^2} }
       \;\;\cdot \Theta(x^0-y^0)
\ \nn \\
 &  &  - \, \frac{i}{(2\pi)^3} \int  
      \frac{{\rm d}^3k}{2 k^0} \, 
            e^{i \vec{k} (\vec{x}-\vec{y})+ i k^0 (x^0-y^0)}  
       \,|\,_{k^0 = \sqrt{\vec{k\,}^{2} + m^2} }
       \;\;\cdot \Theta(y^0-x^0) \nn 
\end{eqnarray}
which can be written in the following way:
\begin{align}
i\, D(x-y) & = \, \frac{1}{(2\pi)^3} 
 \int  \frac{{\rm d}^3k}{2 k^0} \, \left[ 
   e^{-ik(x-y)} \, \Theta(x^0-y^0) \, +\,
     e^{ik(x-y)} \, \Theta(y^0-x^0) 
  \right]_{k^0=\sqrt{\vec{k\,}^{2} + m^2} } \, .
\end{align}
This is 
identical to~(\ref{eq:twopoint}) when the 
Fourier representation~(\ref{eq:scalarfieldFourier})
for $\phi$ is inserted.
Hence one has the identity
\begin{align}
<\!0| T \phi(x) \phi^\dagger (y) |0\!> & = \, i \, D(x-y) \, ,
\end{align}
which connects the Green function of the Klein--Gordon equation
with the 2-point function of the quantized scalar field
and thus with the particle/antiparticle concept obeying causality.
As a byproduct, it also explains the extra factor $i$ in 
the propagator~(\ref{eq:scalarpropagator}).

\newpage

\section{Loop integrals and dimensional regularization}
\label{sec:AppC}

In the calculation of self-energy diagrams the following type of
loop integrals involving two propagators appears when 
the integration is done in $D$ dimensions, denoted by $B_0$
after removing a numerical factor:
\begin{align}
\label{twopointintegral}
     \dkm \,\frac{1}{\Dkk\Dkq} & = \frac{i}{16\pi^2} 
           B_0(q^2,m_1,m_2)  \, .
\end{align}
With help of the Feynman parametrization
\begin{align}
\frac{1}{ab} & = \int^1_0 {\rm d}x \,
   \frac{1}{[ax+b(1-x)]^2} 
\end{align}
and after a shift in the $k$-variable, $B_0$ can be written in the form
\begin{align} 
\label{Bintegral}
     \ipi\, B_0(q^2,m_1,m_2)    & =
 \int^1_0 {\rm d}x \, \frac{\m^{4-D}}{(2\pi)^D} \int
 \frac{{\rm d}^Dk}{[k^2-x^2q^2+x(q^2+m_1^2-m_2^2)-m_1^2+i \varepsilon]^2} \, . 
\end{align}
The advantage of this parametrization is a simpler $k$-integration
where the integrand is only a function of $k^2=(k^0)^2-\vec{k}^2$.
In order to transform it into a Euclidean integral we perform the
substitution
\footnote{The $i\veps$-prescription in the masses ensures that this is
compatible with  the pole structure of the integrand.}
\begin{align}
   k^0 & = i\,k_E^0,\;\, \vec{k} =\vec{k}_E,\;\;
   {\rm d}^D k = i\,{\rm d}^D k_E   
\end{align}
where the new integration momentum $k_E$ has a positive-definite metric:
\begin{align}
   k^2 = -k_E^2, \;\; \;
   k_E^2 = (k^0_E)^2 + \cdots + (k_E^{D-1})^2 \, .   
\end{align}
This leads us to a Euclidean integral over $k_E$,
\begin{align}
\ipi\, B_0 & = i \int^1_0 {\rm d}x \frac{\m^{4-D}}{(2\pi)^D}
\int \frac{{\rm d}^Dk_E}{(k_E^2 + Q)^2 }
\end{align}
where
\begin{align}
\label{Qdef}
 Q & = x^2q^2-x(q^2+m_1^2-m_2^2)+m_1^2 - i\varepsilon
\end{align}
is a constant with respect to the $k_E$-integration.
This $k_E$-integral is of the general type
$$ \int \frac{{\rm d}^Dk_E}{(k_E^2+Q)^n}   $$
of rotational-invariant integrals in a $D$-dimensional Euclidean
space. They can be evaluated using $D$-dimensional polar
coordinates (with the substitution $k_E^2 = R$),
$$ \int\frac{{\rm d}^Dk_E}{(k_E^2+Q)^n}\, =\, \frac{1}{2}
\int {\rm d}\Omega_D \int^{\infty}_0 {\rm d}R\, R^{\frac{D}{2}-1} \,
\frac{1}{(R+Q)^n}  \, ,  $$
yielding
\begin{align}
\label{Dintegral}
\frac{\m^{4-D}}{(2\pi)^D} \int \frac{{\rm d}^Dk_E}{(k_E^2+Q)^n} \,& = \,
\frac{\m^{4-D}}{(4\pi)^{D/2}} \cdot
\frac{\Gamma(n-\frac{D}{2})}{\Gamma(n)}\cdot Q^{-n+\frac{D}{2}} \, .
\end{align}
The singularities of the initially 4-dimensional integrals are now
recovered 
as poles of the $\Gamma$-function for $D=4$ and  values
$n  \leq 2$.

\noi
Although the l.h.s.\
of~(\ref{Dintegral}) as a $D$-dimensional integral is sensible
   only for integer values of $D$, the r.h.s. has
an analytic continuation in the variable $D$: it is well defined for
all complex values $D$ with $n-\frac{D}{2}\neq 0,-1,-2,\dots$,
 in particular for
\begin{align}
D & = 4 -\eps \;\;\; \mbox{ with } \eps >  0 \, . 
\end{align}
For physical reasons we are interested in the vicinity of $D=4$.
Hence  we consider the limiting case $\eps \to 0$
and perform an expansion around $D=4$ in powers of $\eps$.
For this task we need the following properties of the
$\Gamma$-function at  $x\to 0$:
\begin{align}
&  \Gamma(x)   =   \frac{1}{x}\, - \,\g\, +\, {\cal O}(x) \, , \nn \\
&   \Gamma(-1+x)   =   -\,\frac{1}{x} \,+\,\g \,-\, 1 \,+\,{\cal O}(x)\, ,
\end{align}
with Euler's constant
\begin{align}
 \g & = -\,\Gamma'(1) = 0.577\dots 
\end{align}
For the integral $B_0$ we evaluate the integrand of
the $x$-integration in~(\ref{Bintegral}) 
with help of~(\ref{Dintegral}) as follows:
\begin{eqnarray}
\label{eq:epsexpansion}
\frac{\m^{\eps}}{(4\pi)^{2-\eps/2}} \cdot
\frac{\Gamma(\frac{\eps}{2})}{\Gamma(2)} \cdot
 Q^{-\eps/2}   & = & \frac{1}{16\pi^2} \left(
\frac{2}{\eps} -\g + \log 4\pi -\log\frac{Q}{\m^2} \right)
+\, {\cal O}(\eps)  \nn \\
 & = & \frac{1}{16\pi^2} \left( \Delta -\log\frac{Q}{\m^2} \right)
        +\, {\cal O}(\eps) \, .
\end{eqnarray}
Since the $O(\eps)$ terms vanish in the limit $\eps\to 0$ we can skip
them in the following. Insertion into~(\ref{Bintegral})
with $Q$ from~(\ref{Qdef}) yields
\begin{align}
\label{B0}
B_0(q^2,m_1,m_2) & =  \Delta \, -
\int^1_0 {\rm d}x\,\log\frac{x^2 q^2-x(q^2+m_1^2-m_2^2)+m_1^2-i\varepsilon}{\m^2} .
\end{align}
The remaining integration is elementary and the result can be
expressed in terms of logarithms.
The explicit analytic formula can be found, for example, in Ref.~\cite{denneretal}.

In the expression~(\ref{eq:epsexpansion}) above
we have introduced the abbreviation 
\begin{align}
\Delta  = \frac{2}{\eps} -\g + \log 4\pi 
\end{align}
for the pole singularity 
combined with the two purely numerical
terms that always go together in dimensional regularization.
In the $\overline{MS}$ renormalization scheme, the counter terms
required for renormalization
cancel just these $\Delta$ terms appearing in the 
calculation of amplitudes at the loop level.



\begin{thebibliography}{99}


\bibitem{QCD}
D. Gross and F. Wilczek, {\em Phys.\ Rev.\ Lett.}~{\bf 30}  (1973) 1343; 
 \PRD~{\bf D8}  (1973) 3633; \\
H.D. Politzer, {\em Phys.\ Rev.\ Lett.}~{\bf 30}  (1973) 1346;\\
H. Fritzsch, M. Gell-Mann and H. Leutwyler, \PLB~{\bf B47}  (1973) 365; \\
H. Fritzsch and M. Gell-Mann, {Current algebra: quarks and what else?},
in {\em 16th Int.\ Conference on High-Energy Physics,}
Chicago, 1972, Eds. J.~D.~Jackson, A.~Roberts, and R.~Donaldson (National Accelerator Laboratory, Batavia, IL, 1973), p.~135 [arXiv:hep-ph/0208010].



\bibitem{gsw}
S.L. Glashow, \NPB~{\bf B22}  (1961) 579;\\
S. Weinberg, \PRL~{\bf  19}  (1967) 1264; \\
A. Salam, in {\it Proceedings of the  8th Nobel Symposium}, 
Ed.\ N. Svartholm (Almqvist and Wiksell, Stockholm, 1968).

\bibitem{gim}
S.L. Glashow, I. Iliopoulos and L. Maiani, \PRD~{\bf D2}  (1970) 1285.

\bibitem{cabibbo}
N. Cabibbo, \PRL~{\bf 10}  (1963) 531; \\
M. Kobayashi and K. Maskawa, {\it Prog.\ Theor.\ Phys.}\ {\bf 49} (1973) 652 .





\bibitem{textbooks}
F. Halzen and A.D. Martin, {\em Quarks and Leptons: An Introductory Course
in Modern Particle Physics} (Wiley, New York, 1984); \\
M.E. Peskin and D.V. Schroeder, {\em An Introduction to
Quantum Field Theory} (Addison-Wesley, Reading, MA,~1995);\\
A. Denner, M. B\"ohm and H. Joos, {\em Gauge Theories of the
Strong and Electroweak Interactions}  (B.G.~Teubner, 
Stuttgart/Leipzig/Wiesbaden, 2001).

\bibitem{Gavan}
G. Salam, {\em A primer on QCD for hadron colliders}, these proceedings.

\bibitem{Lindner}
M. Lindner, {\em Lectures on Neutrino Physics}, lectures at this school.

\bibitem{PDG08}
C. Amsler {\it et al.}~[Particle Data Group], 
{\em Phys.\ Lett.}~{\bf B667} (2008) 1.  

\bibitem{MSbar}
G. 't Hooft, \NPB~{\bf B61}  (1973) 455;\\
W.A. Bardeen, A.J. Buras, D.W. Duke and T. Muta, \PRD~{\bf D18}  (1978) 3998.


\bibitem{denneretal}
M. B\"ohm, W. Hollik and H. Spiesberger, {\em Fortschr.\ Phys.}~{\bf 34}  (1986) 687;\\
W. Hollik, {\em Fortschr.\ Phys.}~{\bf 38}  (1990) 165;\\
A. Denner, {\em Fortschr.\ Phys.}~{\bf 41}  (1993) 307; 
{Techniques for the calculation of electroweak radiative corrections
at the one-loop level and results for W-physics at LEP200}, 
arXiv:hep-ph/07091075.





\bibitem{DeltaAlpha}
F. Jegerlehner, {\em J. Phys.}~{\bf G29}  (2003) 101; \\
{\tt http://www-com.physik.hu-berlin.de/ $\tilde{}\!$ fjeger/hadr5n09.f}; \\
H. Burkhardt and B. Pietrzyk, \PRD~{\bf D72}  (2005) 057501.


\bibitem{teubneretal}
T. Teubner, K. Hagiwara, R. Liao, A.D. Martin and D. Nomura,
{\em Update of g-2 of the muon and $\Delta\alpha$},
arXiv:1001.5401 [hep-ph].


\bibitem{ewgr} D. Bardin {\it et al.}, {\em Electroweak Working Group Report},  
              arXiv:hep-ph/9709229; \\ 
              {\em Reports of
             the Working Group on Precision Calculations
             for the $Z$ Resonance}, 
              Eds. D.~Bardin, W.~Hollik and G.~Passarino, CERN 95-03 (1995).

\bibitem{HOdeltar}
A. Freitas, W. Hollik, W. Walter and G. Weiglein,
\PLB~{\bf B495} (2000) 338;
\NPB~{\bf B632}  (2002) 189; \\
M. Awramik and M. Czakon, \PRL~{\bf 89}  (2002) 241801;
                       \PLB~{\bf 568}  (2003) 48; \\
A.~Onishchenko and O.~Veretin, \PLB~{\bf B551}  (2003) 111; \\
M.~Awramik, M.~Czakon, A.~Onishchenko and O.~Veretin, 
\PRD~{\bf D68}  (2003) 053004.

\bibitem{beyondtwoloop}
J.~van der Bij, K.~Chetyrkin, M.~Faisst, G.~Jikia and T.~Seidensticker,
\PLB~{\bf B498}  (2001) 156; \\
M.~Faisst, J. K\"uhn, T.~Seidensticker and O. Veretin, 
           \NPB~{\bf B665}  (2003) 649; \\
Y. Schroder and M. Steinhauser, \PLB~{\bf B622}  (2005) 124;\\
K.~Chetyrkin, M.~Faisst, J.~K\"uhn and P.~Maierhofer,
\PRL~{\bf 97}  (2006) 102003; \\
R.~Boughezal and M.~Czakon, \NPB~{\bf B755}  (2006) 221; \\
R.~Boughezal, J.B.~Tausk and J.J. van der Bij, 
{\it Nucl.\ Phys.}~{\bf B725} (2005) 3;  {\bf B713}  (2005) 278.

\bibitem{lepewwg}
The LEP Collaborations, the LEP Electroweak Working Group, 
the Tevatron Electroweak Working Group,
the SLD Electroweak and Heavy Flavour Working Groups,
{Precision electroweak measurements and constraints on the
Standard Model},
CERN-PH-EP/2009-023; \\
{\tt http://www.cern.ch/LEPEWWG}.


\bibitem{HOsineff}
  M.~Awramik, M.~Czakon, A.~Freitas and G.~Weiglein,
  {\it Phys.\ Rev.\ Lett.}~{\bf 93} (2004) 201805;\\
  M.~Awramik, M.~Czakon and A. Freitas, \PLB~{\bf B642}  (2006) 563; 
  {\em JHEP} {\bf 0611}  (2006) 048; \\
  W. Hollik, U. Meier and S. Uccirati, 
  \NPB~{\bf B731}  (2005) 213; \PLB~{\bf B632}  (2006) 680;  
  \NPB~{\bf B765}  (2007) 154.

\bibitem{BNL} H.N. Brown {\it et al.}, \PRL~{\bf 86}  (2001) 2227; \\
G.W. Bennett {\it et al.}, \PRL~{\bf 89} (2002) 101804; 
\PRL~{\bf 92}  (2004) 161802.

\bibitem{JN}
F. Jegerlehner, 
{\em Springer Tracts in Modern Physics} {\bf 226} (2007); \\
F. Jegerlehner and A. Nyffeler, {\em Phys. Rep.}~{\bf 477}  (2009) 1.


%

\bibitem{lepreports}
{\em LEP2 Monte Carlo Workshop: Report of the Working Groups on Precision Calculations
for LEP2 Physics}, Eds.\ S. Jadach, G.~Passarino and R. Pittau,
CERN 2000-009 (2000).

\bibitem{GFITTER}
H. Fl\"acher, M. Goebel, J. Haller, A. H\"ocker, K. Moenig and J. Stelzer,
{\em Eur.\ Phys.\ J.}~{\bf C60}  (2009) 543.


\bibitem{LHCworkshop}
S. Haywood {\it et al.}, {Electroweak physics}, arXiv:hep-ph/0003275, 
in:
{\em Standard Model Physics (and more) at the LHC}, 
Eds.\ G. Altarelli and M. Mangano,
CERN 2000-004 (2000), pp. 117--230.


\bibitem{TESLA}
A. Accomando {\it et al.}, {\em Phys. Rep.} {\bf 299}  (1998) 1; \\ 
J.A. Aguilar-Saavedra {\it et al.},
{\em TESLA Technical Design Report, 
Part 3, Physics at an $e^+e^-$ Linear Collider}, arXiv:hep-ph/0106315.

\bibitem{gigaZ} J. Erler, S. Heinemeyer,  W. Hollik, G. Weiglein and P.M. Zerwas,
                \PLB~{\bf B486}  (2000) 125.


\bibitem{TevHiggsWG}
The TEVNPH Working Group, 
{Combined CDF and D0 upper limits on Standard-Model Higgs-Boson
     production with 2.1-5.4 fb$^{-1}$ of data},
     arXiv:0911.3930 [hep-ex];\\
T. Aaltonen {\it et al.}\ [CDF and D0 Collaborations],
\PRL~{\bf 104}  (2010) 061802. 

\bibitem{lindner} 
L. Maiani, G. Parisi and R. Petronzio,
                 \NPB~{\bf B136}  (1979) 115; \\
                  N. Cabibbo, L. Maiani, G. Parisi and R. Petronzio,
                   \NPB~{\bf B158}  (1979) 259; \\ 
                 R. Dashen and H. Neuberger, \PRL~{\bf 50}  (1983) 1897; \\
                 D.J.E. Callaway, \NPB~{\bf B233} (1984) 189; \\
                 M.A. Beg, C. Panagiotakopoulos and A. Sirlin,
                 \PRL~{\bf 52}  (1984) 883; \\
                 M. Lindner, \ZPC~{\bf C31}  (1986) 295.

\bibitem{higgsbounds}
         M. Lindner, M. Sher and H. Zaglauer, 
                 \PLB~{\bf B228}  (1989) 139; \\
        G. Altarelli and G. Isidori, \PLB~{\bf B337}  (1994)141; \\
        J.A. Casas, J.R. Espinosa and M. Quiros,
        \PLB~{\bf B342}  (1995) 171 and {\bf B382} (1996) 374 .

\bibitem{hambye} T. Hambye and K. Riesselmann, 
            \PRD~{\bf D55}  (1997) 7255.

\bibitem{lattice} 
J.~Kuti, L. Lin and Y. Shen, \PRL~{\bf 61}  (1988) 678;\\
        P. Hasenfratz {\it et al.}, \NPB~{\bf B317}  (1989) 81;\\
        M. L\"uscher and P. Weisz, \NPB~{\bf B318}  (1989) 705; \\
        M. G\"ockeler, H. Kastrup, T. Neuhaus and F. Zimmermann,
        \NPB~{\bf B404}  (1993) 517.

\bibitem{decaylattice}   
         M. G\"ockeler, H. Kastrup, J. Westphalen and F. Zimmermann,
          \NPB~{\bf B425}  (1994) 413.

\bibitem{ghinculov}  A. Ghinculov, \NPB~{\bf B455}  (1995) 21; \\
                   A. Frink, B. Kniehl and K. Riesselmann, 
                  \PRD~{\bf D54}  (1996) 4548.

\bibitem{riess} K. Riesselmann, 
{Limitations of a Standard Model Higgs boson}, 
arXiv:hep-ph/9711456.


\bibitem{binoth} T. Binoth, A. Ghinculov and J.J. van der Bij,
             \PRD~{\bf D57}  (1998) 1487; 
             \PLB~{\bf B417}  (1998) 343.
\bibitem{Hff} L. Durand, B.A. Kniehl and K. Riesselmann,
             \PRL~{\bf 72}  (1994) 2534 [Erratum-ibid. {\bf 74}  (1995) 1699]; \\
          A. Ghinculov, \PLB~{\bf B337}  (1994) 137 [Erratum-ibid. {\bf B346}  (1995) 426]; \\
          V. Borodulin and G. Jikia, \PLB~{\bf B391}  (1997) 434. 


\bibitem{TeVtoLHC}
U. Aglietti {\it et al.}, {Tevatron-for-LHC report: Higgs},
arXiv:hep-ph/0612172.

\bibitem{BRs}
{\tt http://diablo.phys.northwestern.edu/pc/brs.html}

\bibitem{MSSM}
H.P. Nilles, {\em Phys. Rep.}  {\bf 110}, 1 (1984);\\
H. Haber and G. Kane, {\em Phys. Rep.}  {\bf 117}  (1985) 75.

\bibitem{djouadi}
A. Djouadi, {\em Phys. Rep.}  {\bf 459}  (2008) 1.


\bibitem{feynhiggs}
S. Heinemeyer, W. Hollik and G. Weiglein, \PLB~{\bf B440}  (1998) 296;
{\em Eur.\ Phys. J.}~{\bf C9}  (1999) 343; 
{\em Comput.\ Phys.\ Commun.}~{\bf 124} (2000) 76 ; \\
G. Degrassi, S. Heinemeyer, W. Hollik, P. Slavich and G. Weiglein,
{\em Eur.\ Phys. J.}~{\bf C28} (2003) 133 ; \\
M. Frank, T. Hahn, S. Heinemeyer, W. Hollik, H. Rzehak and G. Weiglein, 
{\em JHEP}~{\bf 0702} (2007) 047; \\
T. Hahn, S. Heinemeyer, W. Hollik, H. Rzehak and G. Weiglein, 
{\em Comput.\ Phys.\ Commun.}~{\bf 180} (2009) 1426.

\bibitem{mssm_ren}
W. Hollik, E. Kraus, M. Roth, C. Rupp, K. Sibold and D. St\"ockinger,
\NPB~{\bf B639} (2002) 3.


\bibitem{Heinemeyer:2004gx}
  S.~Heinemeyer, W.~Hollik and G.~Weiglein,
  {\em Phys. Rep.}~{\bf 425} (2006) 265;\\
 Heinemeyer, W. Hollik, D. St\"ockinger, A.M. Weber and G. Weiglein,
{\em JHEP} {\bf 0608} (2006) 052;\\ 
 Heinemeyer, W. Hollik, A.M. Weber and G. Weiglein,
{\em JHEP} {\bf 0804} (2008) 039.




\bibitem{stoeckinger}
D. St\"ockinger, {\em J. Phys.}~{\bf G34} (2007) 45. 


\bibitem{deboer} 
W. de Boer, A. Dabelstein, W. Hollik and W. M\"osle and U. Schwickerath, 
                 \ZPC~{\bf C75} (1997) 627; \\
W. de Boer and C. Sander,
{\it Phys.\ Lett.}  {\bf B585}  (2004) 276.

\bibitem{buchmulleretal}
O. Buchmueller {\it et al.}, \PLB~{\bf B657}  (2007) 87; 
{\em JHEP} {\bf 0809}  (2008) 117.



\end{thebibliography}
\end{document}